\documentclass[showpacs,preprintnumbers,amsmath,amssymb,floatfix,nofootinbib]{revtex4}

\usepackage{graphicx}
\usepackage{dcolumn}
\usepackage{bm}

\usepackage{pstricks}

\newcommand{\sherpa}{{\tt SHERPA}}
\newcommand{\m}{{-}}

\begin{document}
\preprint{CERN-PH-TH/2004-168}
\title{Simulating $W$/$Z$+jets production at the Tevatron}
 \homepage{http://www.physik.tu-dresden.de/~krauss/hep/}

\author{Frank Krauss}
\email{krauss@theory.phy.tu-dresden.de}
\altaffiliation[also at ]{Physics Dept., Theory Div., CERN, Geneva, Switzerland}
\author{Andreas Sch{\"a}licke}
 \email{dreas@theory.phy.tu-dresden.de}
\affiliation{Institute for Theoretical Physics, D-01062 Dresden, Germany}
\author{Steffen Schumann}
 \email{steffen@theory.phy.tu-dresden.de}
\affiliation{Institute for Theoretical Physics, D-01062 Dresden, Germany}
\author{Gerhard Soff}
 \email{soff@physik.tu-dresden.de}
\affiliation{Institute for Theoretical Physics, D-01062 Dresden, Germany}

\date{\today}
\begin{abstract}
\noindent
The merging procedure of tree-level matrix elements and the subsequent parton shower 
as implemented in the new event generator \sherpa\ will be validated for the example 
of $W$/$Z$+jets production at the Tevatron. Comparisons with results obtained from 
other approaches and programs and with experimental results clearly show that the
merging procedure yields relevant and correct results at both the hadron and parton 
levels.
\end{abstract}

\maketitle

\psset{linewidth=2pt}
\psset{unit=1pt}

\section{Introduction}
\noindent
The production of electroweak gauge bosons, e.g.\ $W^\pm$ and $Z$ bosons, is one of the most 
prominent processes at hadron colliders. Especially through their leptonic decays they leave a 
clean signature, namely either one charged lepton accompanied by missing energy for $W$ bosons 
or two oppositely charged leptons for the $Z$ bosons. The combination of clear signatures and 
copious production rates allows a measurement of some of their parameters, e.g.\ the 
$W$ mass and width, with a precision comparable with that reached at LEP2 at the Tevatron
\cite{Abachi:1995xc,Affolder:2000bp,Abazov:2002bu,unknown:2003sv,Albajar:1990hg,Alitti:1991dm,
Abe:1995wf,Abbott:1999tt,Affolder:2000mt,Abazov:2002xj}, or even better at the LHC 
\cite{ATLAS_TDR,Haywood:1999qg}. The same combination, clear signature and large production 
rate, renders them a good candidate process for luminosity measurements, especially at the LHC
\cite{Khoze:2000db,Dittmar:1997md,Martin:1999ww,Giele:2001ms}. 
This holds true in particular for $W$-bosons, since their production rate is enhanced by roughly 
an order of magnitude with respect to $Z$ production. Furthermore, especially in combination 
with additional jets, the production of gauge bosons represents a serious background to many 
other interesting processes, leading to multi particle final state topologies. The production and 
decay of pairs of top quarks or of SUSY particles may serve as illustrative examples for such 
signal processes. The special interest in this classic production process is reflected by the 
fact that it was one of the first to be calculated at next-to leading order
(NLO) \cite{Altarelli:1979ub,Kubar-Andre:1978uy,Harada:1979bj,Abad:1978ke,Humpert:ux}
and next-to-next-to leading order (NNLO) \cite{Hamberg:1990np,Harlander:2002wh}
in QCD. Recently, the first distribution related to these processes, namely the boson rapidity, 
has been calculated at NNLO \cite{Anastasiou:2003ds}. In addition to such fixed-order calculations, 
programs such as RESBOS \cite{Balazs:1997xd} have been made available, which resum 
soft gluon effects. Cross sections and distributions for $W$ or $Z$ bosons being produced together 
with jets can be evaluated at the parton level through a number of different computer codes:
specialised ones such as VECBOS \cite{Berends:1990ax}, and multi-purpose parton-level generators 
such as COMPHEP \cite{Pukhov:1999gg}, GRACE/GR@PPA \cite{Ishikawa:1993qr,GRAPPA}, 
MADGRAPH/MADEVENT \cite{Stelzer:1994ta,Maltoni:2002qb}, ALPGEN \cite{Mangano:2002ea}, and 
AMEGIC \cite{Krauss:2001iv}. All of them operate at the tree level,
at NLO the program MCFM \cite{Campbell:2002tg,Campbell:2003hd} provides cross sections and 
distributions for $W/Z+$jets for up to 2 jets.

\noindent
Apart from such techniques, based on analytical methods, event generators play a major role in 
the experimental analysis of collider experiments. In the past years, programs such as PYTHIA 
\cite{Sjostrand:1993yb,Sjostrand:2003wg} or HERWIG \cite{Corcella:2000bw,Corcella:2002jc}
proved to be successful in describing global features of boson production processes, such as 
the bosons transverse momentum or rapidity distribution. Apart from the parton shower,
which takes proper care and resums the leading and some of the subleading Sudakov logarithms, 
these programs include the first-order matrix element for the emission of an extra parton, 
implemented through a correction weight on the parton shower. Because of the different approximations 
made for their parton shower, this is realized in different manners inside the two programs, 
cf.\ \cite{Miu:1999ju} and \cite{Seymour:1995df,Corcella:2000gs}.

\noindent
In view of the need for more precise simulations, both in terms of total rates and in the 
description of exclusive final states, two quite orthogonal approaches have been developed 
recently, which aim at a systematic combination of higher-order matrix elements with the 
parton shower. The first one, called MC@NLO, provides a method to consistently match NLO 
calculations for specific processes with the parton shower. It has been implemented for the 
production of colour-singlet final states, such as $W$ and $Z$ bosons, or pairs of these bosons 
\cite{Frixione:2002ik}, or the Higgs boson, and for the production of heavy quarks 
\cite{Frixione:2003ei}. The implementations are available as a code called MC@NLO 
\cite{Frixione:2004wy} residing on top of HERWIG. The idea of this approach is to
organise the counter-terms necessary to technically cancel real and virtual infrared 
divergencies in such a way that the first emission of the parton shower is recovered. This 
allows the generation of hard kinematics configurations, which can eventually be fed into a parton 
shower Monte Carlo. The alternative approach is to employ matrix elements at the tree level 
(at leading order) for different jet multiplicities and to merge them with the parton shower. 
The basic idea in this approach is to define a region of jet production, i.e.\ hard emissions, 
and a region of jet evolution, i.e.\ soft emissions, divided by a $k_\perp$-type of jet measure 
\cite{Catani:1991hj,Catani:1992zp,Catani:1993hr}. Leading higher-order effects are added to the 
various matrix elements by reweighting them with appropriate Sudakov form factors. Formal 
independence at leading logarithmic order of the overall result on the jet measure is achieved 
by vetoing hard emissions inside the parton shower, and suitable starting conditions. This approach 
was presented for the first time \cite{Catani:2001cc} for $e^+e^-$ collisions; it has 
been reformulated for dipole cascades in \cite{Lonnblad:2001iq} and extended to hadronic collisions 
in \cite{Krauss:2002up}. The method is one of the cornerstones of the new event generator \sherpa\ 
\cite{Gleisberg:2003xi}, written entirely in C++, where it has been consistently implemented 
for nearly arbitrary processes. A case study of this method on the basis of PYTHIA and HERWIG, 
focusing on single-boson production, has been presented in \cite{Mrenna:2003if}. It should be 
noted also that a somewhat related approach has been taken by \cite{Mangano:2001xp}. There, a 
systematic mapping and tracing of partons stemming from the leading order matrix element through 
the parton shower takes proper care of leading logarithmic effects. 

\noindent
The goal of this paper is to validate the merging procedure implemented in \sherpa\ for single-boson 
production and to compare the results with those of other approaches. After a short reminder of the
merging procedure and a brief introduction to some of the implementation details in Sec.~\ref{ReMP}, 
the focus will shift on results obtained by \sherpa. The observables, that will be studied, 
are inclusive, like the transverse momentum and rapidity distribution of the bosons, and more 
exclusive, like the transverse momentum distribution of additional jets. In a first step, the 
self-consistency of the method will be checked by analysing the dependence of different observables 
on the separation cut and on the maximal number of extra jets provided by the matrix elements, see
Sec.~\ref{Self-C}. Following this the results of the merging method will be contrasted with those 
of other approaches: on the matrix element level, the jet transverse momentum distributions of 
\sherpa s reweighted matrix elements will be compared with those of a full-fledged NLO calculation 
provided by MCFM; see Sec.~\ref{ssec_vsMC}. Then, the results 
after parton showering and hadronisation will be compared with those of other event generators in 
Sec.~\ref{ssec_vsMCatNLO}, specifically with those obtained from PYTHIA and MC@NLO. Finally, the 
ability of the method to describe inclusive observables that have been measured, such as the bosons 
transverse momentum, will be exhibited in Sec.~\ref{ssec_vsData}.

\section{Realization of the merging procedure\label{ReMP}}
\subsection{Basic concepts}
\noindent
The key idea of the merging procedure \cite{Catani:2001cc,Krauss:2002up} is to separate the 
phase space for parton emission into a hard region of jet production accounted for by 
suitable tree-level matrix elements and the softer region of jet evolution covered by the 
parton showers. Then, extra weights are applied on the former and vetoes on the latter, 
so that the overall dependence on the separation cut is minimal. The separation is achieved 
through a $k_\perp$-measure; for hadron collisions a longitudinal invariant form is used
\cite{Catani:1992zp,Catani:1993hr}: two final state particles $i$ and $j$ are defined to be 
within two different jets, if their relative $k_\perp$-measure $Q^2_{ij}$, given by
\begin{equation}
  Q^2_{ij} = 2 \min\{p_{\perp,i}^2, p_{\perp,j}^2\} \,
  \left[\cosh(\eta_i-\eta_j) - \cos(\phi_i-\phi_j) \right]\,,
\end{equation}
is larger than the jet resolution scale $Q^2_{\rm cut}$. In the context of the merging procedure,
this scale $Q^2_{\rm cut}$ serves as the separation scale between the regimes of jet production and 
jet evolution. In the above equation, $\eta$ and $\phi$ denote the pseudo-rapidities and azimuthal 
angles of the two particles, respectively. In addition, each jet has to fulfil the constraint that
\begin{equation}
  Q^2_{i}  = p_{\perp,i}^2 
\end{equation}
is larger than the jet resolution scale. Apart from details concerning the merging of two jets into
one, variations of this jet definition exist. For instance, within the $k_\perp$-scheme for Tevatron, 
Run II, the $Q_{ij}$ are eventually rescaled by a ``cone-like'' factor $D$, and instead of the 
cosines just the squares of the differences are taken \cite{Blazey:2000qt}.

\noindent
The weight attached to the matrix elements takes into account the terms that would appear in a 
corresponding parton shower evolution. Therefore, a ``shower history'' is reconstructed by 
clustering the initial and final state particles stemming from the tree-level matrix element 
according to the $k_\perp$-formalism. This procedure yields nodal values, namely the different 
$k_\perp$-measures $Q^2$ where two jets have been merged into one. These nodal values can be interpreted 
as the relative transverse momentum describing the jet production or the parton splitting. The 
four-vectors of the mergers are given by the sum of their two offsprings, leading to increasingly 
massive jets. The first ingredients of the ME weight are the strong coupling constants evaluated 
at the respective nodal values of the various parton splittings, divided by the value of the strong
coupling constant as used in the evaluation of the matrix element. In general, in the matrix element 
calculation, the jet resolution scale is taken to be the renormalisation scale as well as the factorisation  
scale, guaranteeing that the weight is always smaller than 1. This allows a simple hit-or-miss method 
to yield unweighted events. The other part of the correction weight is provided by Sudakov form factors. 
At next-to leading logarithmic (NLL) accuracy the quark and gluon Sudakov form factors are given by 
(see \cite{Catani:1991hj})
\begin{align} 
  \Delta_q(Q,Q_0) &= 
  \exp\left\{-\int\limits_{Q_0}^{Q} dq \,\Gamma_q(q,Q)\right\} \;,\\
  \Delta_g(Q,Q_0) &= 
  \exp\left\{-\int\limits_{Q_0}^{Q} dq 
    \left[ \Gamma_g(q,Q) + \Gamma_f(q) \right]\right\} \;,
\end{align}
where $\Gamma_{q,g,f}$ are $q\to qg$, $g \to gg$ and $g \to q\bar q$
branching probabilities
\begin{align}
  \Gamma_q(q,Q) &= \frac{2 C_F}{\pi}\frac{\alpha_{\rm S}(q)}{q}
  \left( \ln \frac Q q - \frac 3 4 \right) \;,\\
  \Gamma_g(q,Q) &= \frac{2 C_A}{\pi}\frac{\alpha_{\rm S}(q)}{q}
  \left( \ln \frac Q q - \frac{11}{12} \right) \;,\\
  \Gamma_f(q) &= \frac{N_f}{3 \pi}\frac{\alpha_{\rm S}(q)}{q} \;.
\end{align}
A Sudakov form factor yields the probability for no emission (resolvable at scale $Q_0$) 
during the evolution from a higher scale $Q$ to a lower scale $Q_0$. The ratio of two Sudakov 
form factors $\Delta(Q,Q_0)/\Delta(q,Q_0)$ then gives the probability for no emission resolvable 
at a scale $Q_0$ during the evolution from $Q$ to $q$. 

\noindent
Having reweighted the matrix element,  a smooth transition between this and
the parton shower region must be guaranteed. This is achieved by choosing suitable starting 
conditions for the shower evolution of the parton ensemble. In particular, the starting scale 
of the shower evolution of a certain parton is not the jet resolution scale $Q_{\rm cut}$, but
rather the production scale of that parton. Then a veto on any emission harder than 
$Q_{\rm cut}$ properly separates the shower regime from the matrix element region.

\subsection{The algorithm in general}\label{ssec_algo}
\noindent
The extension of the merging algorithm to hadronic initial states has been proposed in 
\cite{Krauss:2002up} for the first time. Here, this algorithm will be briefly reviewed, with 
special emphasis on details of its implementation in the \sherpa\ framework. The
description of the preferred scale choice for different configurations, details of the treatment 
of matrix elements with the highest jet multiplicity, and the solution to the problem of how to
introduce off-shellness to on-shell matrix element particles will be considered in the 
following sections. 

\noindent
The merging algorithm proceeds as follows:
\begin{enumerate}
\item \label{begin_algo}
  One process out of all processes under consideration is selected according to the probability
\begin{equation}
     P^{(0)}_i=\frac{\sigma^{(0)}_i}{\sum_i \sigma^{(0)}_i}\;.
\end{equation}
  This choice provides the initial jet rates, subject to an additional Sudakov and 
  coupling weight rejection. For instance, a typical selection of processes for $W^\m$-boson 
  production would include:
  \begin{align*}
    p \bar{p}&\to jet \, jet \to  e^- \bar{\nu}_e \,, \\
    p \bar{p}&\to jet \, jet \to  e^- \bar{\nu}_e + jet \,, \\
    p \bar{p}&\to jet \, jet \to  e^- \bar{\nu}_e + jet \, jet \,, \\
    p \bar{p}&\to jet \, jet \to  e^- \bar{\nu}_e + jet \, jet \, jet \,.
  \end{align*}
  The cross sections $\sigma^{(0)}_i$ are calculated using the corresponding tree-level matrix 
  elements; the only phase-space restriction is given by the $k_\perp$-measure. The renormalisation scale
  $\mu_R$ and the factorisation scale  $\mu_F$ are fixed to the cut-off scale $Q_{\rm cut}$, 
  with an exception only for the process with the highest number of jets, cf.\ Sec.\ 
  \ref{sec_highest_multi}. 
\item
  Having chosen a single process, the respective momenta are distributed according to the
  corresponding differential cross section.
\item \label{step_cluster}
  The nodal values $q_i$ are determined. In doing so a corresponding parton shower history is 
  reconstructed. The backward clustering procedure is guided by the $k_\perp$-measure, respecting
  additional constraints:
  \begin{itemize}
    \item Unphysical combinations like ($qq$) and ($\bar q \bar q$) are ignored. Within the 
      \sherpa\ framework this is implemented by employing the knowledge of the Feynman diagrams 
      contributing to the process under consideration. Thus, ``unphysical'' translates
      into the non-existence of a corresponding Feynman amplitude.
    \item When an outgoing parton of momentum $p_j$ is to be clustered with a beam, the 
      $k_\perp$-measure does not provide the information as to which beam it has to be clustered. In
      general the beam with the same sign as its longitudinal momentum component is preferred. 
      In addition the new momentum given by $ p'_i = p_i - p_j $ must exhibit a positive energy in
      the frame where the initial state shower is performed.
  \end{itemize}
\item
  The backward clustering stops with a $2\to2$ process. The hardest scale of this ``core'' 
  process has to be found. It depends both on the process and its kinematics (cf.\ Sec.\
  \ref{sec_scale_choice}).
\item 
  The weight is determined, employing the nodal values $q_i$, according to the following rules:
  \begin{itemize}
    \item For every internal (QCD) line with nodal values $q_i$ and $q_j$ for its production and 
      its decay, a factor $\Delta(q_i, Q_{\rm cut})/ \Delta(q_j,Q_{\rm cut})$ is added. For
      outgoing lines, a factor $\Delta(q_i, Q_{\rm cut})$ is added.
    \item For every QCD node a factor $\alpha_s(q_i)/\alpha_s(Q_{\rm cut})$ is added.
  \end{itemize}
\item 
  The event is accepted or rejected according to this weight. If the event is rejected, the 
  procedure starts afresh, with step \ref{begin_algo}.
\item 
  The initial and final state showers emerges from the core $2 \to 2$ process. Matrix element 
  branchings are included as predetermined branchings within the shower. The starting conditions 
  are determined by the clustering performed before (in step \ref{step_cluster}), i.e.\ the 
  evolution of a parton starts at its production scale\footnote{Since a virtuality-ordered shower 
    is employed within \sherpa, the virtuality of its predecessor, i.e.\ its invariant mass, is 
    used.}. The first emission from the initial state shower has to take the factorisation
  scale $\mu_F$ into account, used during the matrix element
  calculation. 
\item
  Veto any emission with a $k_\perp$ above the jet resolution scale $Q_{\rm cut}$.
\end{enumerate}

\subsection{Special case : $W$-boson production at hadron colliders}
\label{sec_scale_choice}
\noindent
The algorithm described above and, in particular, the incorporated scale choices, will be 
illustrated with a few examples dealing with $W$-boson production:

\noindent
The leading order contributions to $W^\m$ production are of the Drell--Yan type, i.e. 
processes of the form
\[
q \,\bar{q}' \to e \bar{\nu}_e\;.
\]
Clustering does not take place, since this is already a $2 \to 2$ process. Furthermore, there 
is no strong coupling involved, and the rejection weight is given by two quark Sudakov factors
only:
\begin{equation}
  {\cal W} = \Delta_{q}(Q,Q_{\rm cut})\,\Delta_{\bar{q}'}(Q,Q_{\rm cut})\,.
\end{equation}
The hard scale $Q$ is fixed by the invariant mass of the fermion pair
$Q^2 = M_{e \bar{\nu}_e}^2$.
\begin{figure}
\includegraphics[width=230pt]{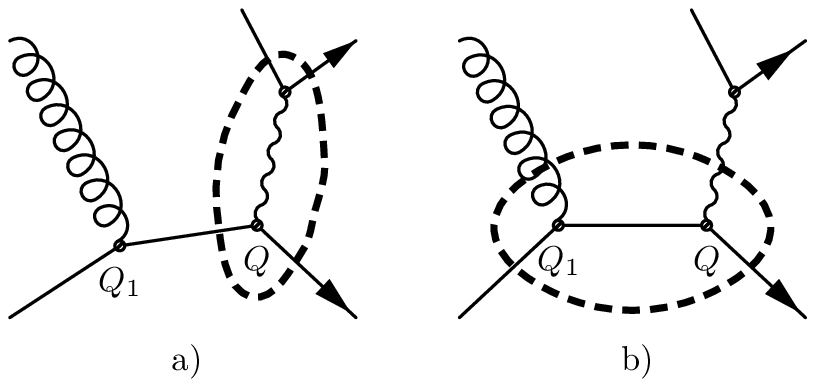}
\caption{\label{fig_w1jet}Two possible cluster configurations of a W+1 jet event. 
  The dashed line highlights the hard $2 \to 2$ process.}
\end{figure}
\noindent
Possible configurations resulting from the clustering of $W+1$jet events are exhibited in 
Fig.\ \ref{fig_w1jet}. The hard $2 \to 2$ process either is again a Drell--Yan process (Fig.\ 
\ref{fig_w1jet}a) or of the type $q\bar{q}' \to gW$ (Fig.\ \ref{fig_w1jet}b). The weight in
the first case reads:
\begin{equation}
  {\cal W} = \Delta_{q}(Q,Q_{\rm cut})\,\Delta_{\bar{q}'}(Q,Q_{\rm cut})
  \,\Delta_{g}(Q_1,Q_{\rm cut}) \frac{\alpha_s(Q_1)}{\alpha_s(Q_{\rm cut})}\;,
\end{equation}
where $Q^2 = M_{e \bar{\nu}_e}^2$ and the nodal value $Q_1$ is given by the $k_\perp$-algorithm. 
For this configuration the gluon jet tends to be soft, i.e.\ $Q_1$ preferentially is close to 
$Q_{\rm cut}$. The second configuration differs from the first only by the result of
the clustering. The transverse momentum of the gluon jet $p_{\perp,g}^2$ now is of the order 
of the $W$-boson mass or larger. The weight looks still the same only the scale definitions 
are altered. In such a case, the hard scale is now given by
\begin{equation}
  Q^2 = p_{\perp,g}^2 + M_{e \bar{\nu}_e}^2 \,,
\end{equation}
i.e.\ the transverse mass of the $W$. Also, the nodal value $Q_1$ has not been determined 
by the cluster algorithm, since it belongs to the (in principle unresolved) core process. 
A natural choice is the transverse momentum of the corresponding jet
\begin{equation}
  Q_1 = p_{\perp,g} \;.
\end{equation}
These scale definitions guarantee a smooth transition between the two regimes, i.e.\ from the 
case where the gluon is soft to a case where the gluon is hard.

\begin{figure}
\includegraphics[width=230pt]{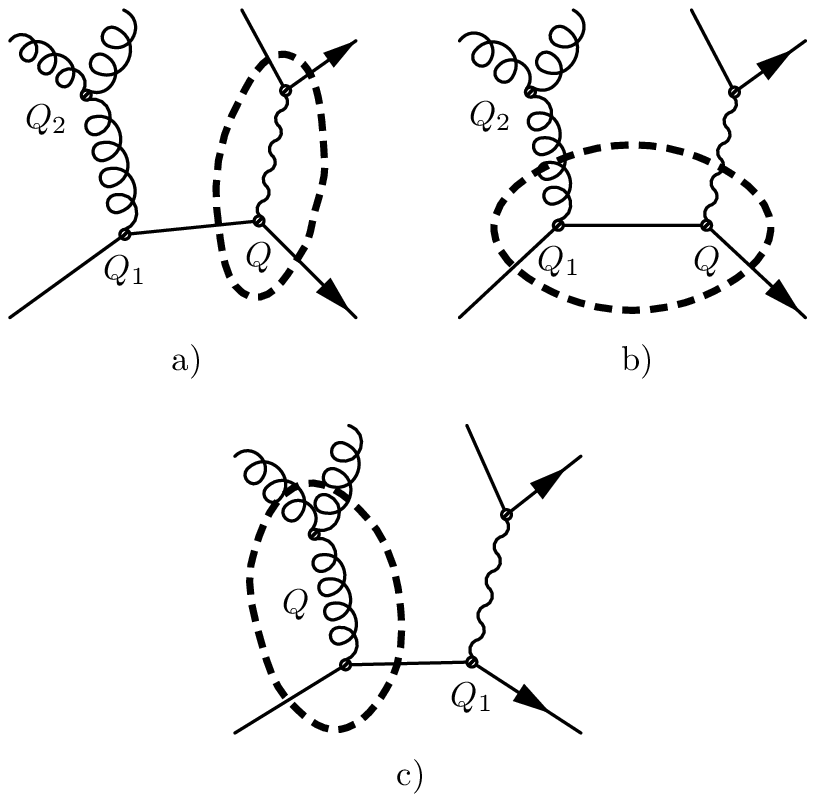}
\caption{\label{fig_w2jet}Three possible cluster configurations of a W+2 jet event. The dashed 
  line highlights the hard $2 \to 2$ process, being either of Drell--Yan type (a), 
  a vector boson production (b) or a pure QCD process (c).}
\end{figure}

\noindent
More complicated processes involve the production of at least two extra jets. There are 
many processes contributing to this category. Some illustrative examples are displayed 
in Fig.\ \ref{fig_w2jet}.
Cases a) and b) of Fig.\ \ref{fig_w2jet} are very similar to the example with one extra jet
only. The corresponding weight reads:
\begin{equation}
  {\cal W} = \Delta_{q}(Q,Q_{\rm cut})\,\Delta_{\bar{q}'}(Q,Q_{\rm cut})\,\Delta_{g}(Q_1,Q_{\rm cut})\,
  \Delta_{g}(Q_2,Q_{\rm cut}) 
  \frac{\alpha_s(Q_1)}{\alpha_s(Q_{\rm cut})}\,\frac{\alpha_s(Q_2)}{\alpha_s(Q_{\rm cut})}\;.
\end{equation}
The nodal value $Q_2$ is given by the $k_\perp$-algorithm. The scales $Q_1$ and $Q$ are chosen 
as in the one-jet case.

\noindent
In contrast a new situation arises when a pure QCD process has been chosen as the ``core'' 
$2 \to 2$ process, see Fig.\ \ref{fig_w2jet}c). Since the ``core'' process is not resolved, 
there is only one scale available, $Q^2=(2 s t u)/(s^2+t^2+u^2)\approx p_\perp^2$, the
transverse momentum of the outgoing jets. The correction weight consequently reads:
\begin{equation}
  {\cal W} = \Delta_{q}(Q,Q_{\rm cut})\,\frac{\Delta_{q}(Q,Q_{\rm cut})}{\Delta_{q}(Q_1,Q_{\rm cut})}\,
  \Delta_{\bar{q}'}(Q_1,Q_{\rm cut})
  \,[\Delta_{g}(Q,Q_{\rm cut})]^2 \left[\frac{\alpha_s(Q)}{\alpha_s(Q_{\rm cut})}\right]^2\,.
\end{equation}

\noindent
The extension to higher multiplicities is straightforward. However, the number of extra jets 
accounted for by matrix elements is limited. This limitation in available MEs enforces a 
specific treatment of the processes with the highest multiplicity.

\subsection{The highest multiplicity treatment}
\label{sec_highest_multi}
In general, the initial cross sections $\sigma^{(0)}_i$ used in step \ref{begin_algo} of the 
merging algorithm above are defined by
\begin{equation}
  \sigma^{(0)}_i = \int dx_1\, dx_2\, d\Omega\; f_1(x_1,\mu_F) f_2(x_2,\mu_F)
       \left|{\cal M}_i\right|^2\,,
\end{equation}
where $d\Omega$ represents the appropriate invariant phase-space element and ${\cal M}_i$ is
the Feynman amplitude for the respective process. The choice $\mu_F=Q_{\rm cut}$ together with the 
Sudakov factors and the coupling weight leads to a modified cross section 
$\sigma_i = {\cal W}\,\sigma^{(0)}_i$. Adding all cross sections with the same number of strong 
particles yields the cross section for production processes accompanied by -- exclusively -- $n$ jets, 
\begin{equation}
  \sigma_{n-jet}^{({\rm excl})}= \sum_{i(n\,jet)} \sigma_i \;.
\end{equation}
Of course the number of extra jets that can be considered in this respect is limited by the 
available matrix elements; in \sherpa, this number is usually in the range of three to four.
In order to compensate for all the omitted processes with more jets, the treatment of processes 
with the highest number of extra jets differs slightly from the handling of lower jet multiplicities.
The changes are as follows:
\begin{itemize}
\item the factorisation scale is set dynamically to $\mu_F = Q_{\rm min}$, i.e.\ to the smallest 
  nodal value as determined by the $k_\perp$-algorithm,
\item the resolution scale $Q_{\rm cut}$ of the Sudakov weights is
  also replaced by $Q_{\rm min}$, and 
\item the shower veto is applied with $Q_{\rm min}$ instead of $Q_{\rm cut}$.
\end{itemize}
This guarantees that parton showers attached to matrix elements with the highest number of jets
are allowed to produce softer jets. In other words: the merging procedure is meant to take into 
account quantum interference effects in jet production at leading order up to a maximal
number of jets; any softer jet is left to the parton shower. For the configuration shown in 
Fig.\ \ref{fig_w2jet}a), the modified Sudakov and coupling weight reads: 
\begin{equation}
 \tilde{\cal W} = \Delta_{q}(Q,Q_{\rm min})\,\Delta_{\bar{q}'}(Q,Q_{\rm min})\,
   \Delta_{g}(Q_1,Q_{\rm min})\, 
  \frac{\alpha_s(Q_1)}{\alpha_s(Q_{\rm cut})}\,\frac{\alpha_s(Q_2)}{\alpha_s(Q_{\rm cut})}\;,
\end{equation}
with lowest scale $Q_{\rm min}= Q_2$. Following this procedure, the sum of all cross sections 
$\tilde\sigma_i = \tilde{\cal W}\,\tilde\sigma^{(0)}_i$  for a
 number of jets $n$ can be interpreted as an inclusive cross section
\begin{equation}
  \sigma_{n-jet}^{({\rm incl})}= \sum_{i(n\,jet)} \tilde\sigma_i \;,
\end{equation}
i.e.\ the probability to find at least $n$ jets. Adding all exclusive
cross sections for multiplicities lower than a maximal multiplicity
$n_{\rm max}$ to the inclusive cross section for the highest
multiplicity results in a fully inclusive cross section.

\subsection{On-shell matrix elements vs. off-shell parton shower kinematics}
One subtle problem when combining matrix elements with parton showers is connected to the 
question of how to translate the matrix element kinematics, determined with on-shell (for light 
quarks usually massless) momenta, into a kinematics suitable for the virtuality-ordered parton
shower, i.e.\ invoking off-shell momenta. Within the \sherpa\ framework, this problem is dealt with 
by the parton shower in the usual fashion: to begin with, the energy fraction $z$ is determined
from on-shell kinematics, and afterwards it is reinterpreted. For details on the virtuality-ordered 
parton in \sherpa, the reader is referred to a forthcoming publication \cite{apacic-2-0}. 
However, if no further emissions are added through the parton shower, the partons stemming from 
the matrix element are left on their mass-shell and the kinematics remains unaltered. On the other 
hand, if the virtuality of one or more partons from the matrix element is increased through
secondary emissions induced by the shower, the kinematics is modified. Usually, the scales 
involved in the showering are much smaller than the scales prevalent in the matrix elements,
which are of the order of or larger than $Q_{\rm cut}$. Consequently, any manipulation of the 
kinematics tends to be mild. Nevertheless in some cases changes in the kinematics may lead 
a posteriori to a considerable change of the nodal values from the $k_\perp$-algorithm. For instance, 
the production of $W$ bosons at the Tevatron exhibits a strong asymmetry, which, to a considerable
fraction, leads to configurations of the initial state, where the emission of jets or extra partons 
is concentrated on one incoming parton only. On rare occasions, such mass effects may alter the 
smallest nodal value in such a way that it becomes smaller than $Q_{\rm cut}$. In these cases, the phase-space
separation underlying the full merging procedure is violated. Within the \sherpa\ framework, these 
events are rejected. The corresponding rejection procedure is such that 
\begin{itemize} 
\item the next event is of the same process as the one being rejected, and
\item the Sudakov weight is only applied to correct the kinematics rather than the rates.
\end{itemize}
Therefore, the jet rates determined by the prescription given in the previous section are not 
altered by the parton shower, and the phase-space separation through the $k_\perp$-algorithm is enforced.

\section{Consistency checks\label{Self-C}}
\noindent
In this section the self-consistency of the results obtained with \sherpa\ is checked by analysing 
the dependence of different observables on the key parameters of the merging procedure, namely the
separation scale $Q_{\rm cut}$ and the highest multiplicity of included matrix elements $n_{\rm max}$. 
All plots in this section correspond to $W^-$ boson production at the Tevatron, Run II; the
parameter settings can be found in the Appendix. If not stated otherwise, the distributions shown 
are inclusive hadron level results, i.e.\ no cuts have been applied. 

\subsection{Variation of the separation cut $Q_{\rm cut}$}
\noindent
In all figures, the black, solid line represents the total inclusive result as obtained by 
\sherpa. A vertical dashed line indicates the respective separation
cut $Q_{\rm cut}$, which has been 
varied between 10~GeV and 50~GeV. To guide the eye, all plots also show the same observable as 
obtained with a separation cut $Q_{\rm cut}=20$~GeV,  
shown as a dashed black curve. The coloured lines give the contributions of different multiplicity 
processes. Note that the separation cut always marks the transition between $n$-jet and $n+1$-jet 
matrix elements.
\begin{figure*}[h]
\begin{center}
\begin{pspicture}(400,275)
\put(250,125){\includegraphics[width=5.5cm]{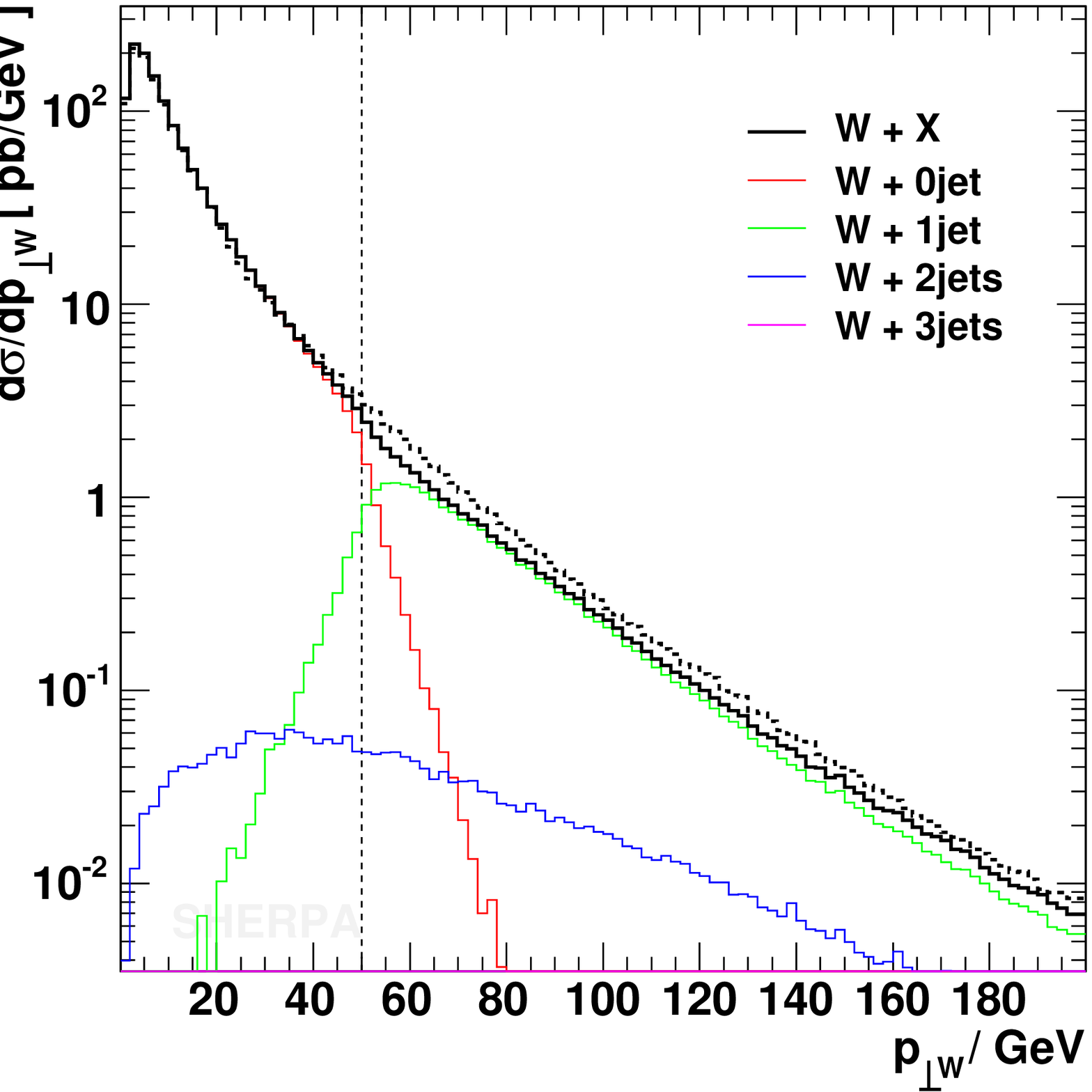}}
\put(125,125){\includegraphics[width=5.5cm]{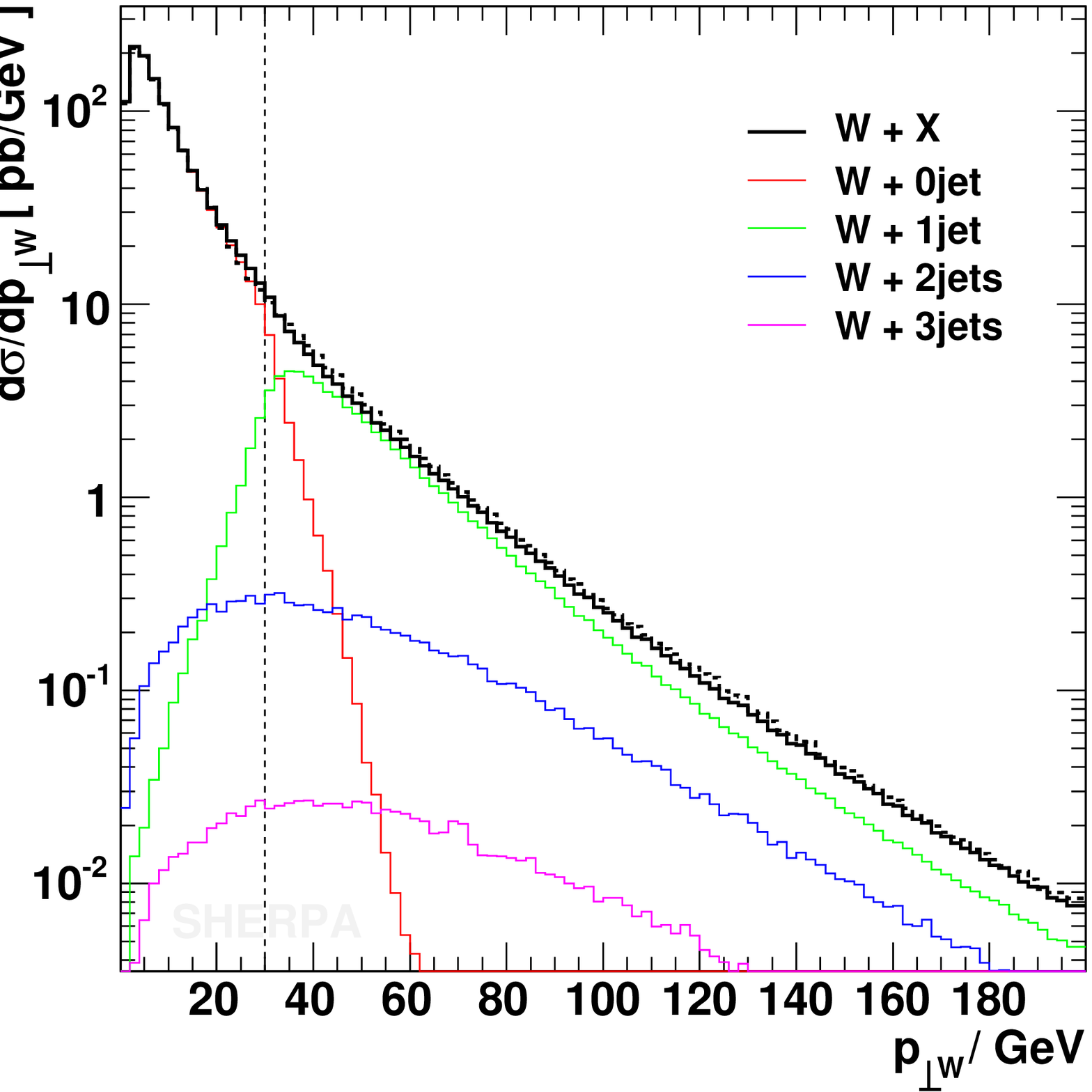}}
\put(0,125){\includegraphics[width=5.5cm]{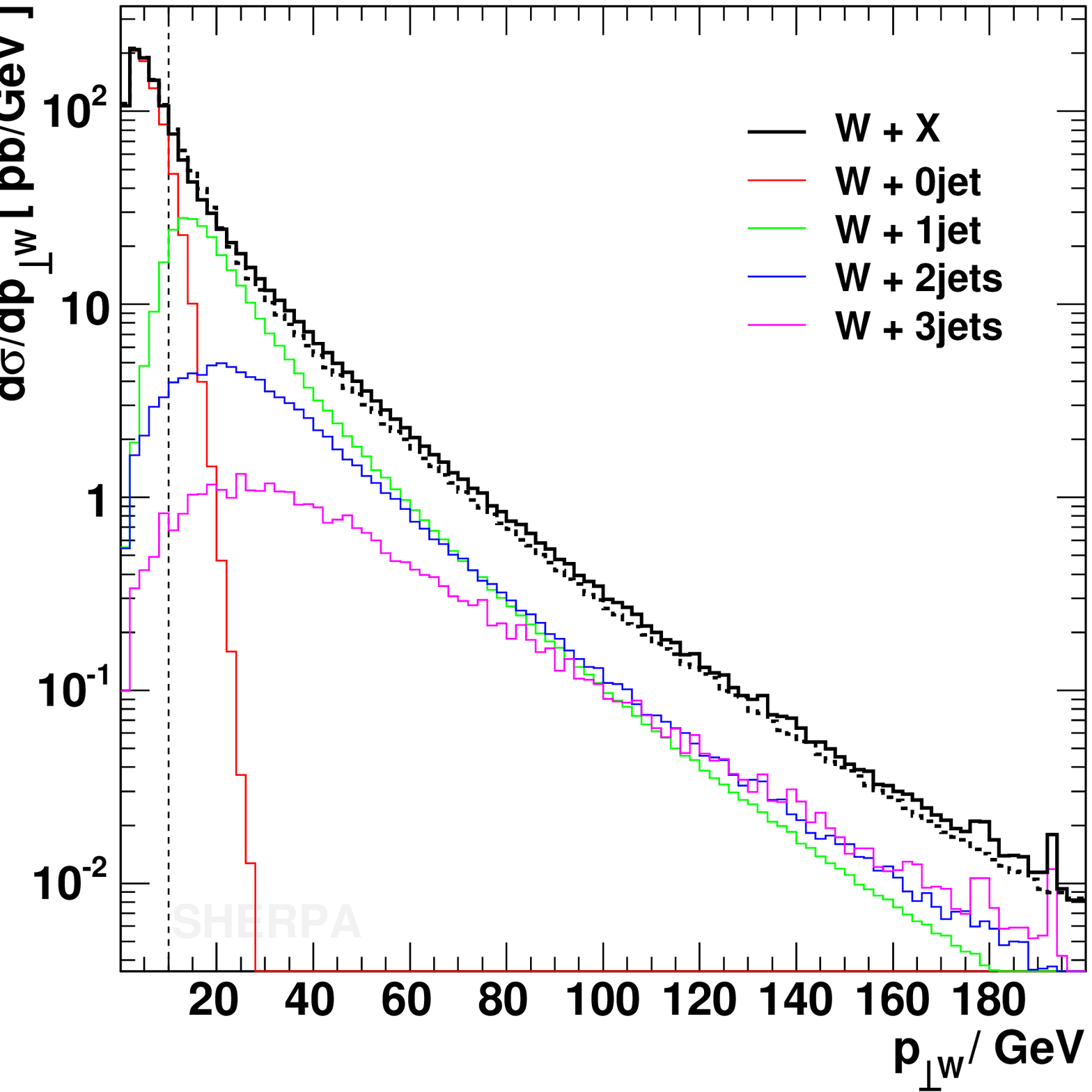}}
\put(250,0){\includegraphics[width=5.5cm]{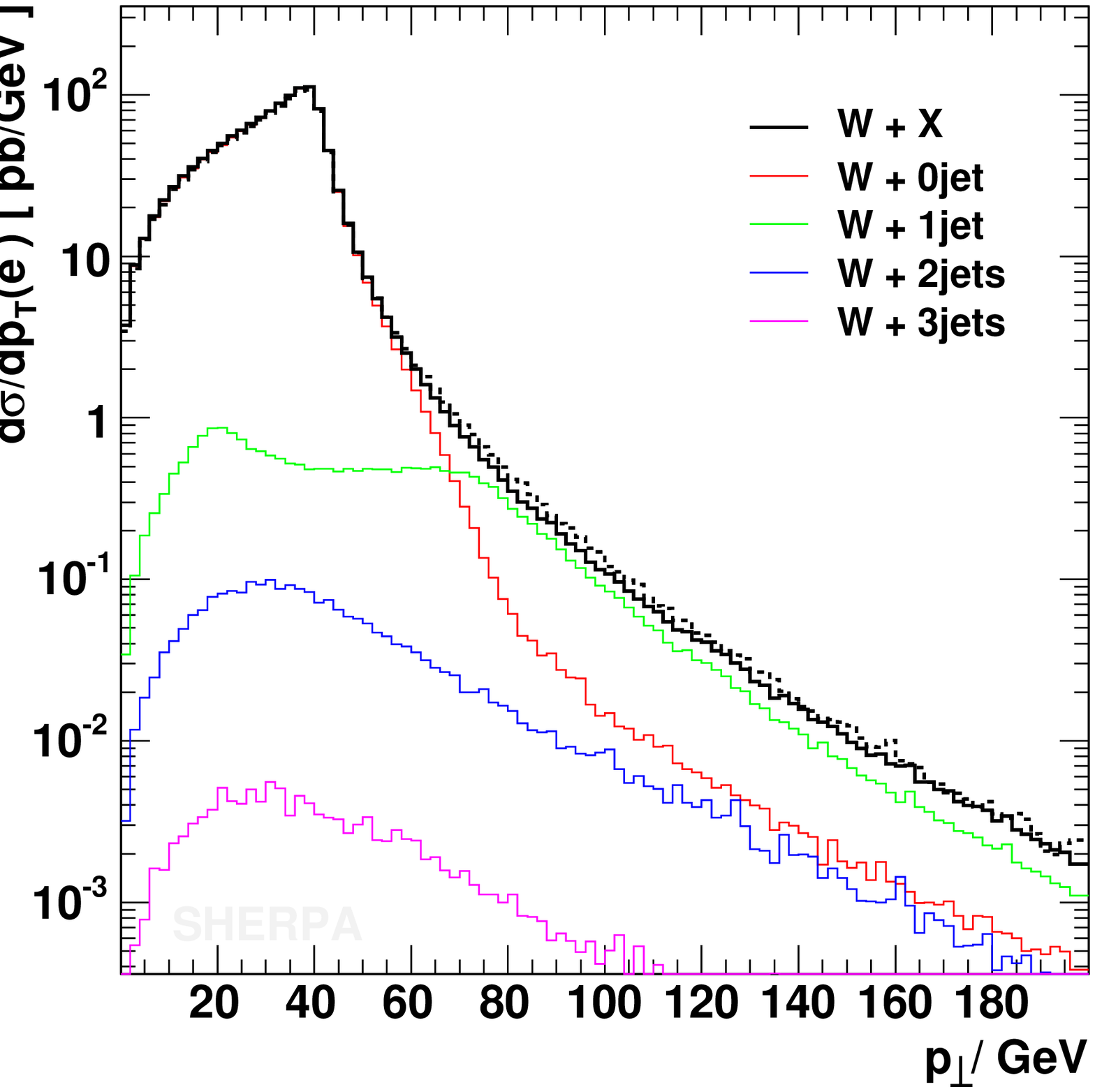}}
\put(125,0){\includegraphics[width=5.5cm]{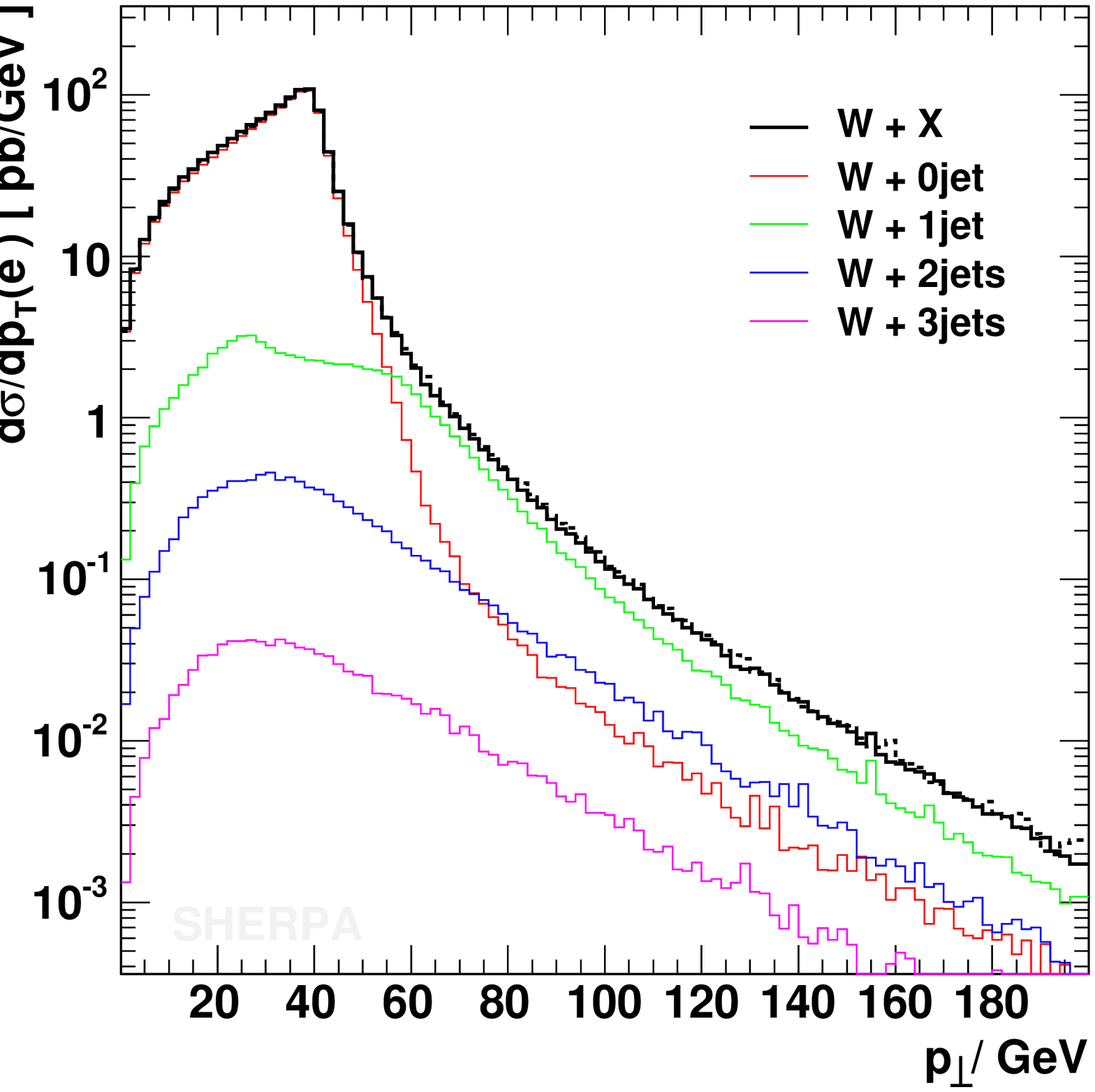}}
\put(0,0){\includegraphics[width=5.5cm]{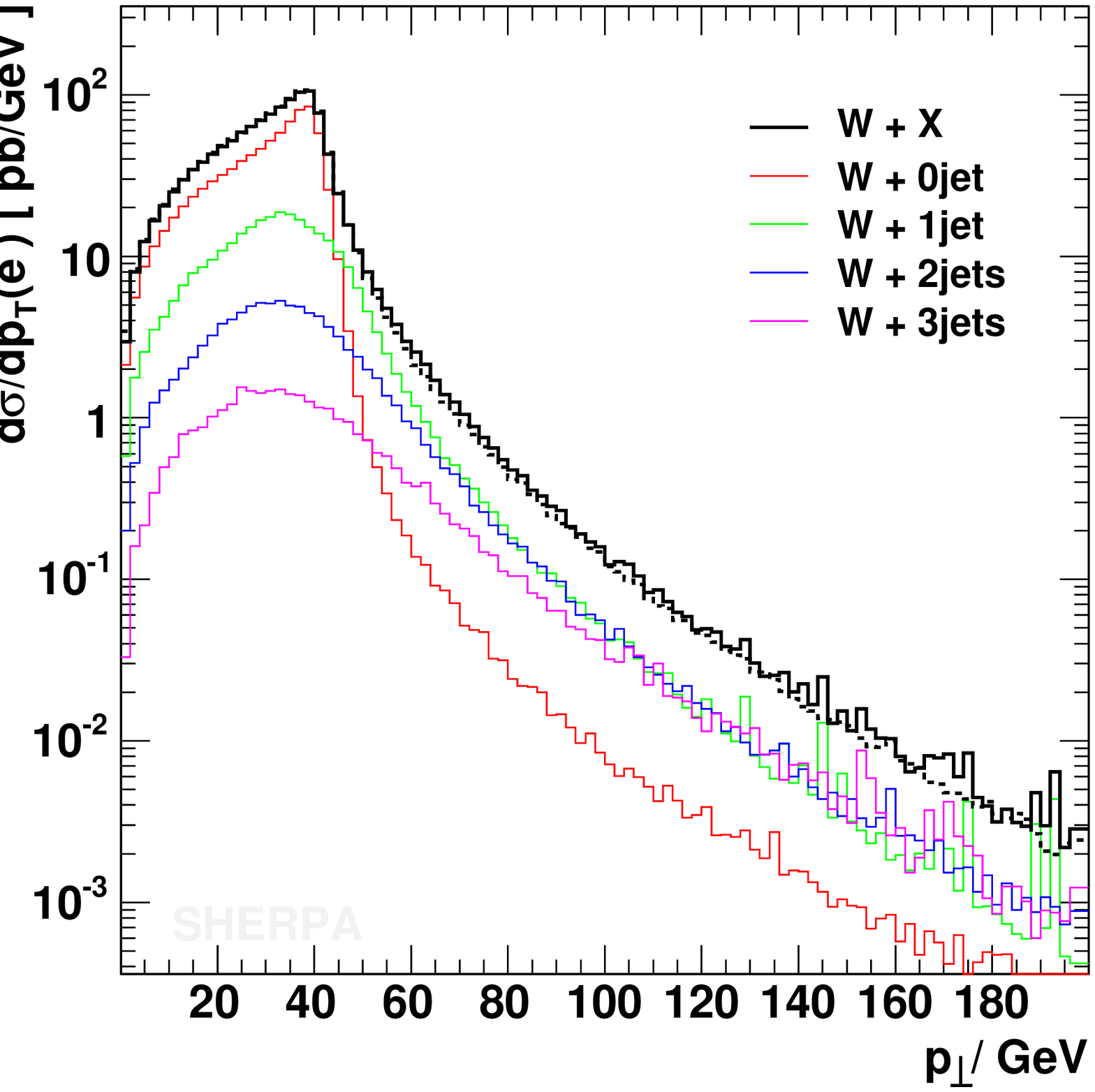}}
\end{pspicture}
\end{center}
\caption{\label{ycut_pt}$p_\perp(W^-)$ and $p_\perp(e^-)$ for 
         $Q_{\rm cut}=10$ GeV, $30$ GeV and $50$ GeV in comparison with 
         $Q_{\rm cut}=20$ GeV. }
\end{figure*}
\begin{figure*}[h]
\begin{center}
\begin{pspicture}(400,275)
\put(250,125){\includegraphics[width=5.5cm]{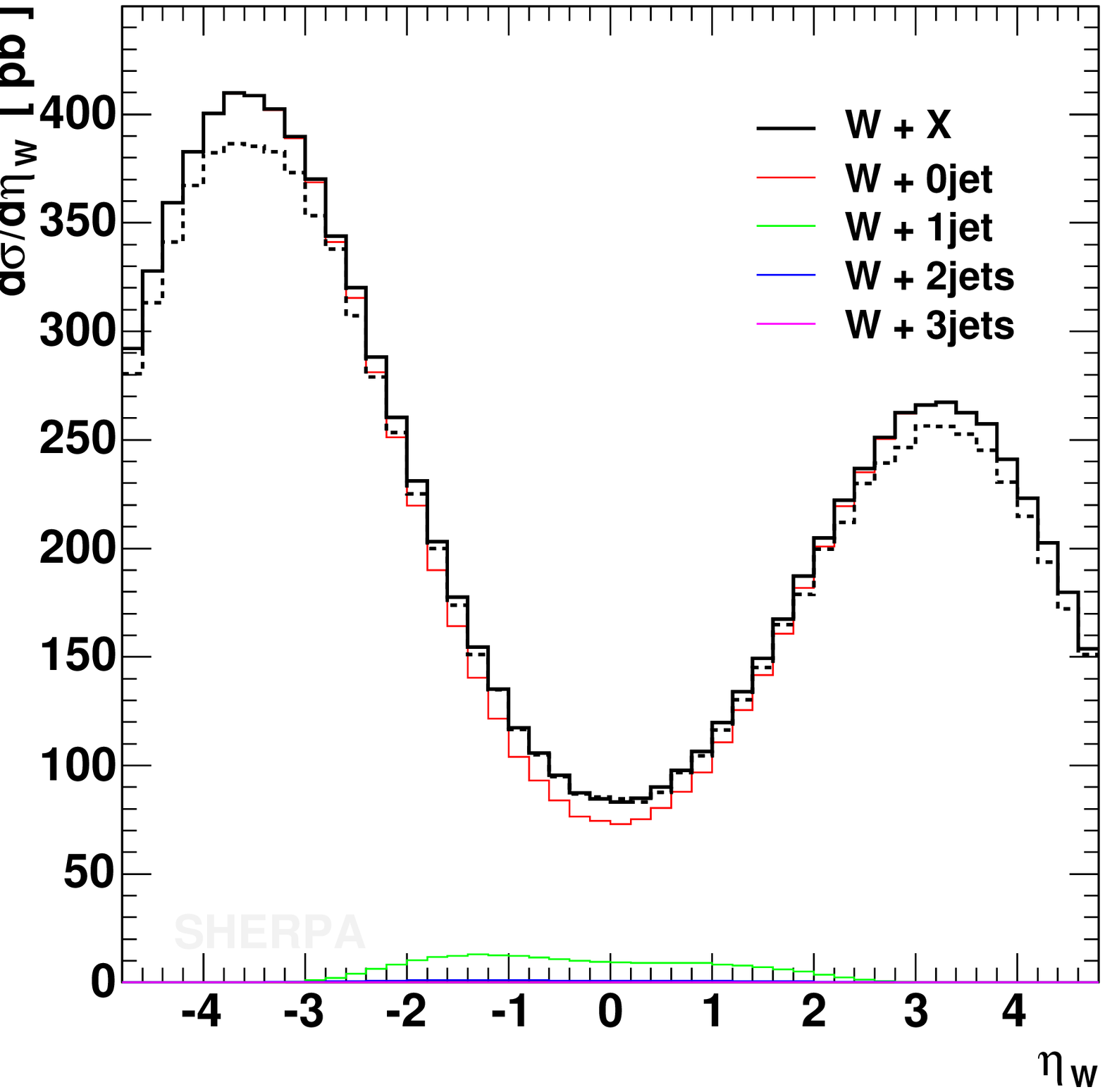}}
\put(125,125){\includegraphics[width=5.5cm]{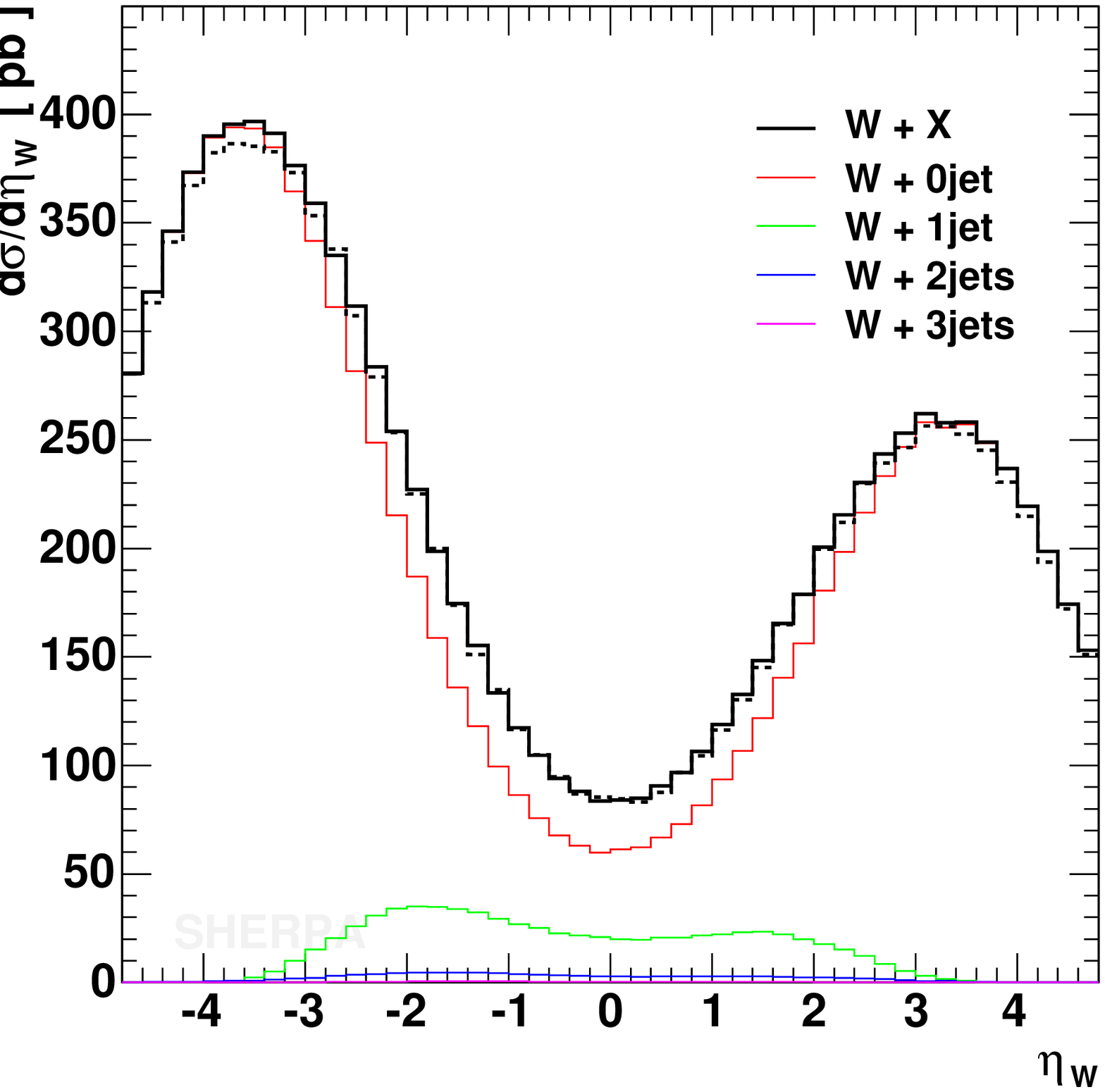}}
\put(0,125){\includegraphics[width=5.5cm]{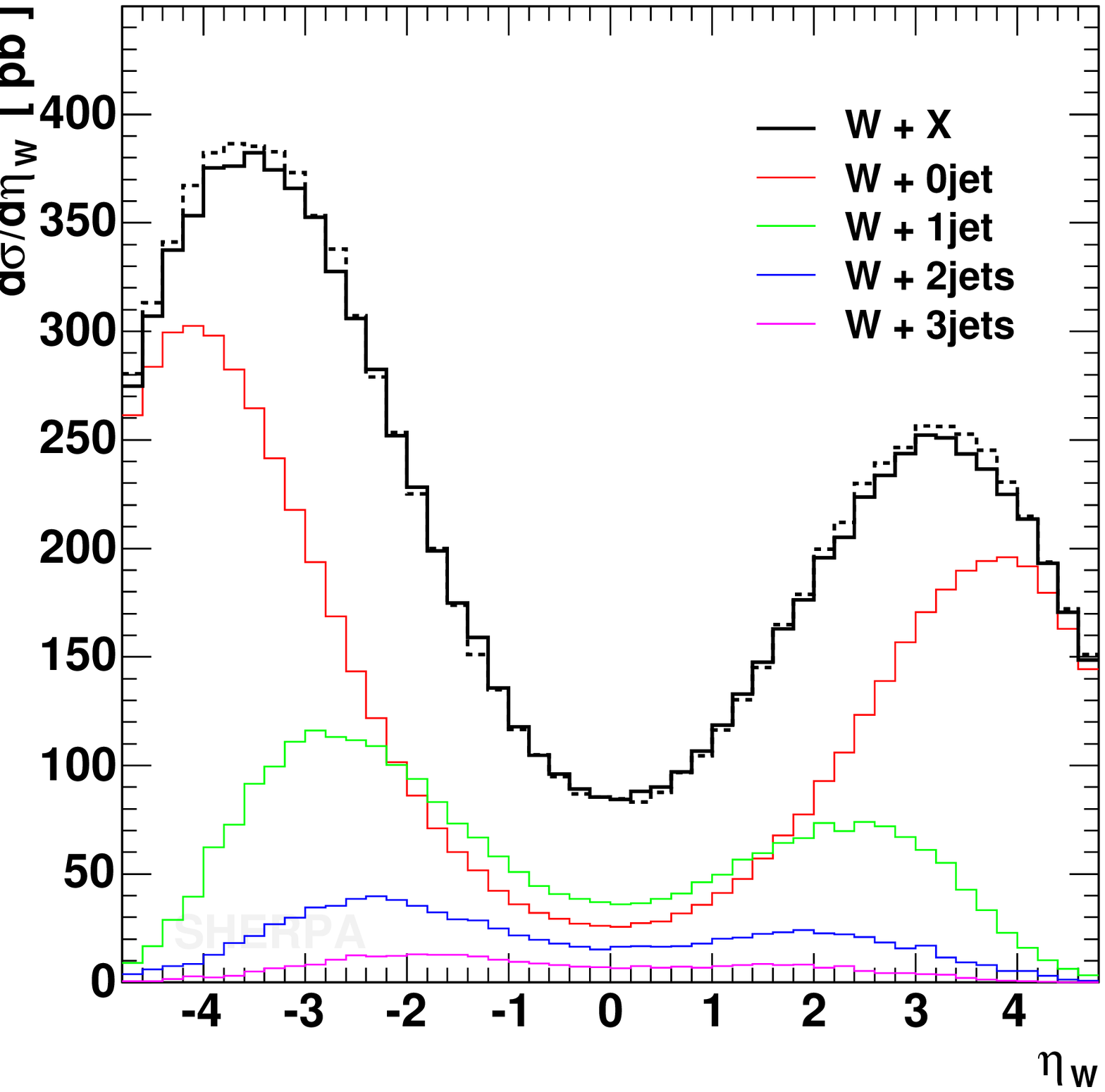}}
\put(250,0){\includegraphics[width=5.5cm]{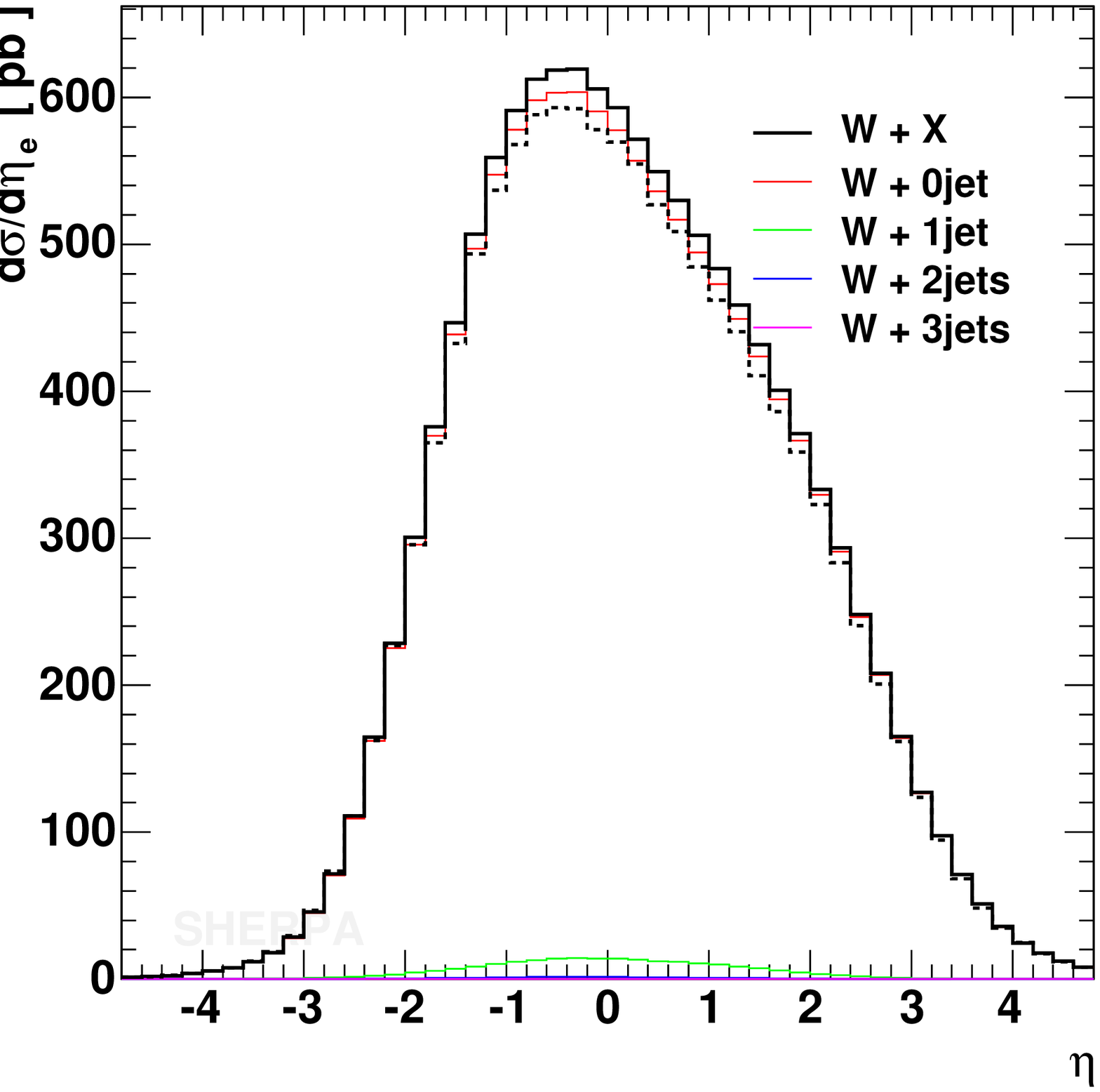}}
\put(125,0){\includegraphics[width=5.5cm]{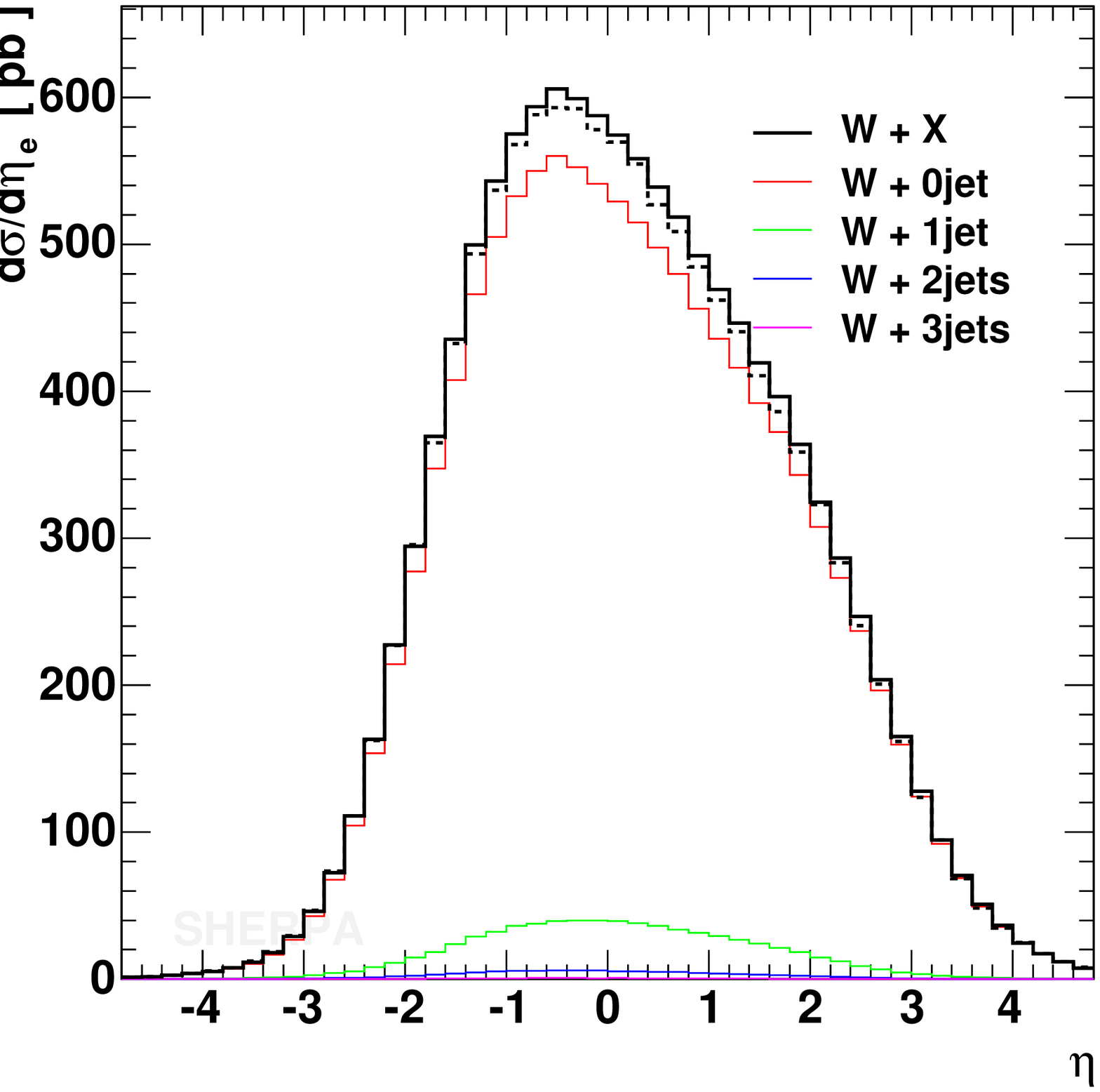}}
\put(0,0){\includegraphics[width=5.5cm]{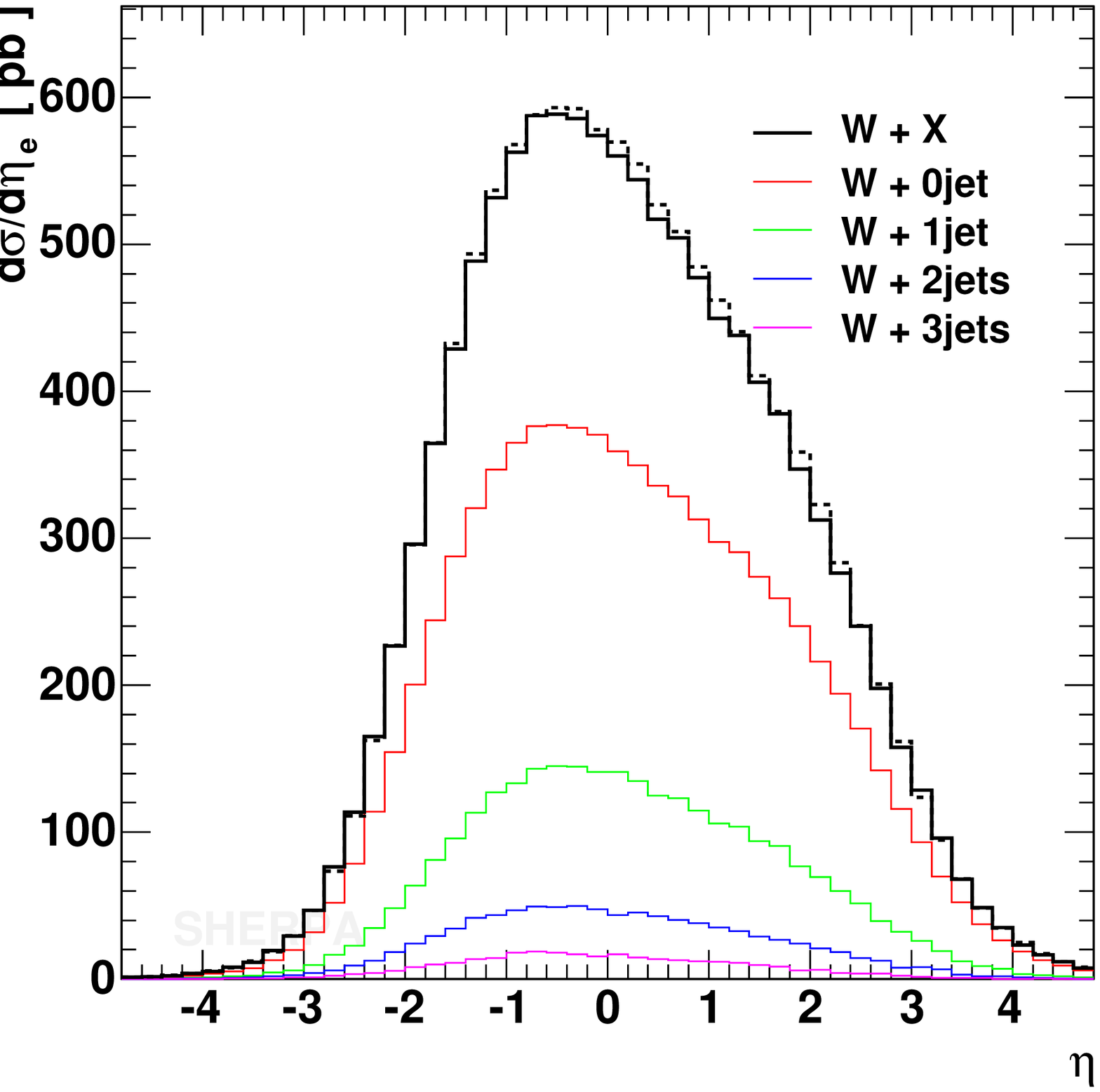}}
\end{pspicture}
\end{center}
\caption{\label{ycut_eta}$\eta(W^-)$ and $\eta(e^-)$ for 
         $Q_{\rm cut}=10$ GeV, $30$ GeV and $50$ GeV in comparison with 
         $Q_{\rm cut}=20$ GeV. }
\end{figure*}
Figs.\ \ref{ycut_pt} and \ref{ycut_eta} show the transverse momentum and the rapidity 
distribution of the $W^-$ boson and the corresponding electron. For the transverse momentum
of the $W$ below the cut, the distribution is dominated by the LO matrix element with no
extra jet, i.e.\ the transverse momentum is generated by the initial state parton shower 
only. Around the cut, a small dip is visible in Fig.\ \ref{ycut_pt}. The $p_\perp$ distribution 
of the electron, in contrast, is hardly altered. The rapidity distributions in Fig.\ \ref{ycut_eta} 
exhibit the asymmetry, which has been anticipated when considering merely the negatively 
charged $W$ boson. The shape of these distributions is very stable under a variation of the 
separation cut. In all observables a small increase of the total cross section of a few percent
when changing $Q_{\rm cut}$ from 10~GeV to 50~GeV is visible. This underlines the fact that the 
dependence on the separation cut is weak. 

\begin{figure*}[h]
\begin{center}
\begin{pspicture}(400,400)
\put(250,250){\includegraphics[width=5.5cm]{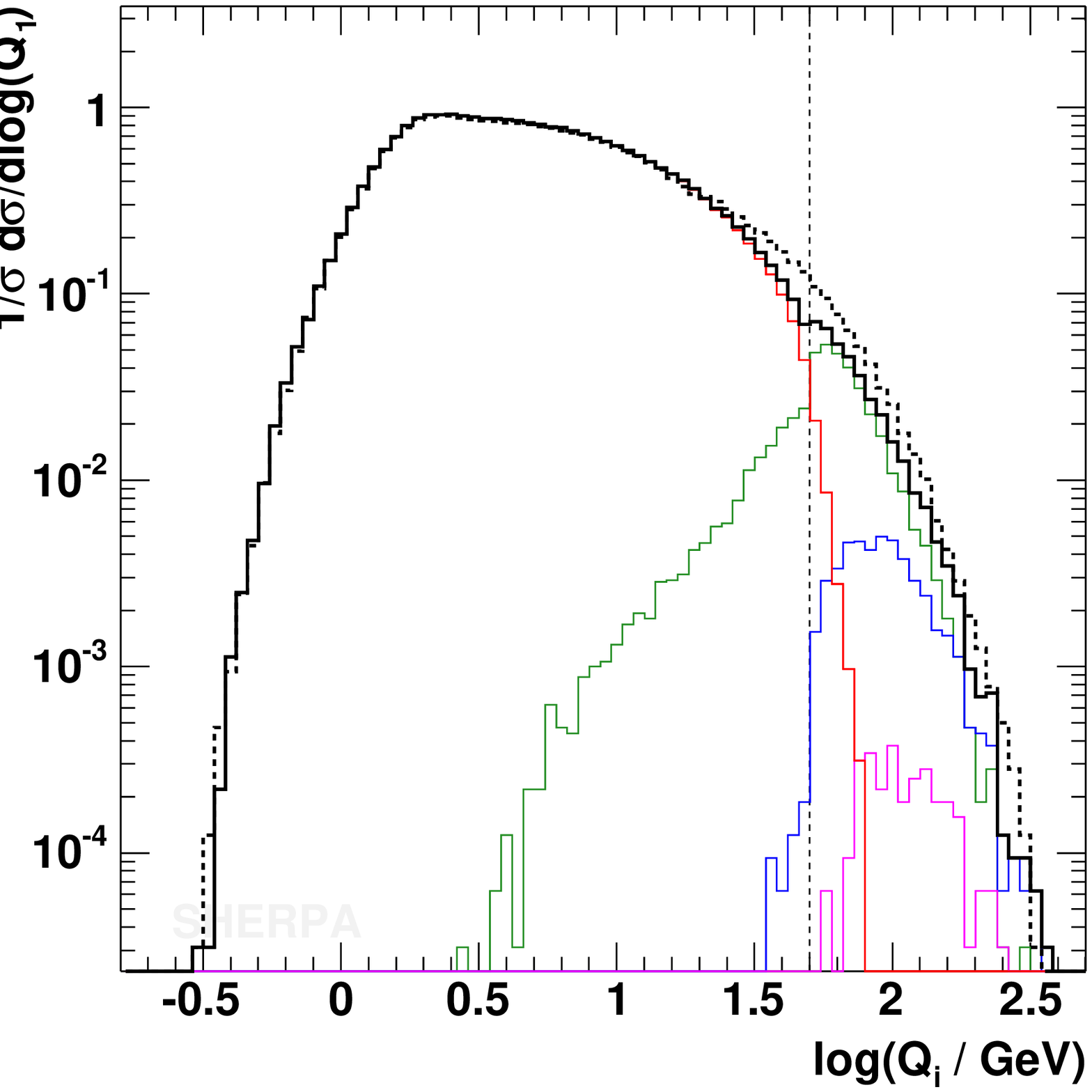}}
\put(125,250){\includegraphics[width=5.5cm]{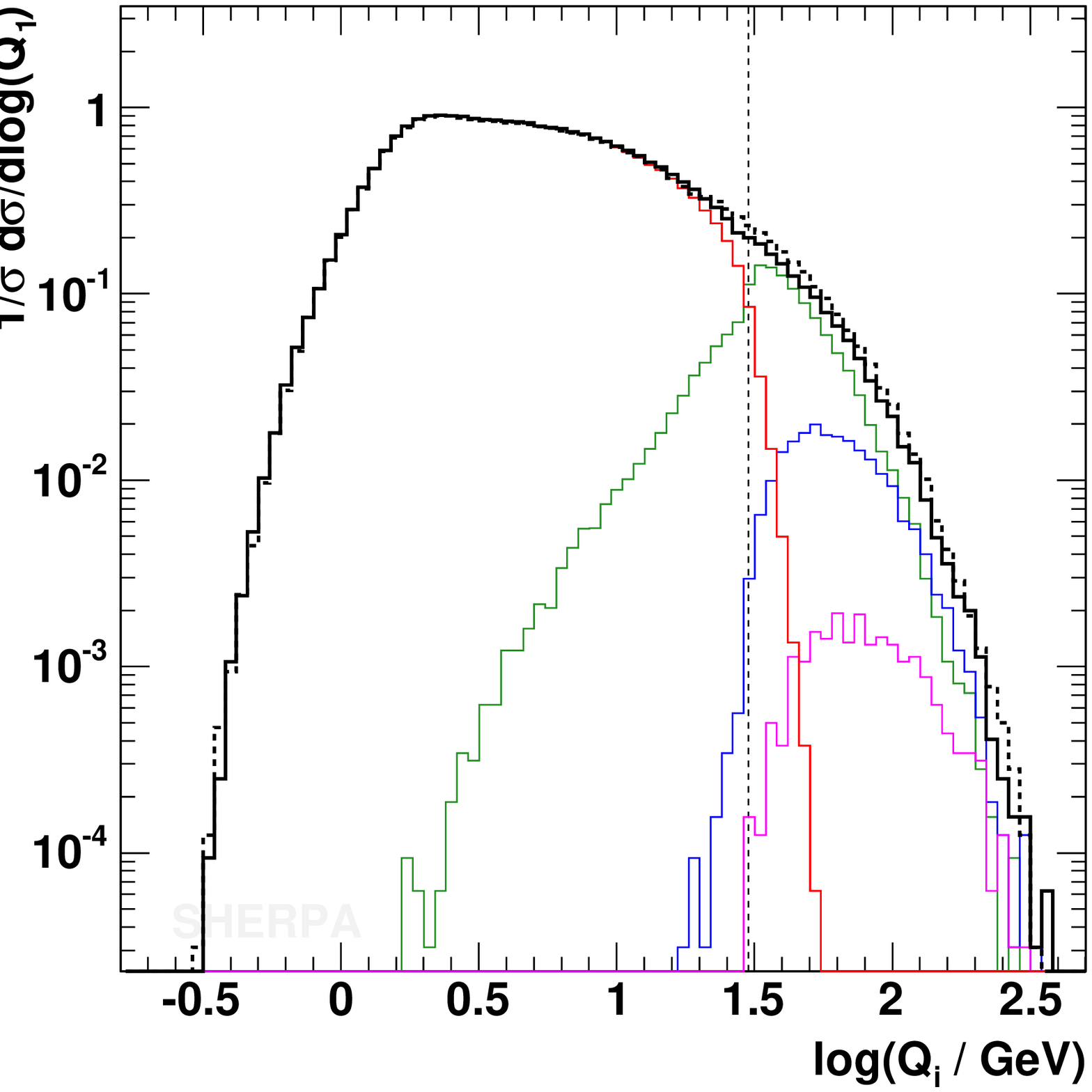}}
\put(0,250){\includegraphics[width=5.5cm]{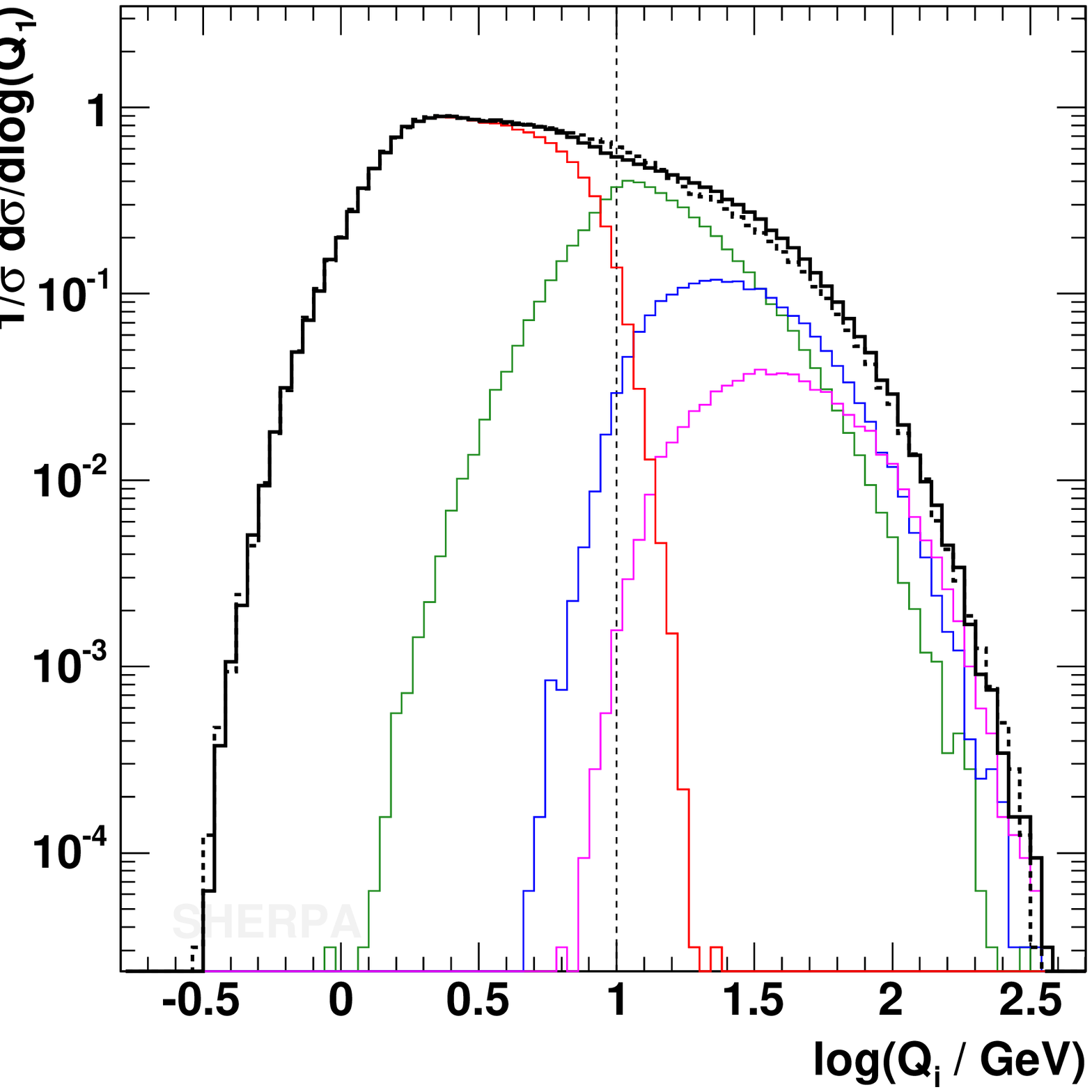}}
\put(250,125){\includegraphics[width=5.5cm]{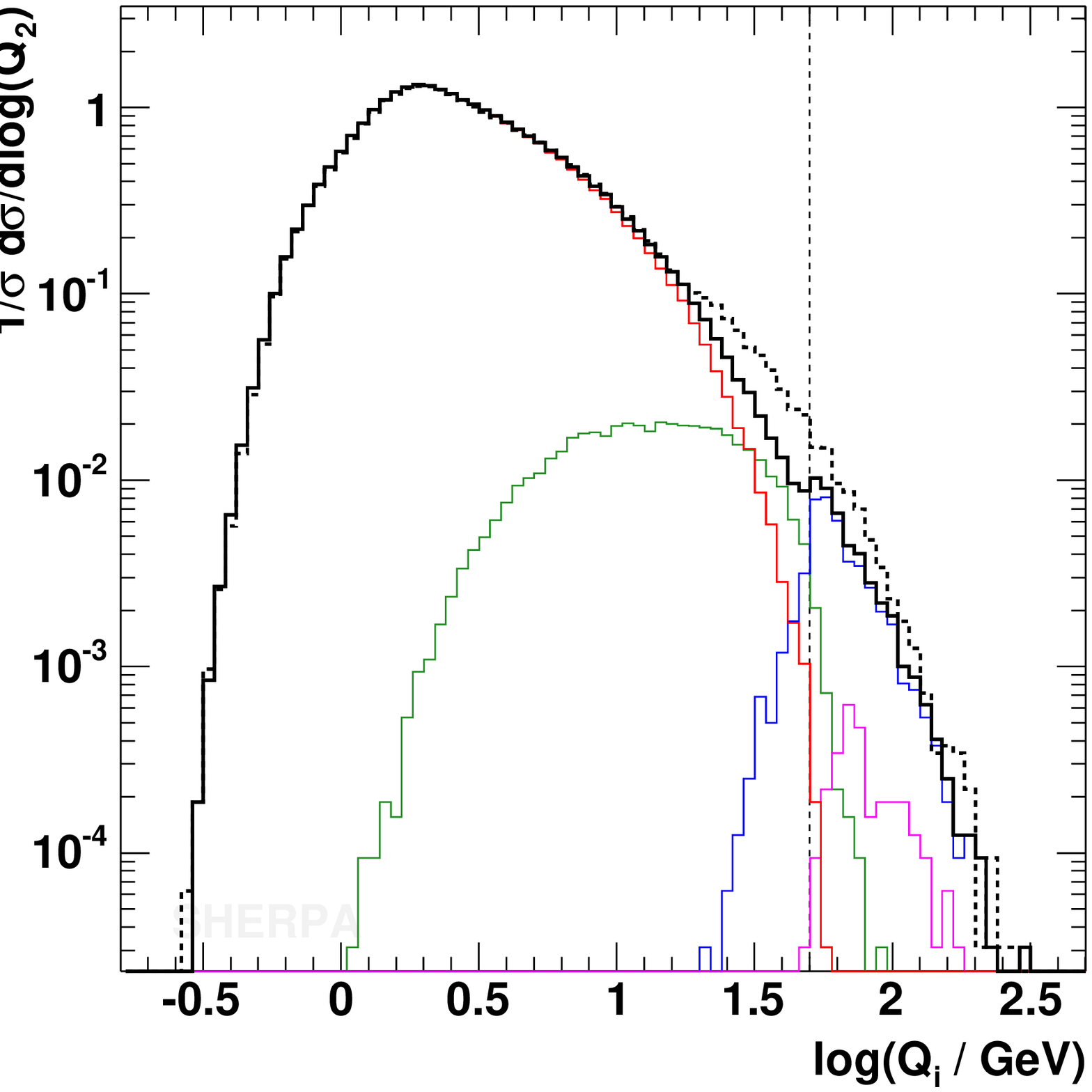}}
\put(125,125){\includegraphics[width=5.5cm]{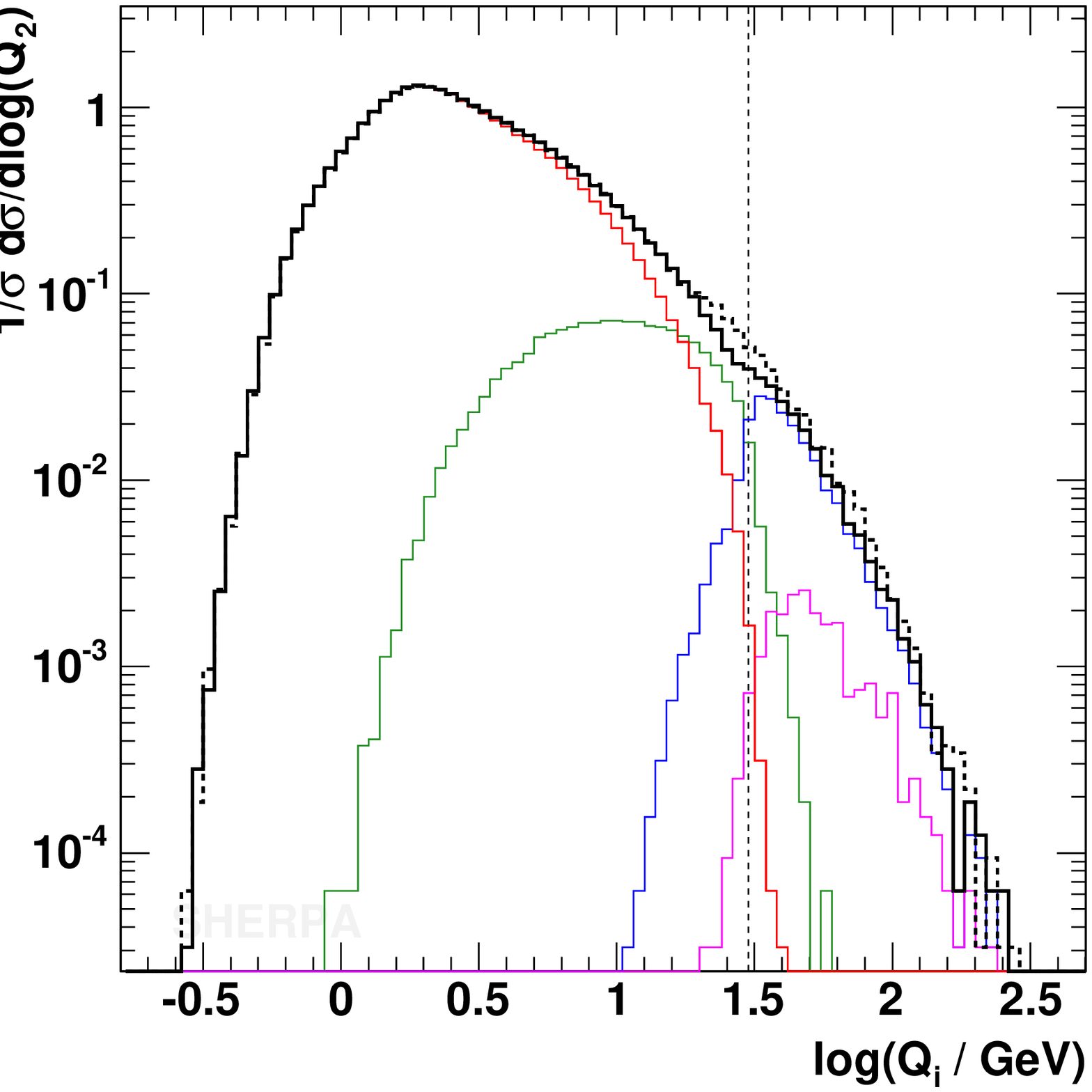}}
\put(0,125){\includegraphics[width=5.5cm]{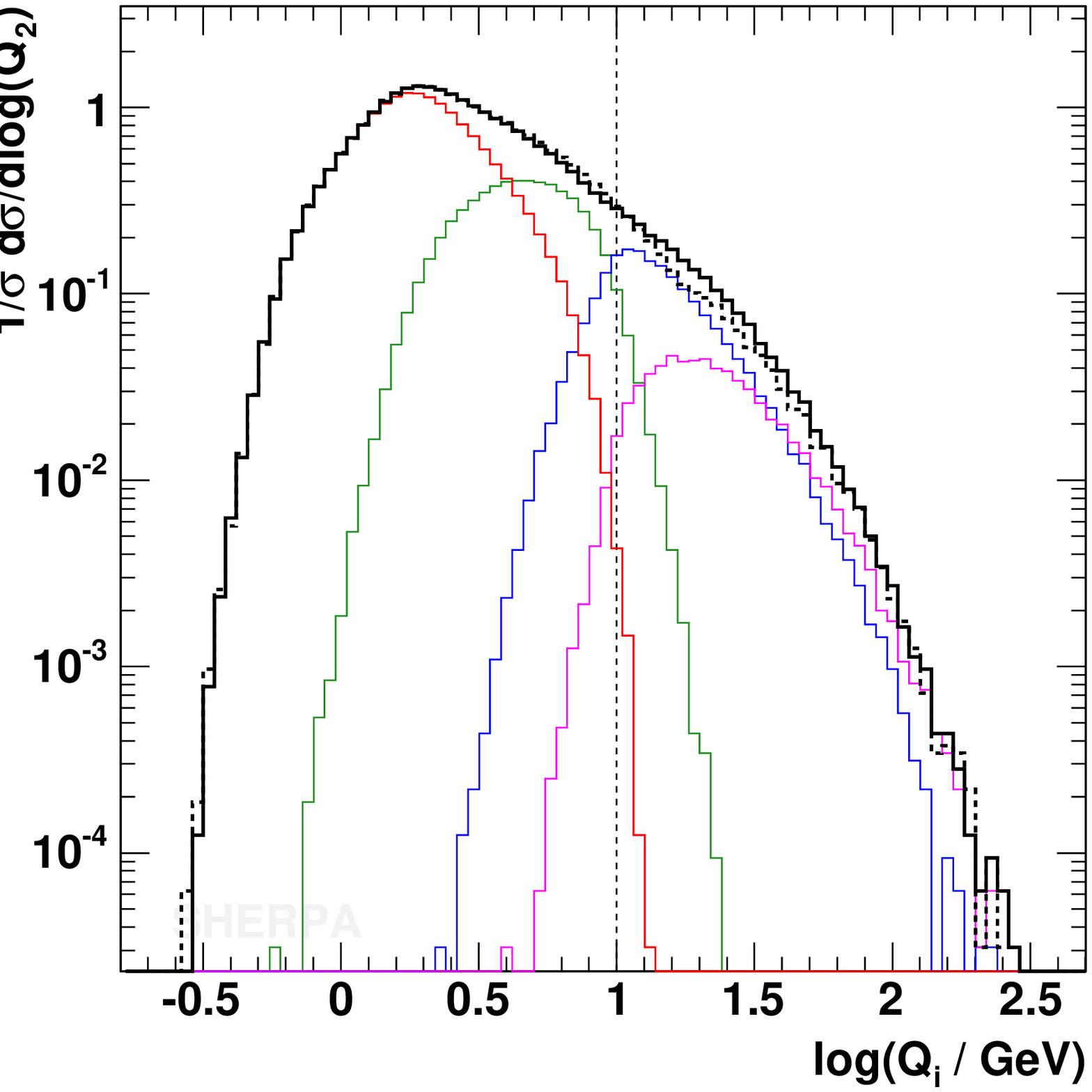}}
\put(250,0){\includegraphics[width=5.5cm]{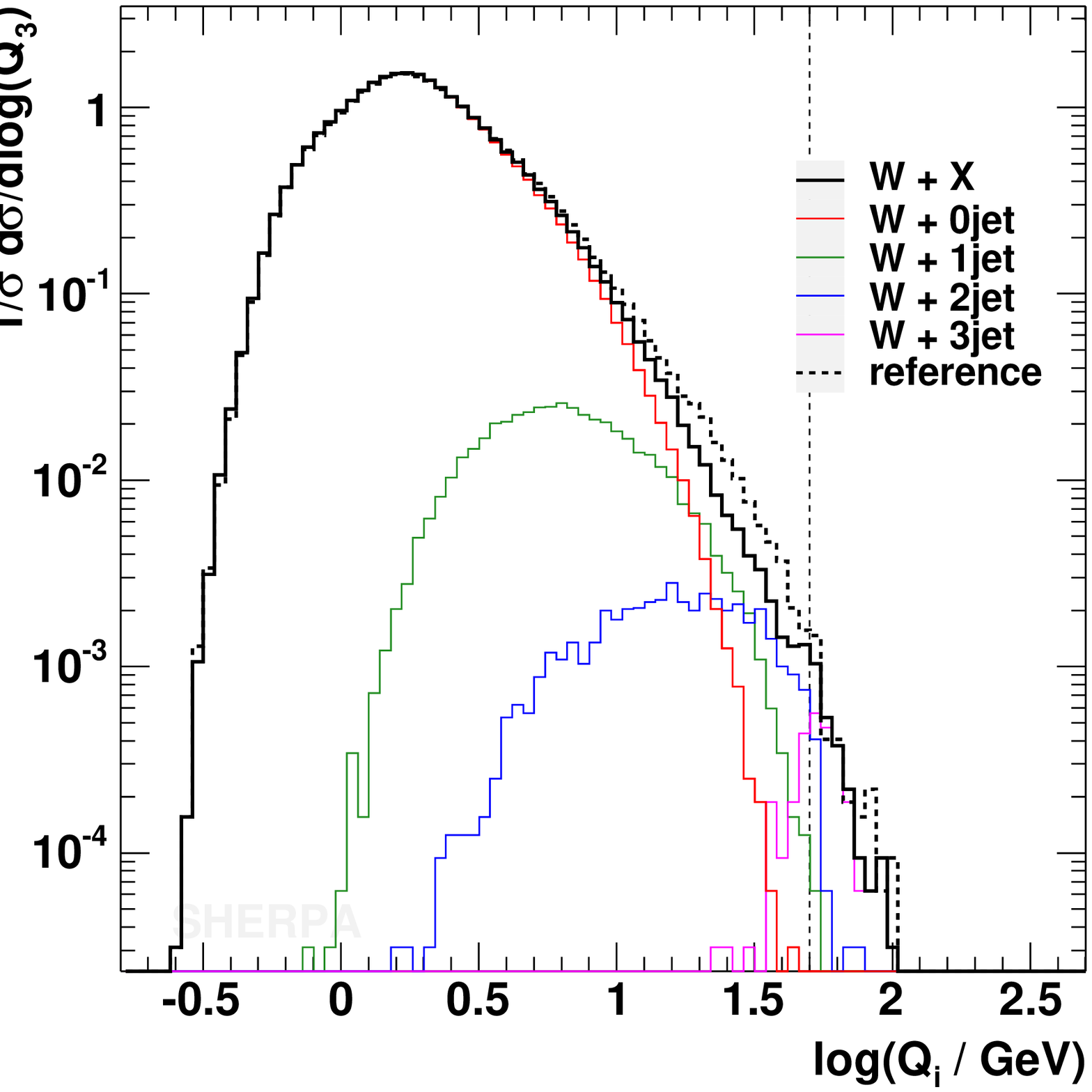}}
\put(125,0){\includegraphics[width=5.5cm]{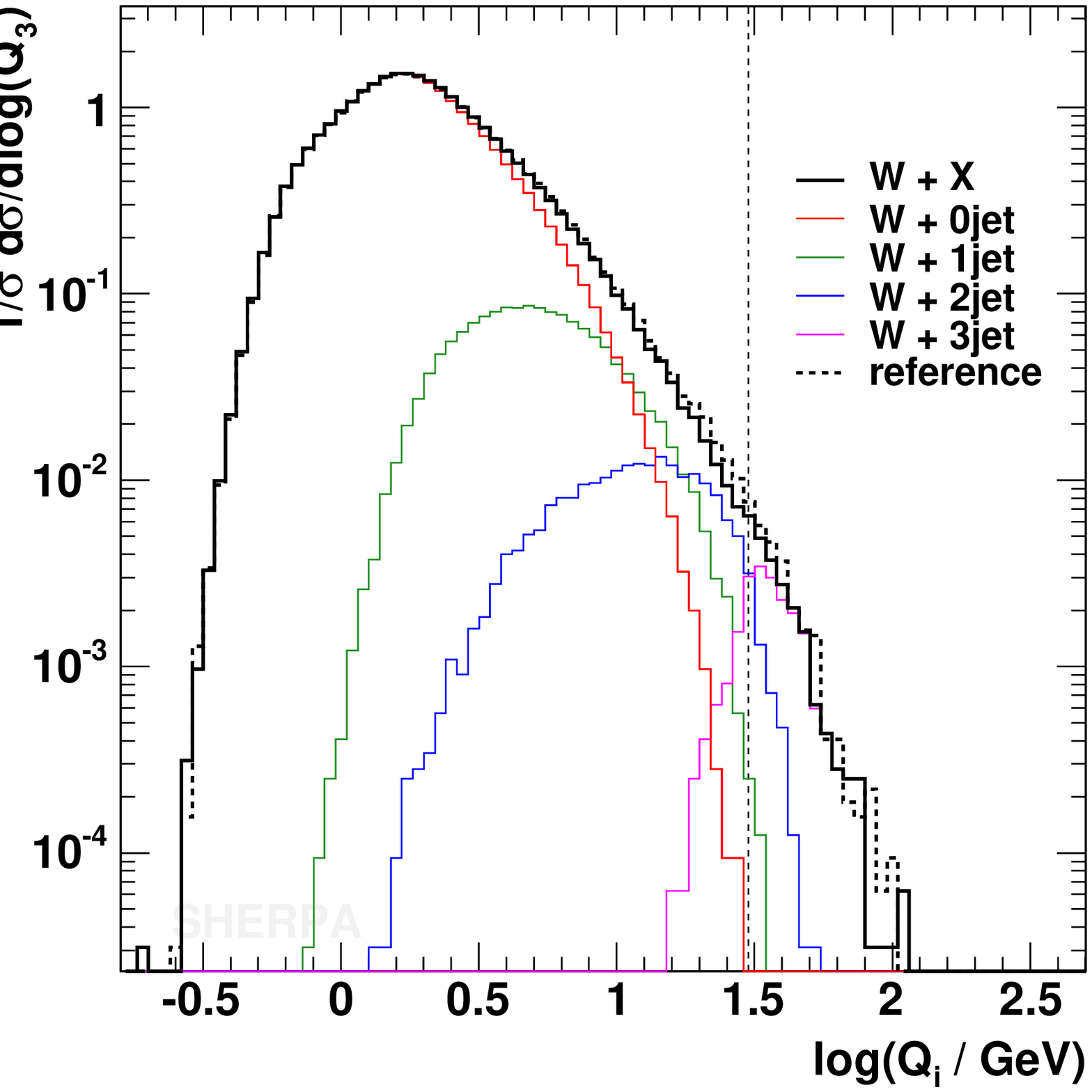}}
\put(0,0){\includegraphics[width=5.5cm]{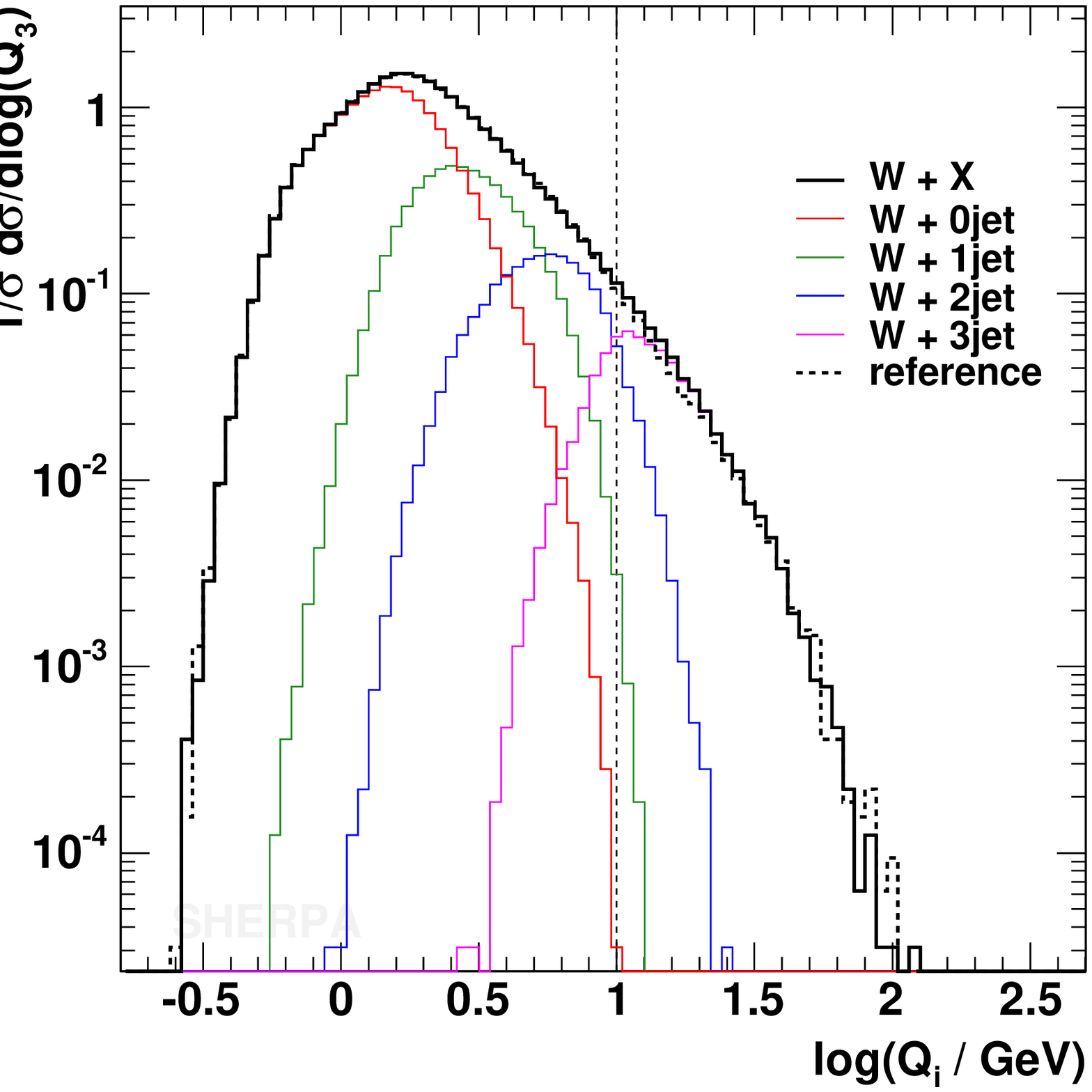}}
\end{pspicture}
\end{center}
\caption{\label{ycut_diff}Differential jet rates for the  $1 \to 0$, 
         $2 \to 1$ and $3 \to 2$ transition (top to bottom), for \\$Q_{\rm cut}=10$ GeV, 
         $30$ GeV, and $50$ GeV (from left to right). In each plot, the results are compared 
         with those for $Q_{\rm cut}=20$ GeV. }
\end{figure*}
\noindent
Differential jet rates with respect to the $k_\perp$-algorithm are interesting observables, 
since they basically exhibit the distributions of nodal values using the cluster algorithm. 
For simplicity the Run II $k_\perp$-algorithm has been used with $D=1$ for the analysis.
Differential jet rates are of special interest, since the nodal values are very close to 
the measure used to separate matrix elements from parton shower emissions. Some minor 
problems with respect to the separation should immediately manifest themselves in these 
distributions. In Fig.\ \ref{ycut_diff} the $1 \to 0$, $2 \to 1$ and $3 \to 2$ differential 
jet rates are shown. Within the given approximation the independence is satisfactory. 

\subsection{Variation of the maximal jet multiplicity $n_{\rm max}$}
\begin{figure*}[h]
\begin{center}
\begin{pspicture}(400,150)
\put(250,0){\includegraphics[width=5.5cm]{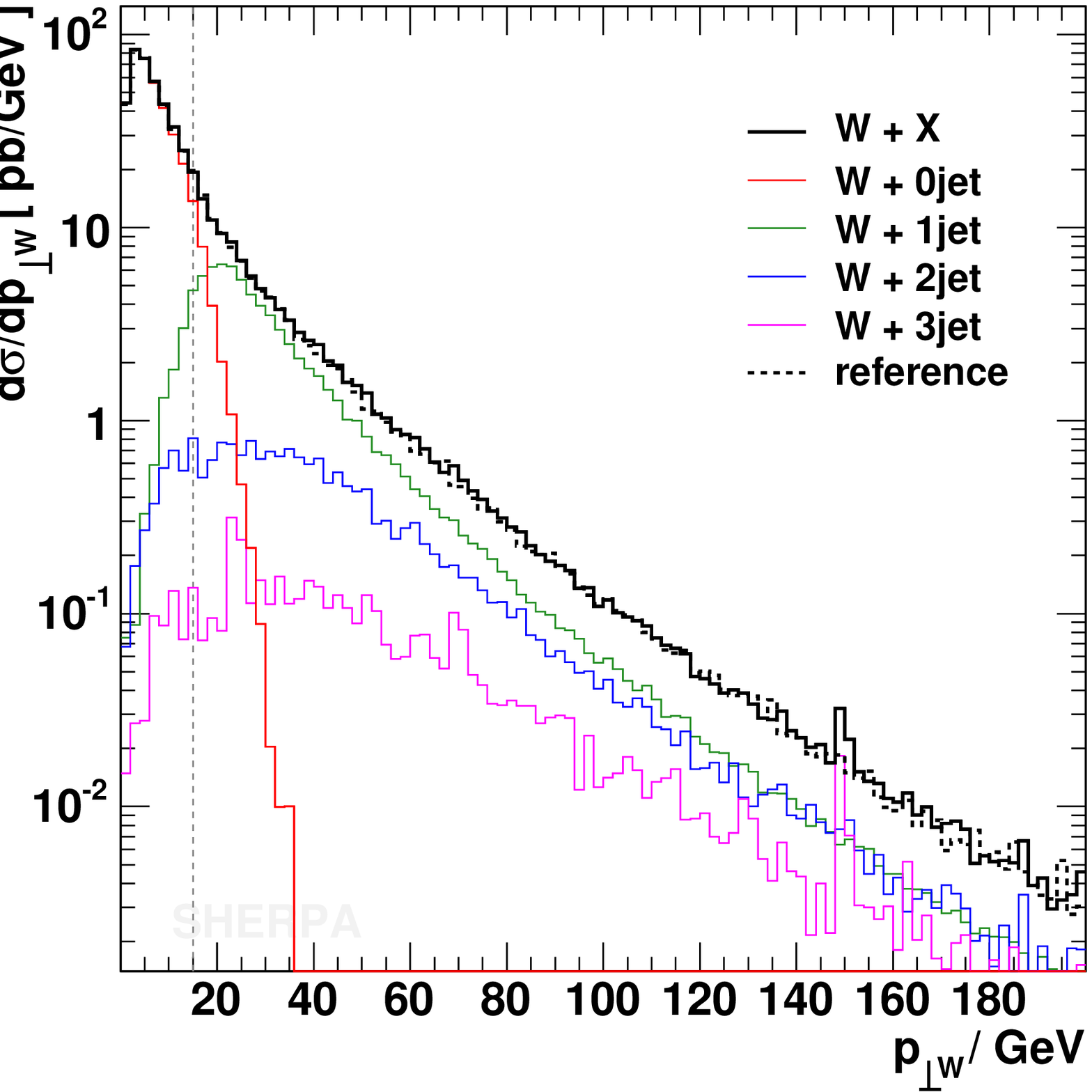}}
\put(125,0){\includegraphics[width=5.5cm]{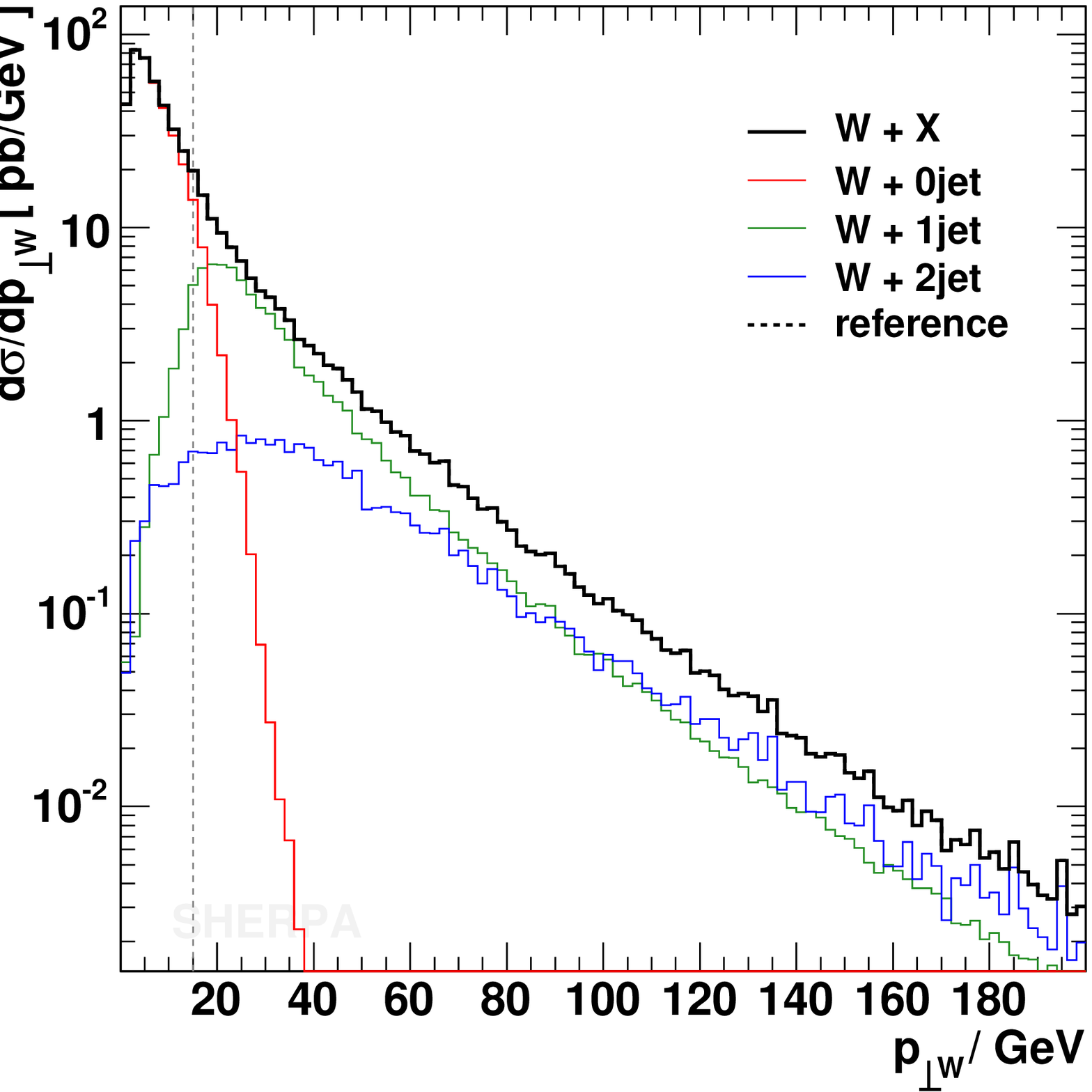}}
\put(0,0){\includegraphics[width=5.5cm]{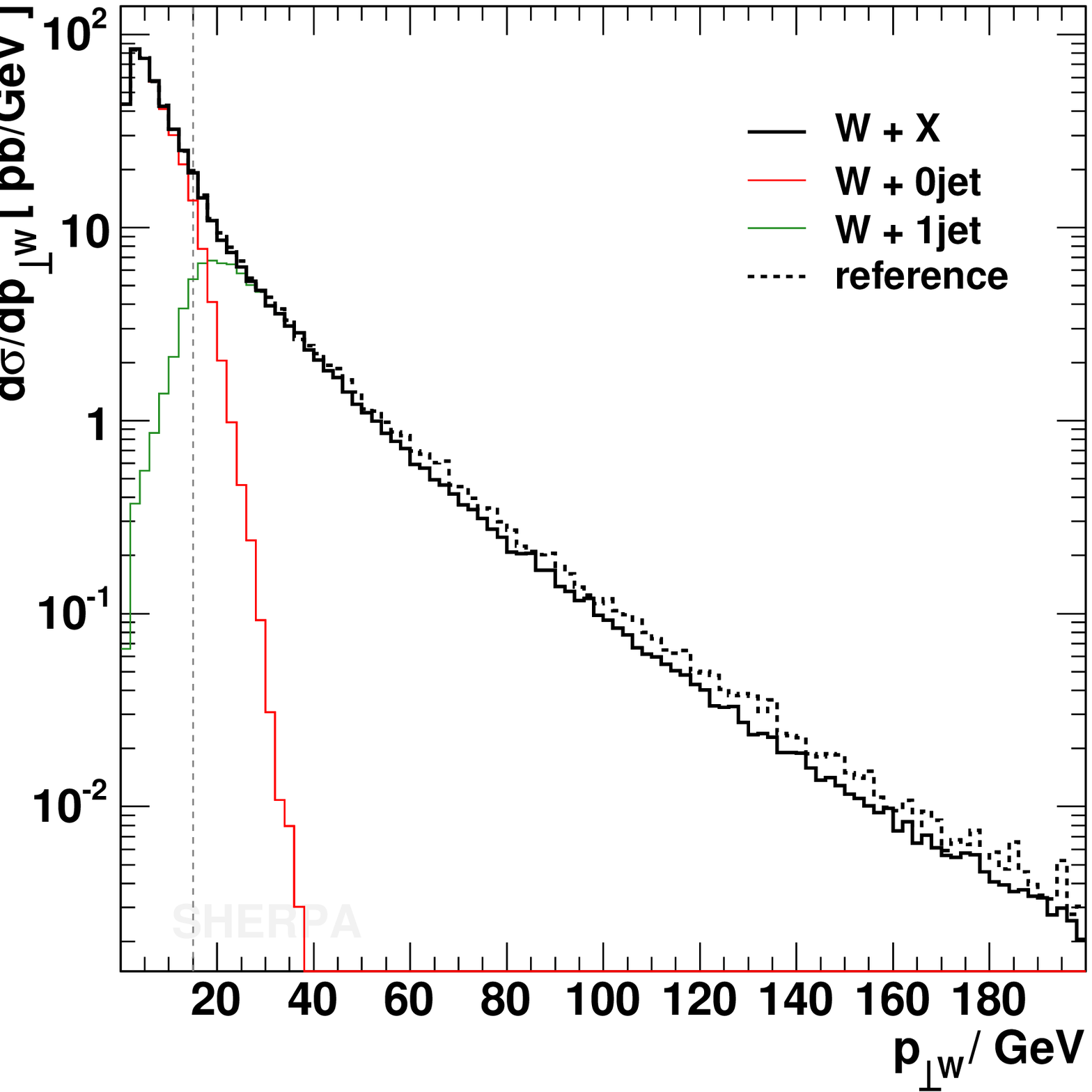}}
\end{pspicture}
\end{center}
\caption{\label{nmax_pt}$p_\perp(W^-)$   for $Q_{\rm cut}=15$
  GeV and different maximal numbers of ME jets included. }
\end{figure*}

\begin{figure*}[h]
\begin{center}
\begin{pspicture}(400,150)
\put(250,0){\includegraphics[width=5.5cm]{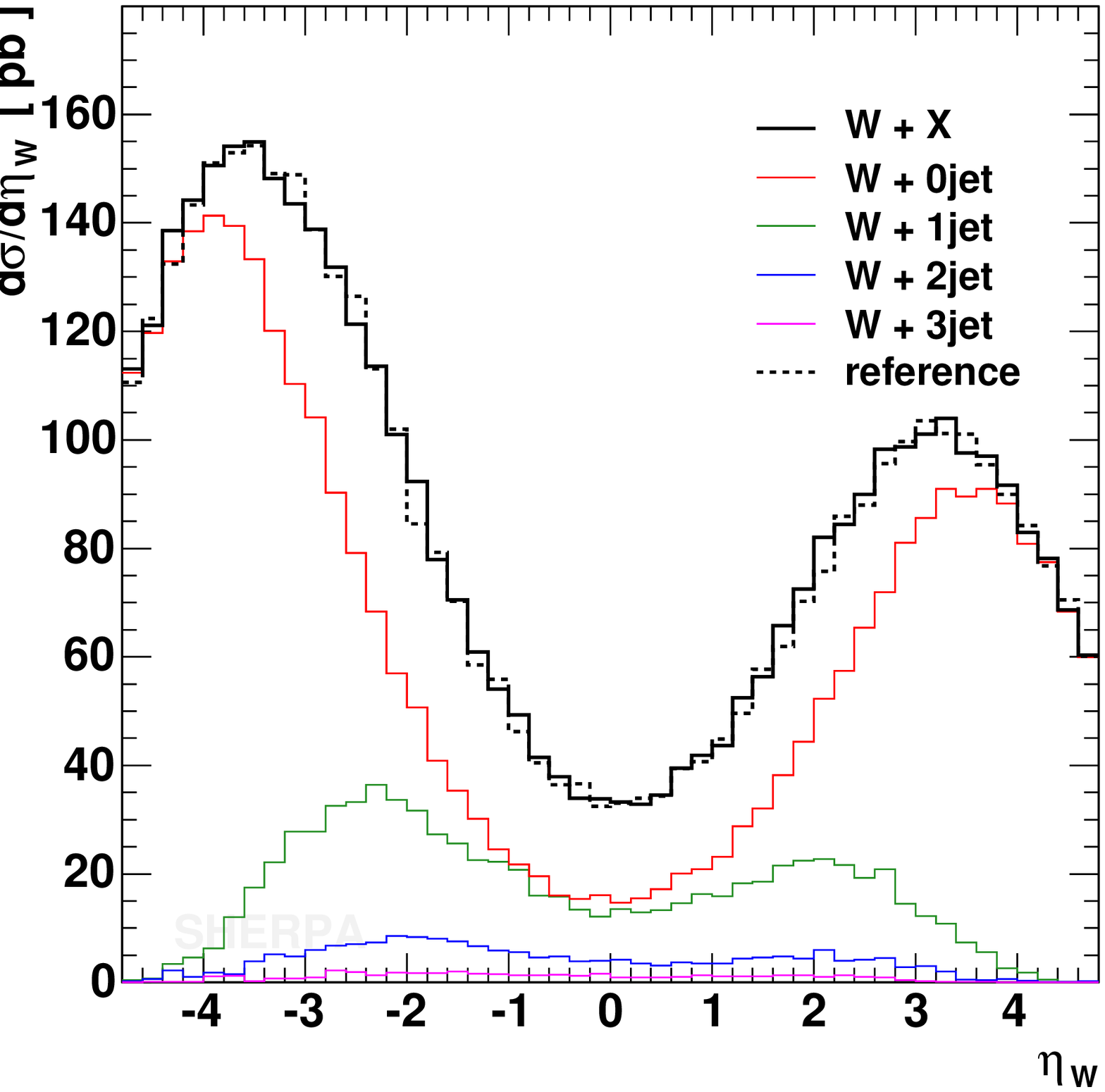}}
\put(125,0){\includegraphics[width=5.5cm]{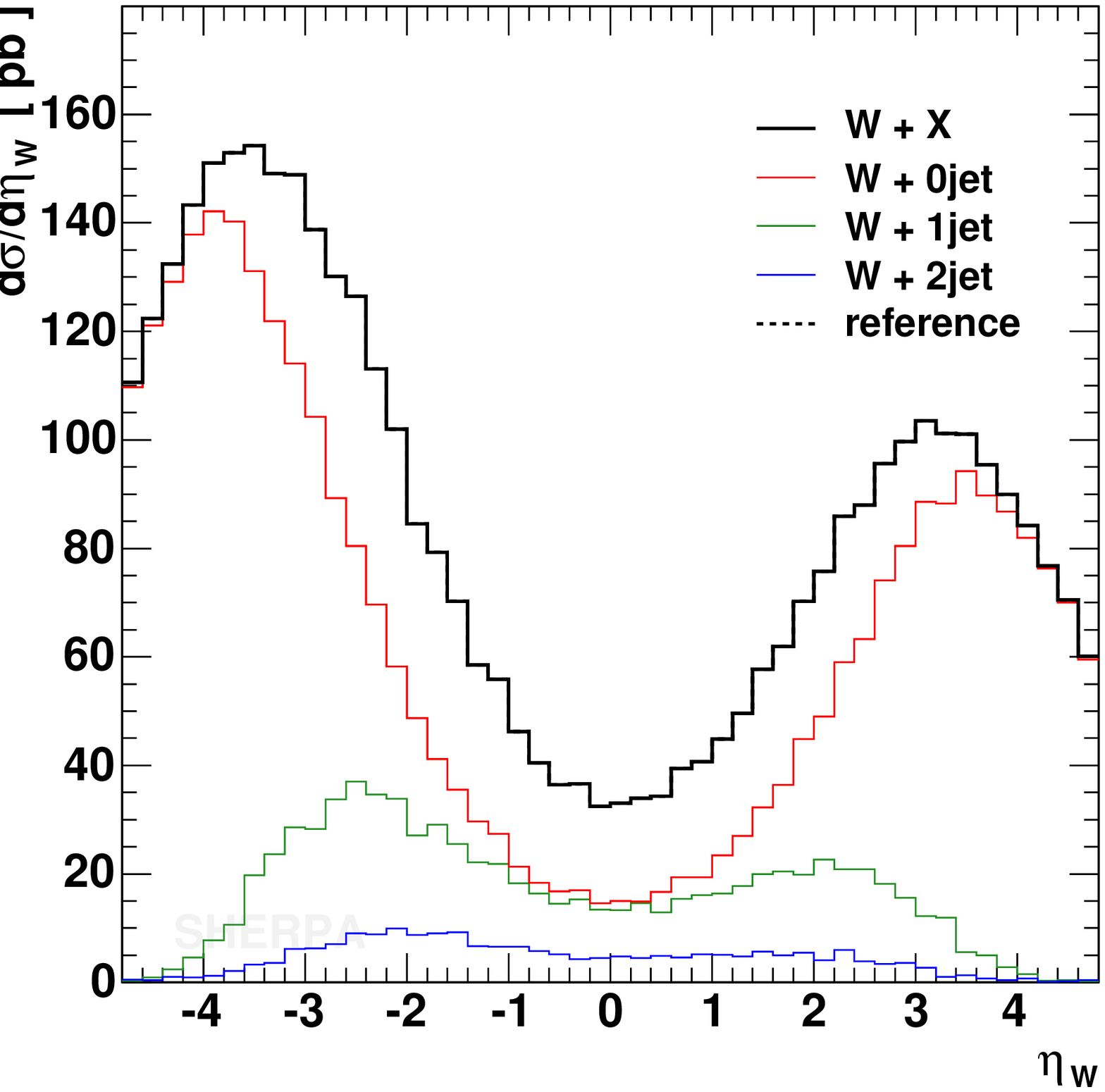}}
\put(0,0){\includegraphics[width=5.5cm]{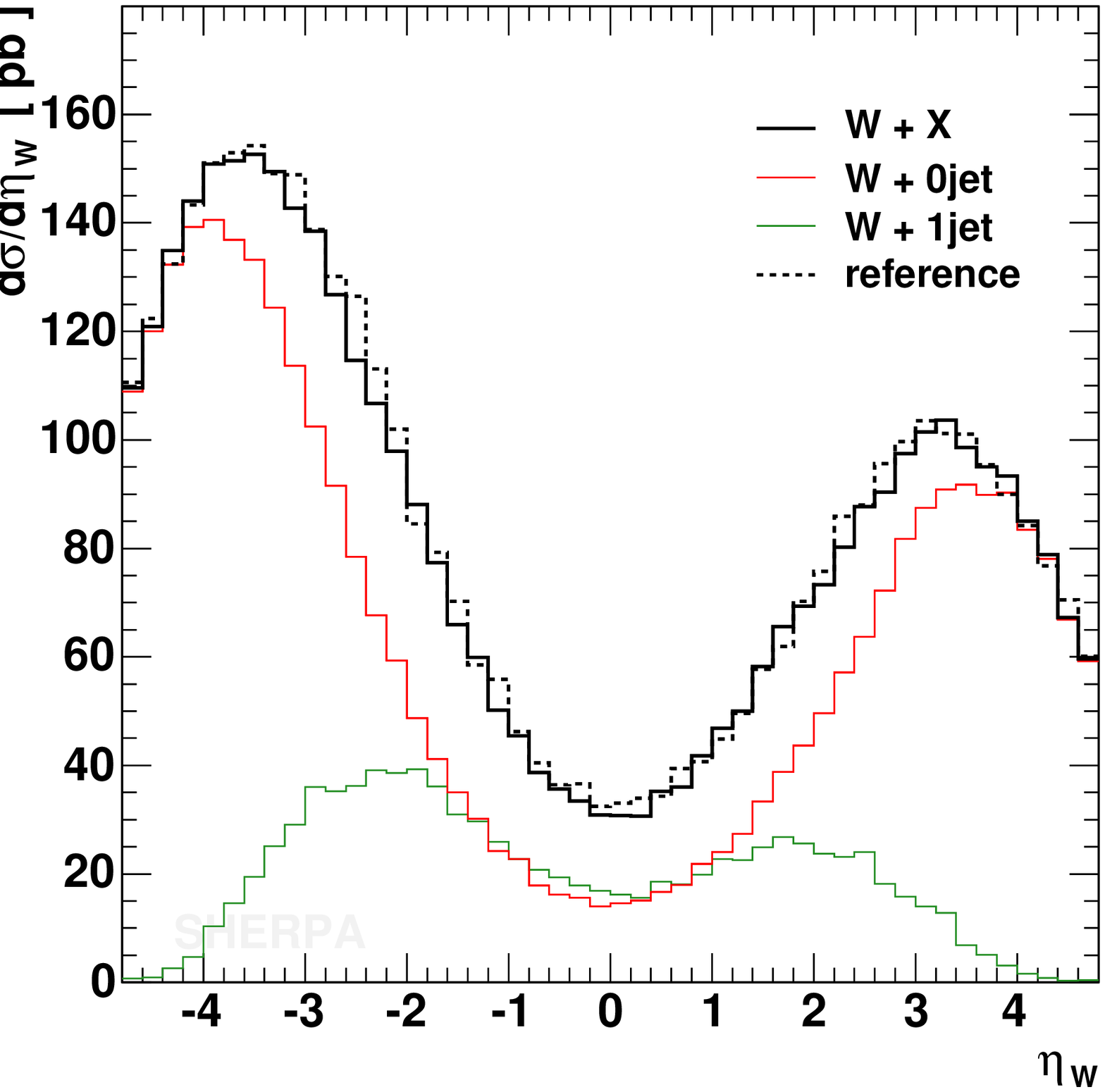}}
\end{pspicture}
\end{center}
\caption{\label{nmax_eta}$\eta(W^-)$  for $Q_{\rm cut}=15$
  GeV and different maximal numbers of ME jets included. }
\end{figure*}

\noindent
For very inclusive observables such as transverse momentum and rapidity of the $W$ boson, it is 
usually sufficient to include the matrix element with only one extra jet in order to obtain a 
reliable prediction. Consequently, the inclusion of matrix elements with more than one extra jet 
in the simulation should not significantly change the result. This can be used as another
consistency check. Figs.\ \ref{nmax_pt} and \ref{nmax_eta} impressively picture the dependence on the
maximal jet number in the matrix elements included. They show that the treatment of the highest 
multiplicity (cf.\ Sec.~\ref{sec_highest_multi}) completely compensates for the missing matrix 
elements, whereas the contribution of the lowest multiplicity is not altered. 

\subsection{Matrix element, parton shower and hadronisation}
\begin{figure}[h]
\begin{center}
\begin{tabular}{c}
\includegraphics[width=8cm]{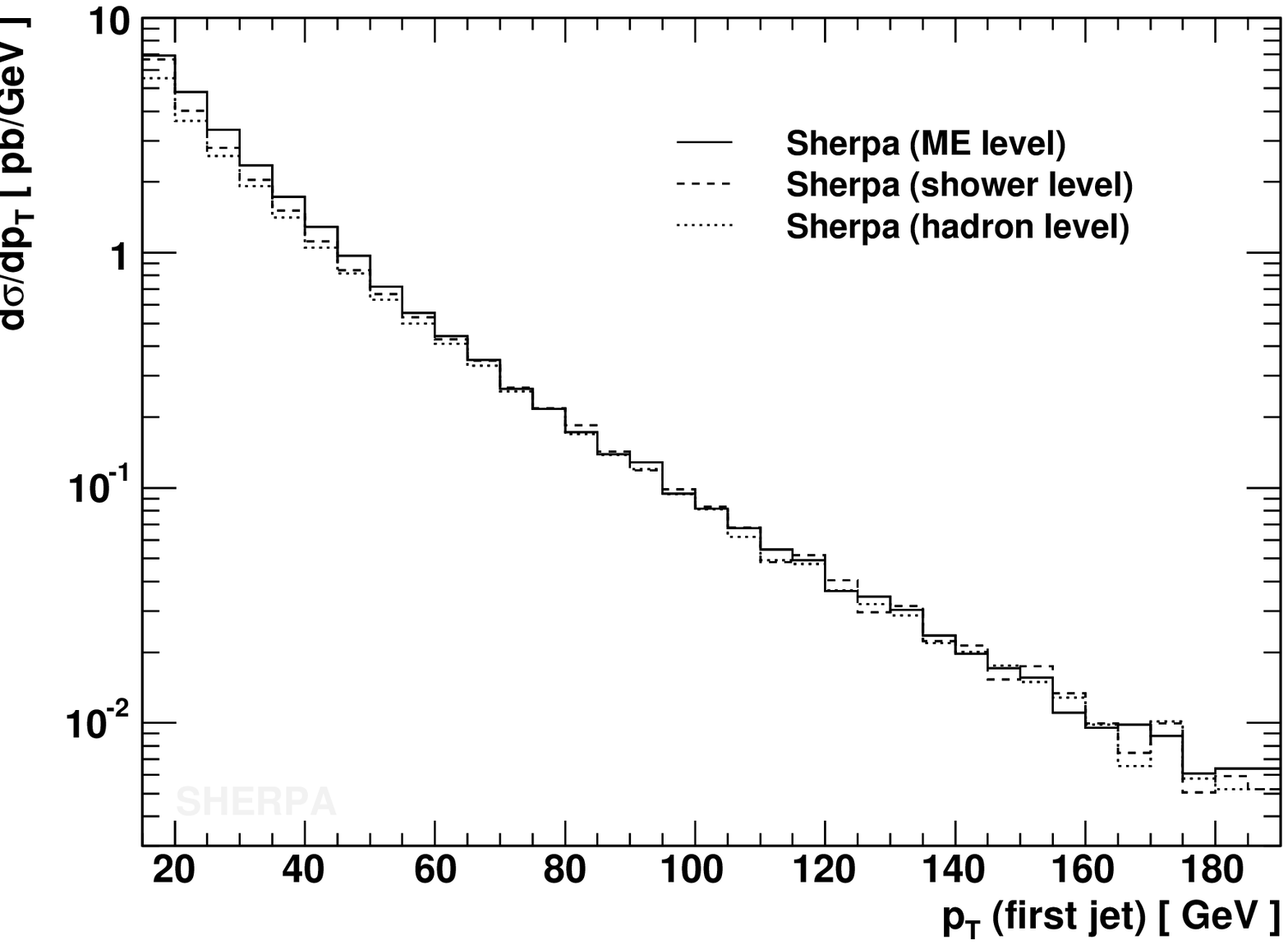}
\end{tabular}
\end{center}
\caption{\label{W1jet_ME_PS_HAD} The $p_\perp$ of the hardest jet in inclusive $W$ production 
  at the Tevatron, Run II. The solid line indicates the distribution as delivered by the matrix
  elements; the dashed line is obtained after parton shower evolution; and the dotted line 
  gives the final result after hadronisation. Here $Q_{\rm cut}=15$ GeV.}
\end{figure}

\noindent
In addition to the self-consistency of the algorithm tested so far at the hadron level,
it is worth while to check that the parton shower and hadronisation do not induce significant 
changes with respect to the initial reweighted matrix element in high-$p_\perp$ regions. 
Fig.\ \ref{W1jet_ME_PS_HAD} proves that the predictions of \sherpa, e.g. the $p_\perp$ distribution
of the hardest jet in $W$ production, are remarkably stable in the region of matrix element
dominance.

\subsection{Variation of factorisation and renormalisation scale}
\begin{figure*}[h]
\begin{center}
\begin{tabular}{cc}
\includegraphics[width=8cm]{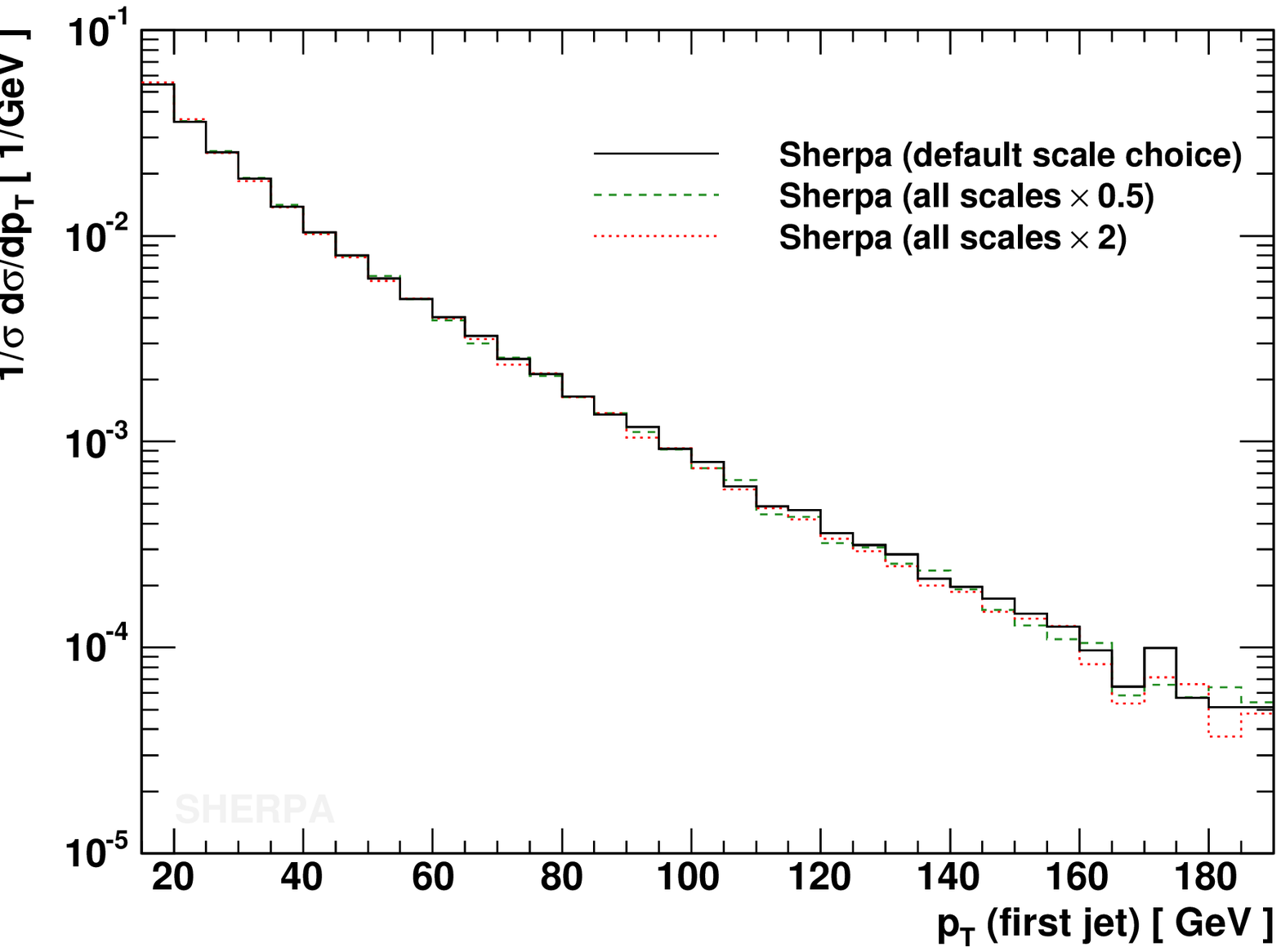}
&
\includegraphics[width=8cm]{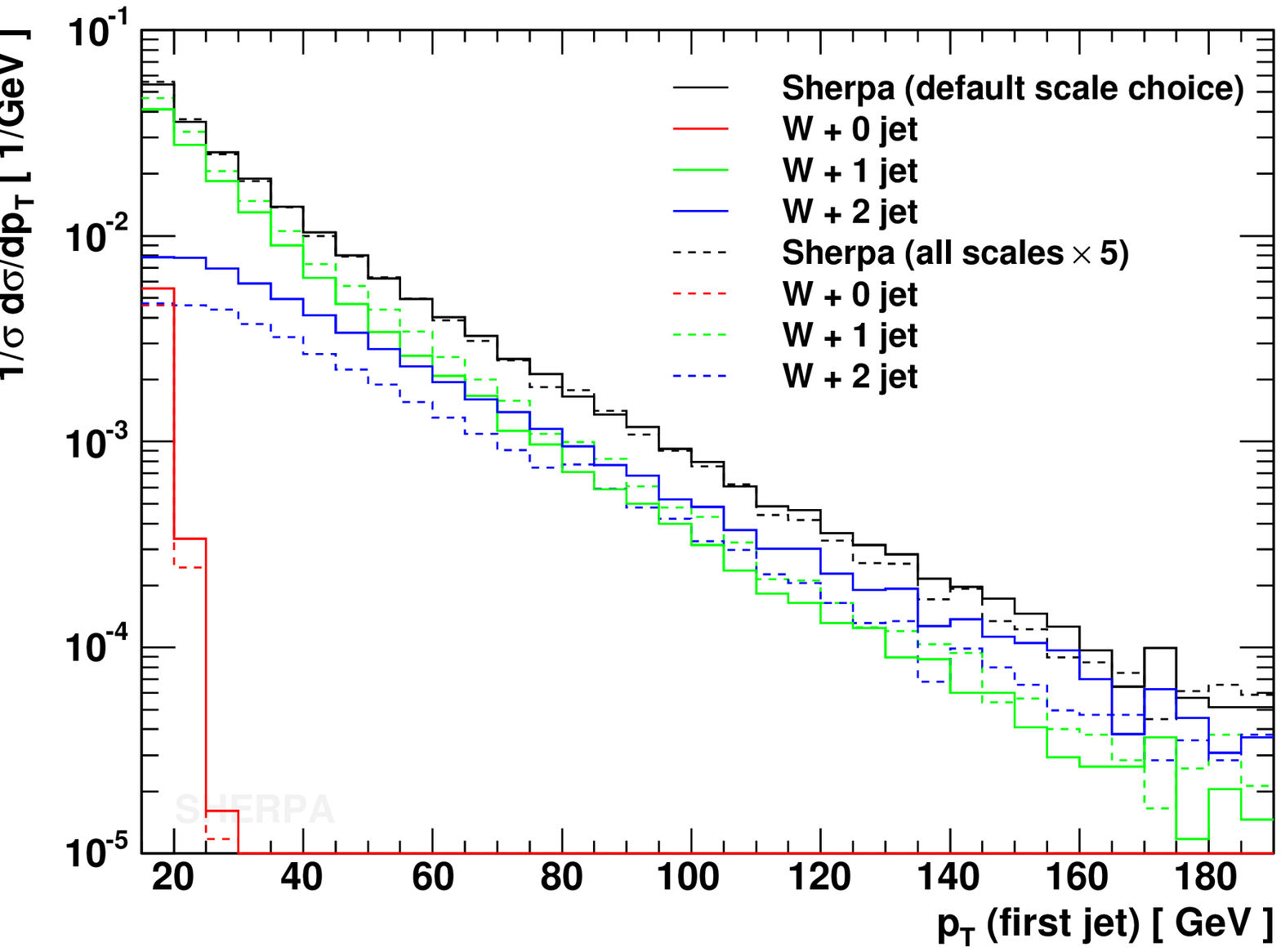}
\end{tabular}
\end{center}
\caption{\label{W1jet_scalevariation} The $p_\perp$ of the hardest jet in inclusive $W$ production 
  at the Tevatron, Run II, and its dependence on different choices in the renormalisation and 
  factorisation scale. In the left plot, the solid line indicates the default hadron level result; 
  the green dashed and the red dotted line are obtained by multiplying all scales in the coupling 
  constants and the PDFs in both the matrix elements and the parton shower by a factor of $0.5$ 
  or $2$, respectively. In the right plot, the same observable is investigated in more detail
  with the default scales (solid lines) and with all scales multiplied by a factor of $5$.
  In addition the different contributions of different jet multiplicities are shown.
}
\end{figure*}

\noindent
Finally, the sensitivity of the previous results with respect to changes in the
renormalisation and factorisation scale are examined. In the following all scales 
occurring in the event generation, both on the matrix element and at the parton
shower level, are multiplied by constant factors, ranging from $0.5$ up to $5$.
It is clear that the total cross section changes with changing scales: starting 
with 930 pb for the default scale choice, 887 pb (959 pb) are obtained when 
using a scale factor of 0.5 (2).  The shape of the 
$p_\perp$ distributions of the jets, however, experiences only mild changes. This is
greatly exemplified by the left panel of Fig.\ \ref{W1jet_scalevariation}, where the
$p_\perp$ spectrum of the hardest jet is displayed. In the right panel of Fig.\ 
\ref{W1jet_scalevariation} the result of the left panel 
is broken down for two different scale prefactors, $1$ and $5$, to the different 
contributions. Clearly, at the individual level different jet multiplicities differ 
also in their shapes; in their interplay, however, these effects cancel in terms of 
the overall shape. 

\section{\label{sec_vsworld}SHERPA vs. data and other MCs}
\noindent
In order to study the impact of the merging prescription, the predictions obtained with \sherpa\ 
may also be compared with other approaches. In a first step, in Sec.\ \ref{ssec_vsMC} the transverse 
momentum distribution of the first and second hardest jets in exclusive and inclusive boson-plus-jet 
production from \sherpa\ are confronted with the NLO QCD predictions of the parton level generator 
MCFM \cite{Campbell:2002tg,Campbell:2003hd}. In Sec.\ \ref{ssec_vsMCatNLO} the full event generators 
MC@NLO \cite{Frixione:2004wy} and PYTHIA \cite{Sjostrand:2003wg} are used to investigate the 
capabilities of \sherpa\ when studying $W/Z$+jet production at the hadron level. Finally Sec.\ 
\ref{ssec_vsData} contains  a comparison of the predictions made for the bosons transverse momentum 
distribution with those measured by the D0 and CDF collaborations at the Tevatron, Run I. 
\subsection{SHERPA vs. MCFM}\label{ssec_vsMC}
In order to compare the \sherpa\ predictions for $W/Z$+1jet and $W/Z$+2jet production, a two-step 
procedure is chosen. In a first step the Sudakov and $\alpha_s$ reweighted matrix elements are 
compared with exclusive NLO results obtained with MCFM. In the case of the next-to-leading order calculation, 
the exclusiveness of the final states boils down to a constraint on the phase space for the real parton 
emission. The exclusive \sherpa\ results consist of appropriate leading order matrix elements with 
scales set according to the $k_\perp$-clustering algorithm and made exclusive by suitable Sudakov form 
factors, cf.\ Sec.\ \ref{ssec_algo}. In a second step, the jet spectra for inclusive production 
processes are compared. For the next-to-leading order calculation, this time the phase space for real
parton emission is not restricted and the \sherpa\ predictions are obtained from a fully inclusive 
sample, using matrix elements with up to two extra jets and the parton showers attached. If not stated
otherwise, all results have been obtained using the input parameters and phase-space cuts summarised 
in the Appendix. Jets are found using the Run II $k_\perp$-clustering algorithm defined in 
\cite{Blazey:2000qt} with a pseudo-cone size of $D=0.7$ and a minimal $p_\perp$ of $15$ GeV.
\subsubsection{Exclusive jet $p_\perp$ spectra}
\begin{figure*}
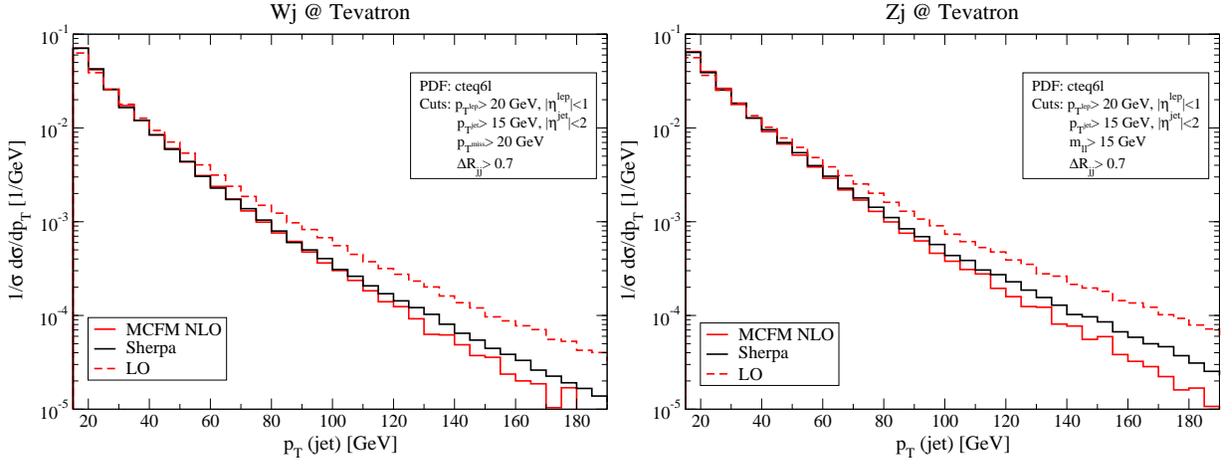

\begin{center}
\begin{tabular}{cc}
\includegraphics[width=8cm]{TeV_W1jet_excl_0.7.eps}
&
\includegraphics[width=8cm]{TeV_Z1jet_excl_0.7.eps}
\end{tabular}
\end{center}
\caption{\label{W1jet_excl_0.7} Jet $p_\perp$ distribution of exclusive $W+1$jet (left)
         or $Z+1$jet (right) events at the Tevatron, Run II.}
\end{figure*}

\noindent
In Fig.\ \ref{W1jet_excl_0.7} the jet $p_\perp$ distribution for the exclusive production of $W$+1jet 
and $Z$+1jet are shown. In both figures, the \sherpa\ prediction is compared with the exclusive NLO 
result obtained with MCFM and with the naive LO prediction, which is the same for the two programs. For 
the fixed-order NLO and LO result, the renormalisation and factorisation scales have been set to 
$\mu_R=\mu_F=80.419\;{\rm GeV}=M_W$. All distributions have been normalised to the corresponding 
total cross section. This allows for a direct comparison of the distributions shape. As stated above,
the \sherpa\ results stem from Sudakov and $\alpha_s$ reweighted $W$+1jet or $Z$+1jet LO matrix elements.
The change between the naive leading order and the next-to-leading order distribution is significant. 
At next-to-leading order the distributions become much softer. For a high-$p_\perp$ jet it is much 
more likely to emit a parton that fulfils the jet criteria and therefore removes the event from the 
exclusive sample. The \sherpa\ predictions show the same feature. The inclusion of Sudakov form 
factors and the scale setting according to the merging prescription improves the LO prediction, 
resulting in a rather good agreement with the next-to-leading order result. 

\begin{figure}
\begin{center}
\begin{tabular}{cc}
\includegraphics[width=8cm]{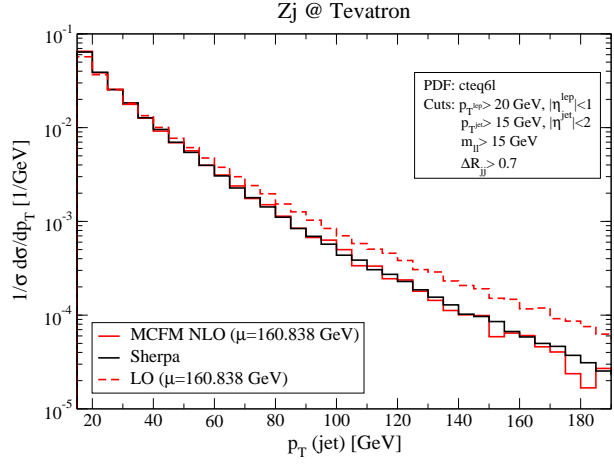}
\end{tabular}
\end{center}
\caption{\label{Z1jet_excl_0.7_scale2} Jet $p_\perp$ distribution of $Z+1$jet
events at the Tevatron where for the NLO and LO calculation the 
renormalisation and factorisation scales have been chosen to be 
$\mu_R=\mu_F=160.838$ GeV.}
\end{figure}

\noindent
In the high-$p_\perp$ tail, however, the NLO calculations from MCFM tend to be a bit below the 
\sherpa\ results. The reason is simply connected to the fact that relevant scales in the 
high-$p_\perp$ tail are much larger than the default choice of $\mu_R=\mu_F=80.419$ GeV. In
order to highlight this, Fig.\ \ref{Z1jet_excl_0.7_scale2} contains the jet $p_\perp$ distribution 
in $Z$+1jet events. In this plot, the renormalisation and factorisation scales have been chosen 
to be $\mu_R=\mu_F=160.838\;{\rm GeV}=2M_W$. Changing the scale in this manner indeed has quite a 
small impact on the total cross section at NLO, but the tail of the distribution becomes 
considerably enhanced. With the above choice of $\mu_R$ and $\mu_F$ the agreement of NLO and 
the \sherpa\ result is impressive. 

\begin{figure*}
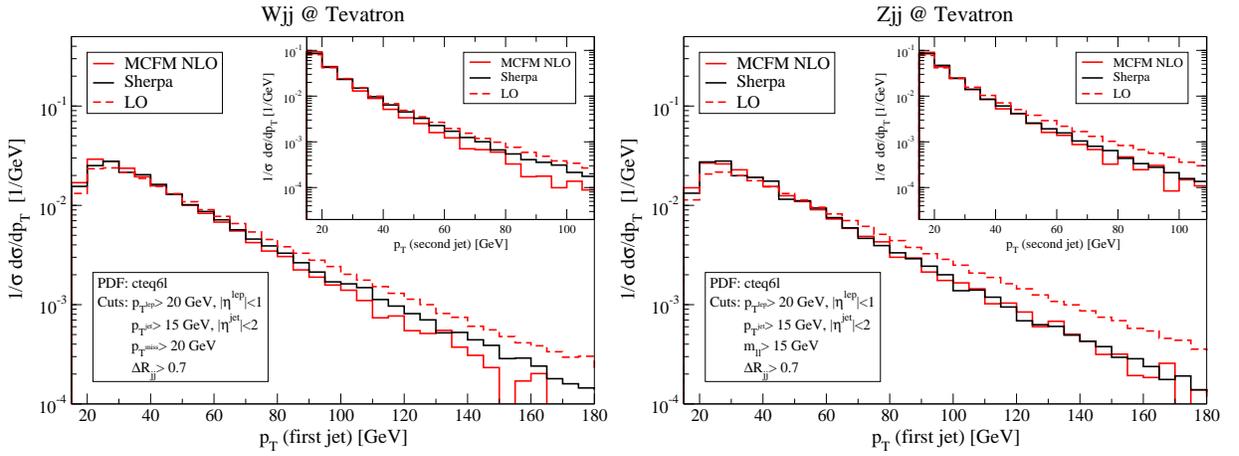

\begin{center}
\begin{tabular}{cc}
\includegraphics[width=8cm]{TeV_W2jet_excl_0.7.eps}
&
\includegraphics[width=8cm]{TeV_Z2jet_excl_0.7.eps}
\end{tabular}
\end{center}
\caption{\label{W2jet_excl_0.7} The $p_\perp$ distribution of the first and second jets in exclusive 
  $W+2$jet (left) and in exclusive $Z+2$jet (right) events at the Tevatron, Run II.}
\end{figure*}

\begin{figure}
\begin{center}
\begin{tabular}{cc}
\includegraphics[width=8cm]{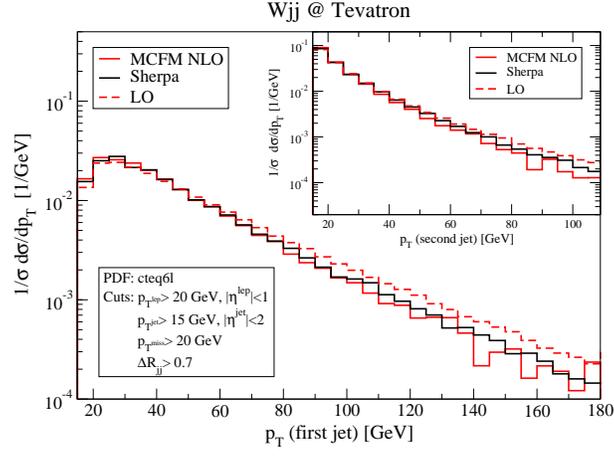}
\end{tabular}
\end{center}
\caption{\label{W2jet_excl_0.7_scale2} The $p_\perp$ distribution of the
  first and second jets in exclusive $W+2$jet events at the
  Tevatron where for the NLO and LO calculation the 
renormalisation and factorisation scales have been chosen to 
$\mu_R=\mu_F=160.838$ GeV.}
\end{figure}

\noindent
The $p_\perp$ distribution of the first and second jets in $W$+2jet and $Z$+2jet production are 
presented in Fig.\ \ref{W2jet_excl_0.7}. Again, the next-to-leading oder distributions are softer 
than the leading order ones, for the same reason as for the 1jet case. In addition, at low-$p_\perp$ 
the leading order result is smaller than the next-to-leading order one. Taken together, the curves 
have a significantly different shape over the whole interval. This
situation clearly forbids the use of  
constant $K$-factors in order to match the leading order with the next-to-leading order result. 
Nevertheless, as before, the \sherpa\ prediction reproduces to a very good approximation the shape of 
the NLO result delivered by MCFM. Fig.\ \ref{W2jet_excl_0.7_scale2} shows that, similar to the $Z$+1jet 
case for $W$+2jet in the high-$p_\perp$ tail, the situation is even better using higher 
renormalisation and factorisation scales (e.g.\ $\mu_R=\mu_F=160.838$ GeV) in the NLO calculation.

\subsubsection{Inclusive jet $p_\perp$ spectra}
\noindent
NLO results for inclusive boson plus jet(s) production obtained with MCFM are compared with fully 
inclusive samples generated with \sherpa. There, the matrix elements for $W/Z$+0,1,2jet production
have been used including the highest multiplicity treatment for the $W/Z$+2jet case. The Sudakov 
and $\alpha_s$ reweighted matrix elements have now been combined with the initial and final 
state parton showers. The hadronisation phase for the \sherpa\ events has been discarded. As for 
the exclusive case the naive leading order prediction is given by the corresponding leading order 
matrix element that is identical to the one in Figs. \ref{W1jet_excl_0.7} and \ref{W2jet_excl_0.7}. For 
the NLO prediction again the renormalisation and factorisation scales have been chosen to coincide, 
namely $\mu_R=\mu_F=80.419$ GeV. 

\begin{figure*}
\begin{center}
\begin{tabular}{cc}
\includegraphics[width=8cm]{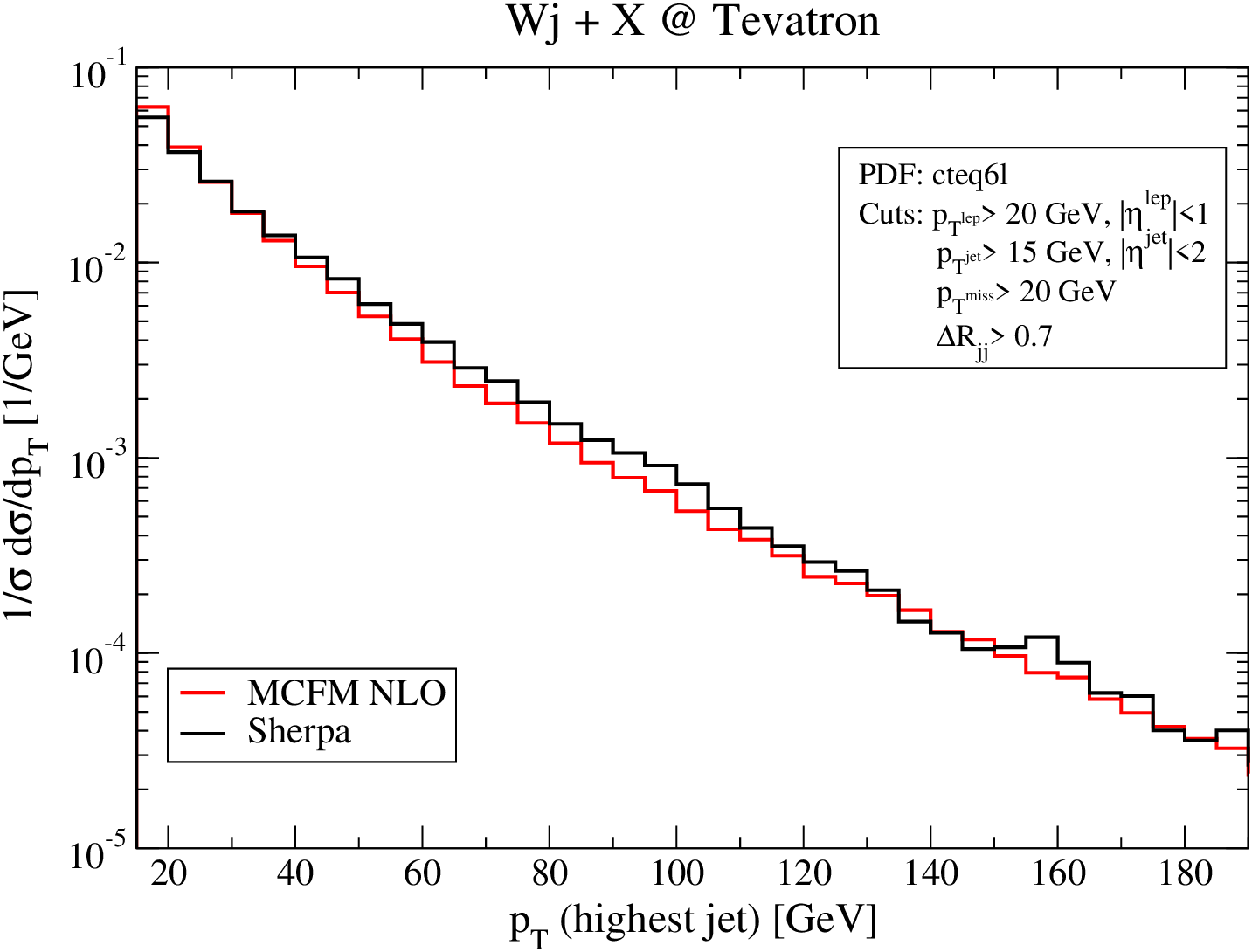}
&
\includegraphics[width=8cm]{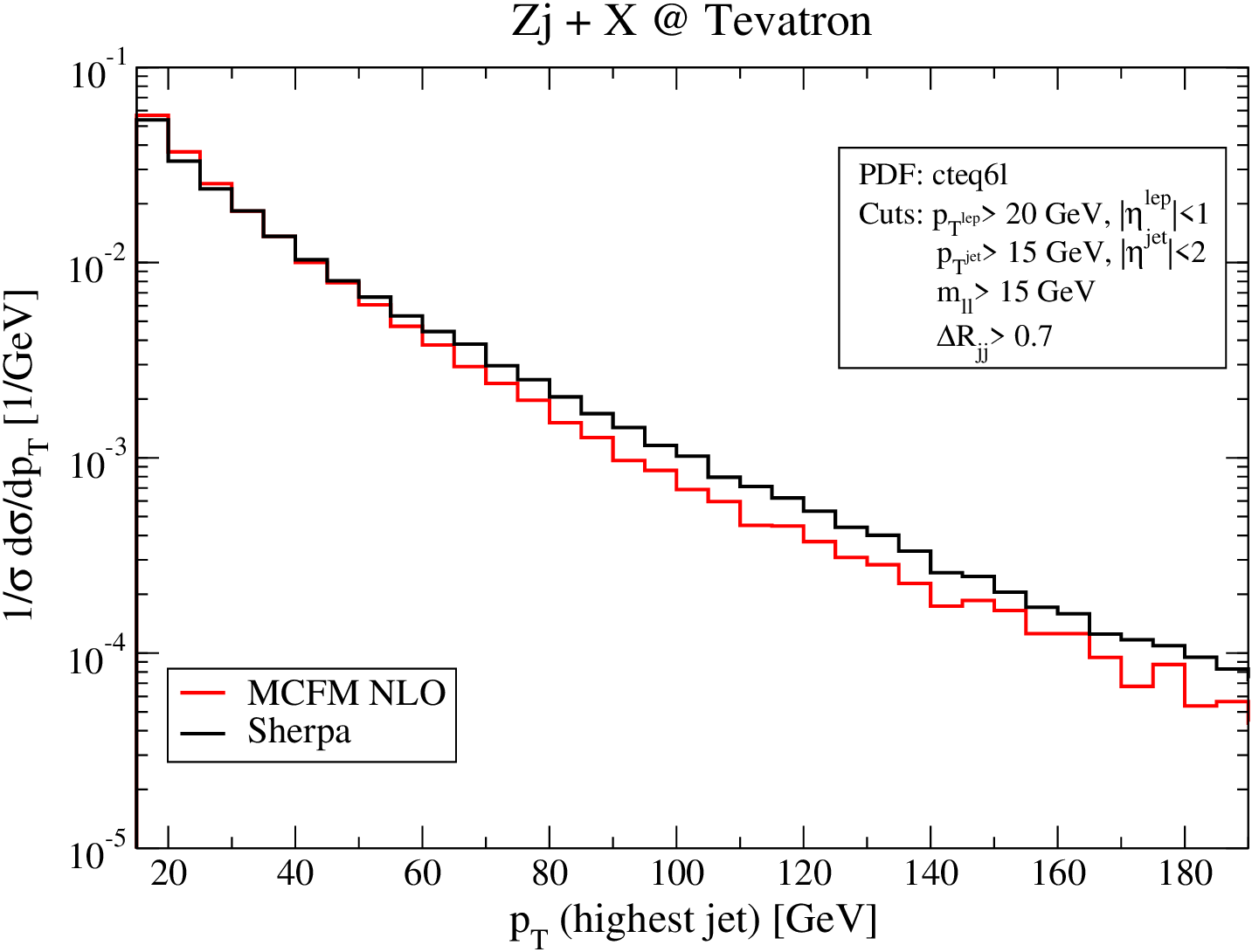}
\end{tabular}
\end{center}
\caption{\label{W1jet_incl} The $p_\perp$ distribution of the hardest jet for inclusive $W+1$jet
(left) and for inclusive $Z+1$jet (right) production at the Tevatron, Run II.}
\end{figure*}

\noindent
In Fig.\ \ref{W1jet_incl}, the $p_\perp$ spectra for the hardest jet in inclusive $W/Z$+1jet production 
are shown. Compared with the exclusive predictions, the high-$p_\perp$ tail is filled again and, hence, the 
differences between the NLO calculations and the LO ones appear to be smaller. For both cases the 
\sherpa\ result and the NLO calculation are in good agreement.

\begin{figure*}
\begin{center}
\begin{tabular}{cc}
\includegraphics[width=8cm]{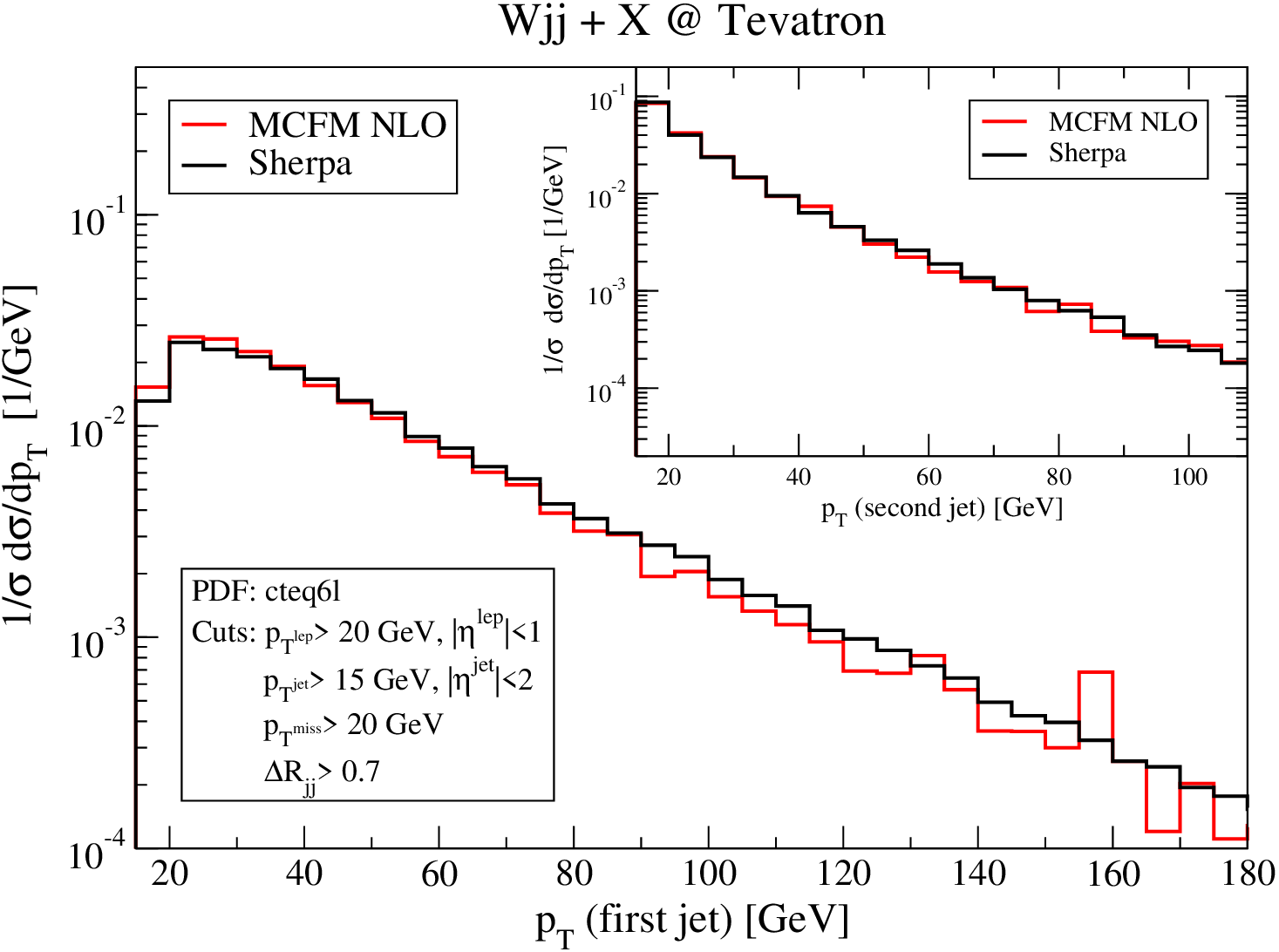}
&
\includegraphics[width=8cm]{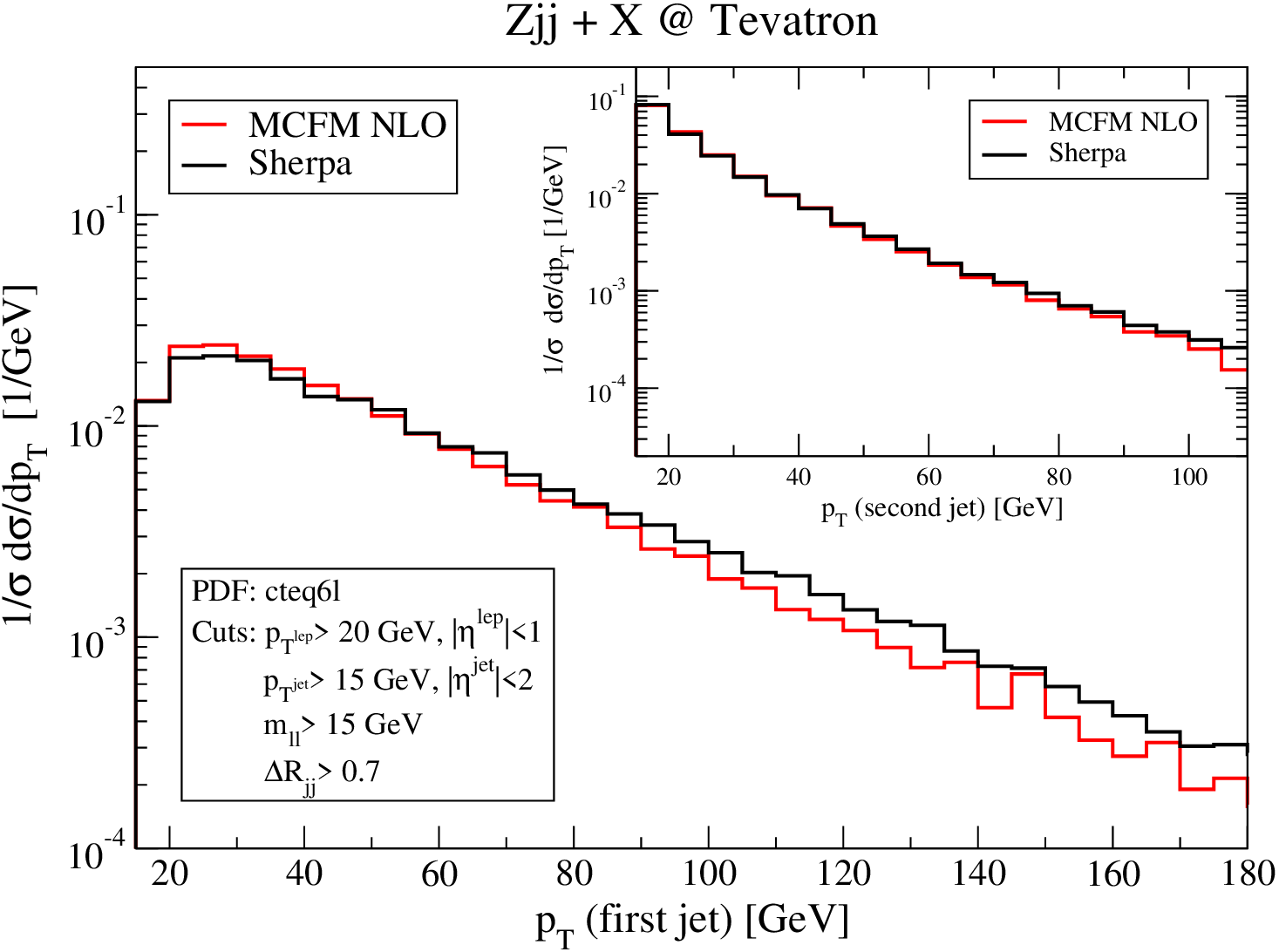}
\end{tabular}
\end{center}
\caption{\label{W2jet_incl}  The $p_\perp$ distribution of the hardest two jets for inclusive $W+2$jet (left) 
and for inclusive $Z+2$jet (right) production at the Tevatron, Run II.}
\end{figure*}

\noindent
In Fig.\ \ref{W2jet_incl} the $p_\perp$ spectra for the first and second hardest jets in inclusive 
$W/Z$+2jet production are presented. Considering the scale dependence of the next-to-leading order 
result in the high-$p_\perp$ region, as already studied in Fig.\ \ref{W2jet_excl_0.7_scale2} 
for the exclusive result, the curves are in pretty good agreement.

\noindent
Altogether, the merging procedure in \sherpa, including the scale-setting prescription of 
the approach and the Sudakov reweighting of the LO matrix elements, proves to lead to a
significantly improved leading order prediction. Seemingly, it takes proper care of the
most relevant contributions of higher order corrections. Although it should 
be stressed that the rate predicted by \sherpa\ is still a leading order value only, a 
constant $K$-factor is sufficient to recover excellent agreement with a full next-to-leading 
order calculation for the distributions considered. Furthermore, by looking at the inclusive 
spectra it is obvious that this statement still holds true after the inclusion of parton 
showers and the merging of exclusive matrix elements of different jet multiplicities.

\subsection{SHERPA vs. MC@NLO and PYTHIA}\label{ssec_vsMCatNLO}

\begin{figure*}
\begin{center}
\begin{tabular}{cc}
\includegraphics[width=8cm]{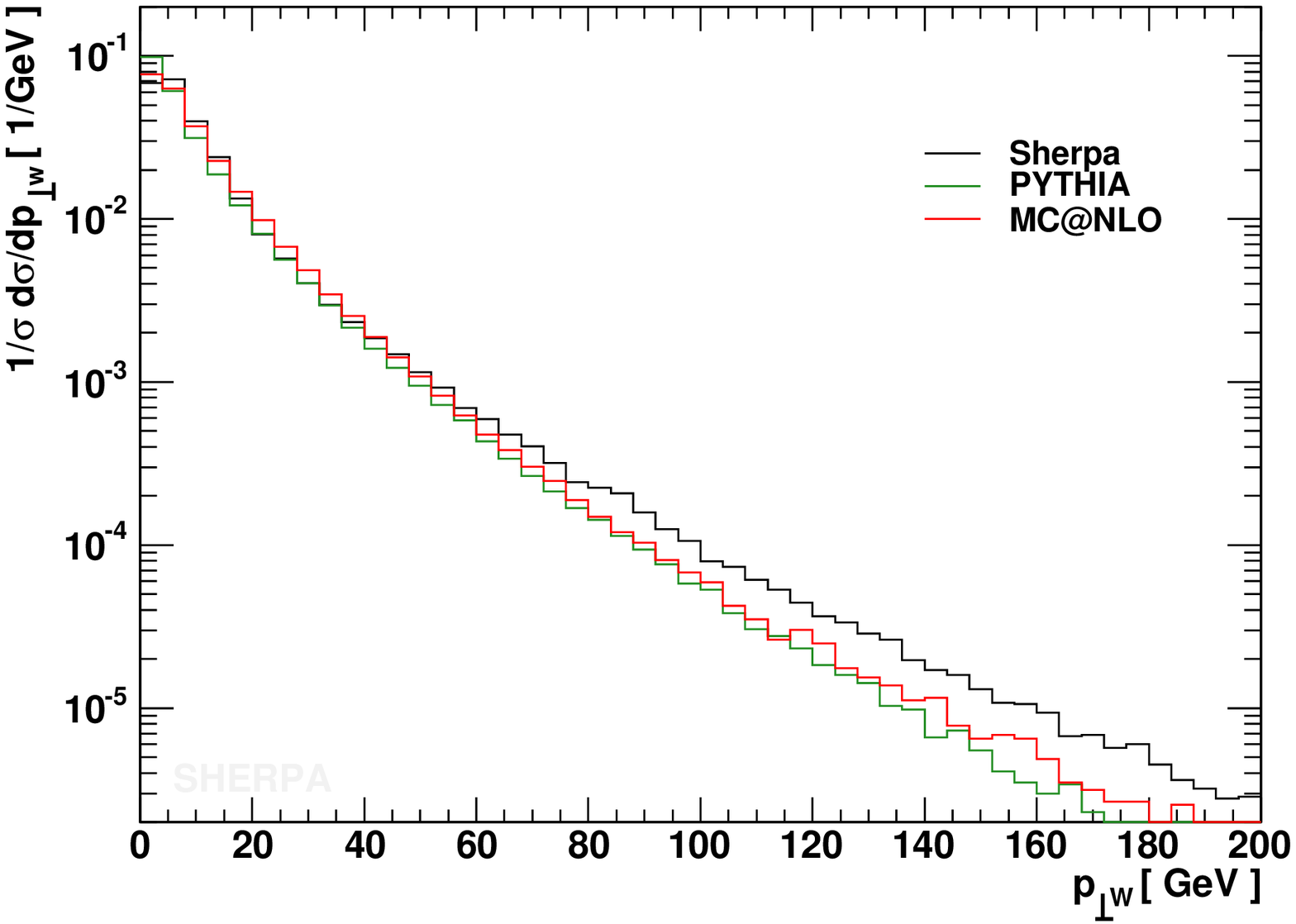}&
\includegraphics[width=8cm]{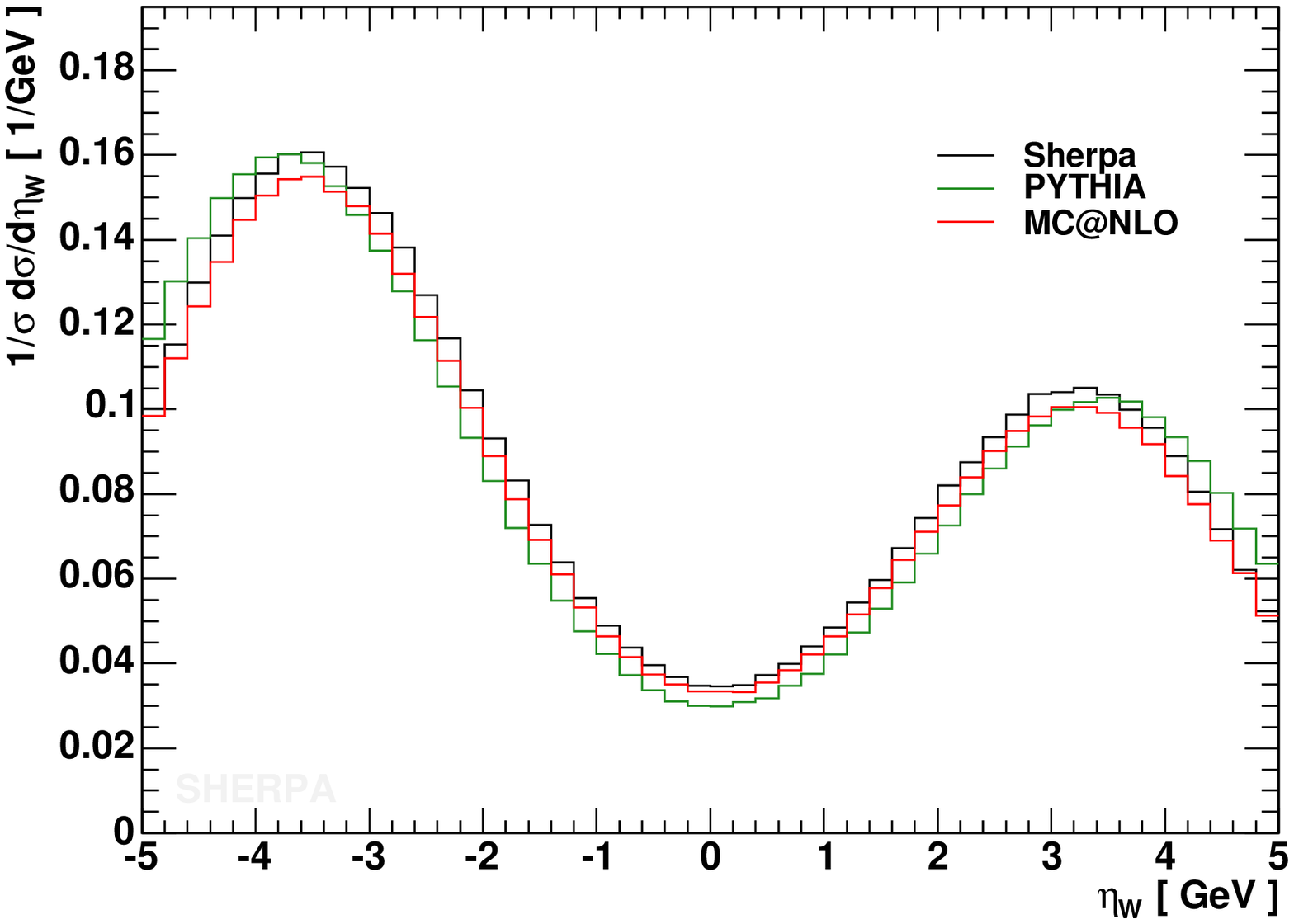}
\end{tabular}
\end{center}
\caption{\label{Wpt_eta_sh_mcs}  The $p_\perp$ (left) and $\eta$ (right) distribution of the 
$W^-$ boson for inclusive production at the Tevatron, Run II. Plotted are the results as obtained 
with the generators PYTHIA (green), MC@NLO (red) and \sherpa\ (black).}
\end{figure*}

\begin{figure*}[h]
\begin{center}
\begin{tabular}{ccc}
\includegraphics[width=6cm]{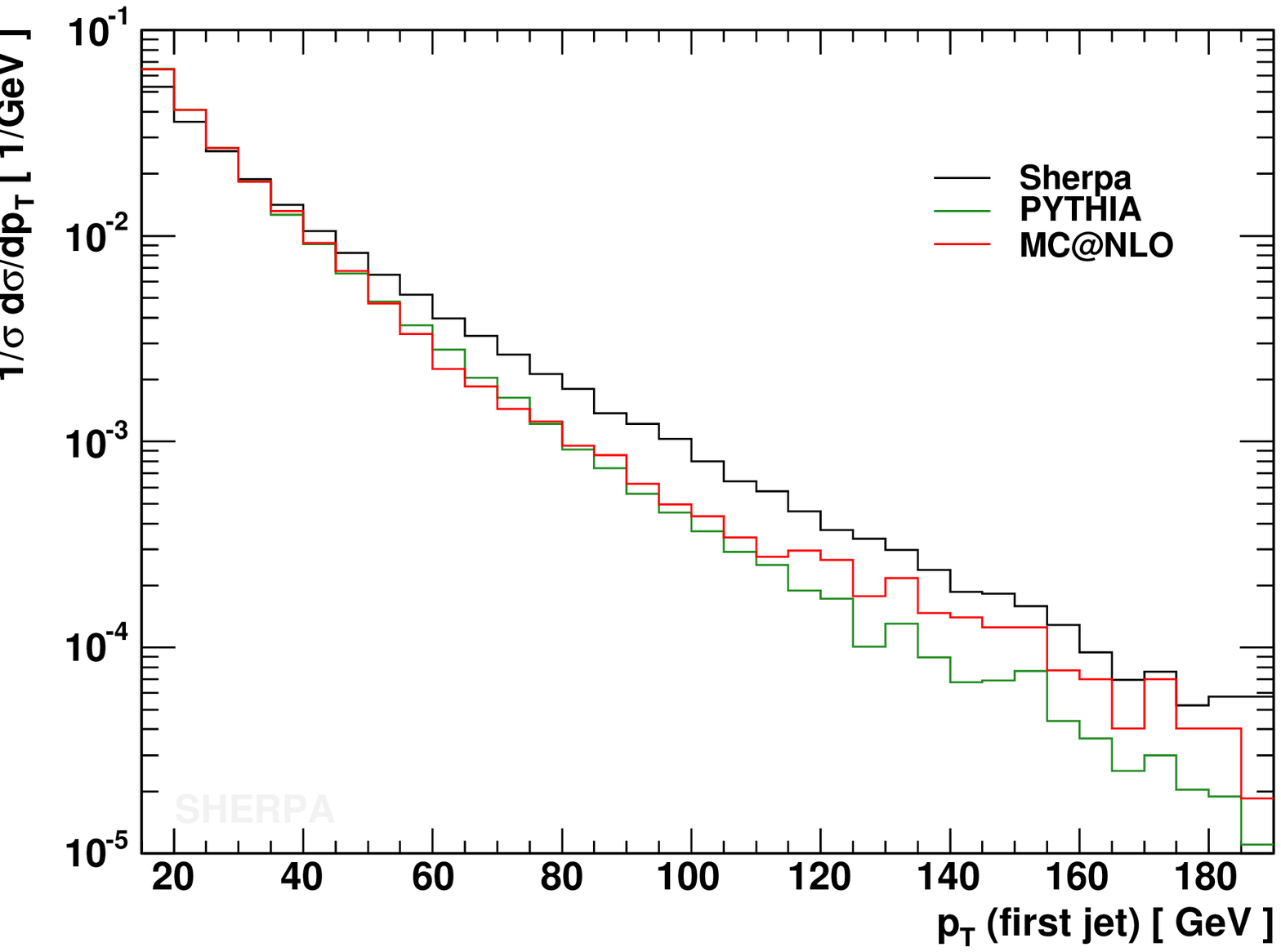}
&
\includegraphics[width=6cm]{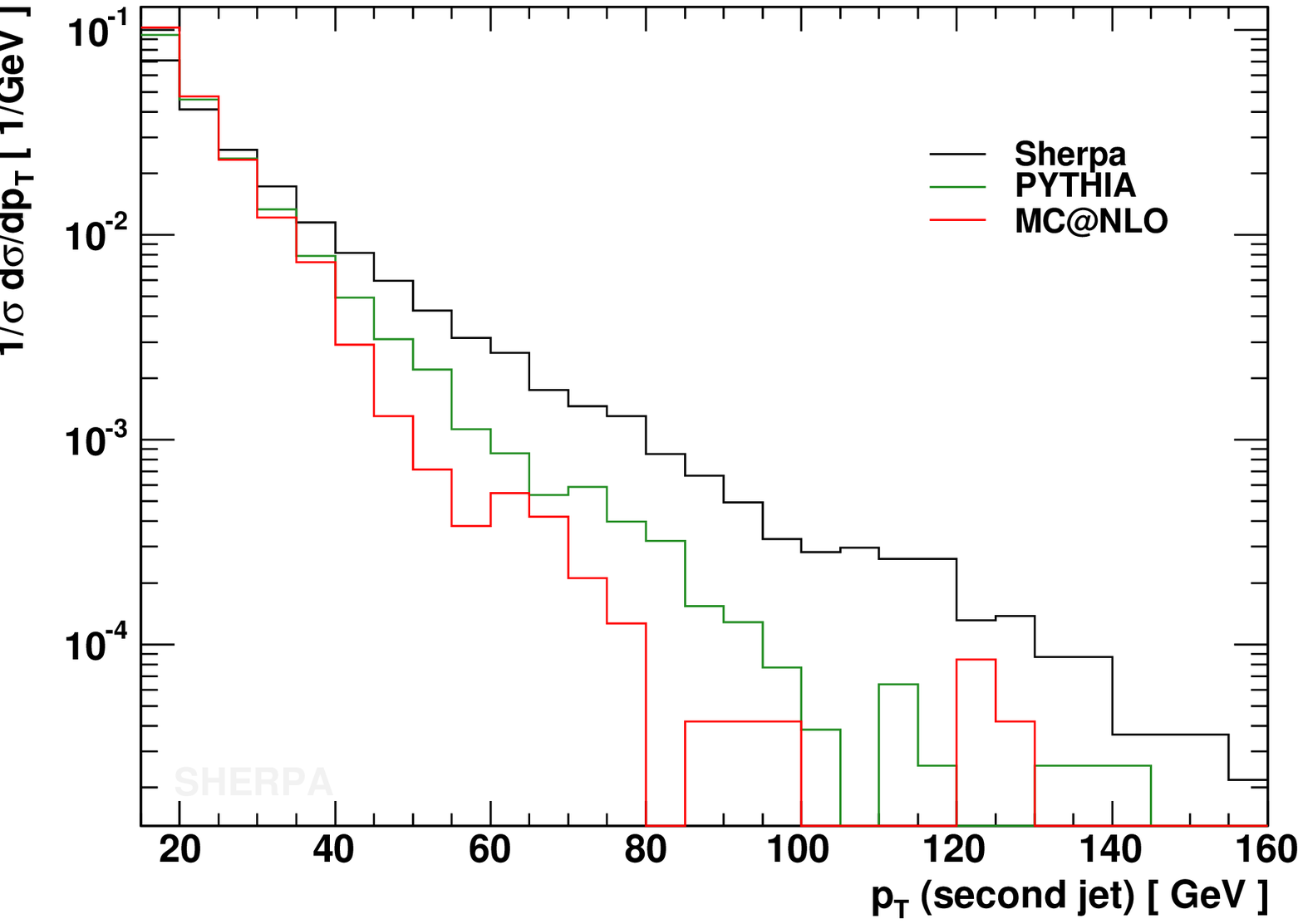}
&
\includegraphics[width=6cm]{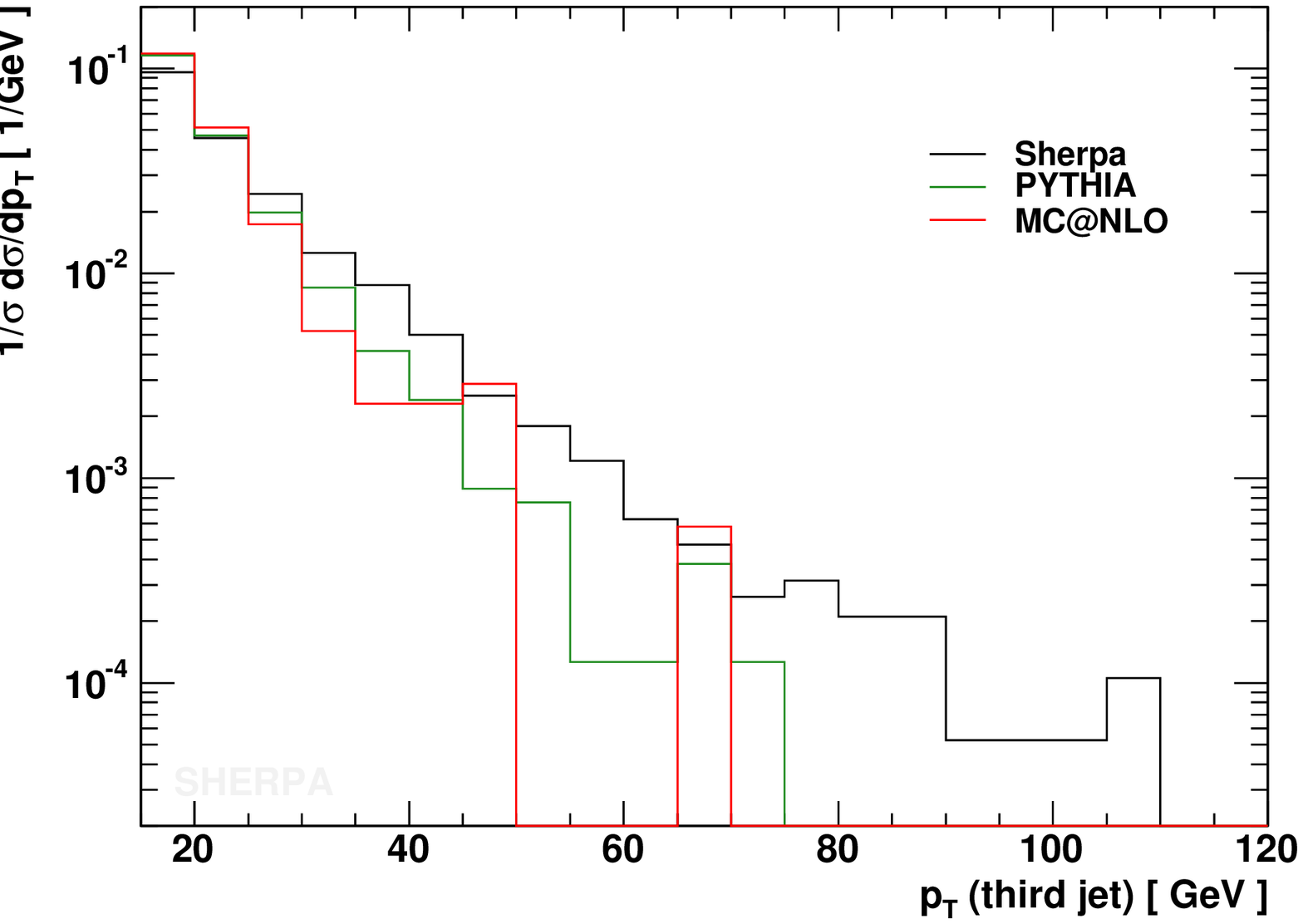}
\end{tabular}
\caption{\label{W1jet_py} Jet $p_\perp$ distribution of the three hardest jets in
  inclusive $W$ production at the Tevatron, Run II. Compared are the hadron level results 
of \sherpa\ (black), PYTHIA (green) and MC@NLO (red) after 2.5 million events. }
\end{center}
\end{figure*}

\noindent
In a next step, results obtained with \sherpa\ are compared with those obtained from two other 
event generators, namely PYTHIA and MC@NLO. In both, standard 
settings have been used for inclusive $W$ production at the Tevatron and the underlying
event has been switched off. The corresponding process number in PYTHIA is {\tt MSEL=12}, the 
relevant MC@NLO process is {\tt IPROC=-1471}. Inclusive quantities like, for instance, 
the $p_{\perp}$ and $\eta$ distributions of the $W$ are in good agreement, see
Fig.\ \ref{Wpt_eta_sh_mcs}, and only in the high-$p_\perp$ tail of the distribution some small
deviations become visible. However, more exclusive quantities such as the $p_{\perp}$-distributions 
of the first three jets show differences that increase with the increasing order of the jet. 
This can clearly be seen from Fig.\ \ref{W1jet_py}. The predictions for the hardest jet
start to disagree with a factor of roughly 2 at jet-$p_\perp$s of the order of 100 GeV, reaching 
up to nearly an order of magnitude at $p_\perp$ around 200 GeV. This trend is greatly enhanced 
for the second and third jets, where discrepancies are of the order of one magnitude for the 
second jet at $p_\perp \approx 100$ GeV or even higher for the third jet.

\noindent
These discrepancies, however, were to be expected since the other two programs do not 
include any higher order correction beyond first order in the strong coupling constant. In 
the case of MC@NLO, predictions have been compared with those obtained from MCFM; after 
a careful calibration of input parameters such as CKM elements etc., inside the code 
both programs coincided in all observables tested \cite{Frixione:private}. Therefore, 
differences in the $p_\perp$ distribution of the hardest jet have to be attributed to
a combination of distinct parameter settings and of differences in parton showering and
hadronisation. The latter type of difference should be taken as some kind of theoretical 
uncertainty. For higher jet configurations, however, the remaining discrepancies are due
to different physics input. 

\subsection{SHERPA vs. data}\label{ssec_vsData}

\noindent
Having compared the \sherpa\ predictions for the case of the jet transverse momentum distributions 
in exclusive and inclusive $W/Z$+1jet and $W/Z$+2jet production against other Monte Carlo programs, 
a comparison with experimental data provides an ultimate test of \sherpa's ability to describe such 
processes. Unfortunately, so far only the inclusive $W$- and $Z$-boson transverse momentum 
distribution measured in Tevatron, Run I, have been published, which allows for an overall check only.
In both cases, matrix elements with up to four ($W$) or three ($Z$) extra jets have been taken 
into account -- as indicated by the different colours - to generate the \sherpa\ sample. The black 
line represents the sum of all contributions. For this sample the required separation cut has been 
chosen to $Q_{\rm cut}=20$ GeV.

\begin{figure}[h]
\begin{center}
\includegraphics[width=8cm]{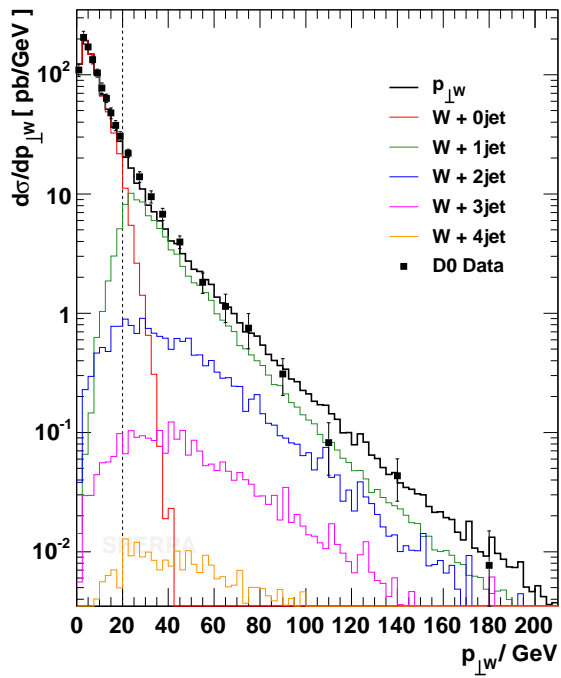}
\end{center}
\caption{\label{W_data} The $p_\perp$ distribution of the $W$-boson in comparison with data from D0 at
the Tevatron, Run I \cite{Abbott:2000xv}. The total result is indicated by the black line. 
The coloured lines show the contributions of the different multiplicity 
processes. Here matrix elements with up to four extra jets have been considered. 
The applied separation cut is $Q_{\rm cut}=20$ GeV. }
\end{figure}

\noindent
In Fig.\ \ref{W_data}, the (inclusive) $p_\perp$ distribution of the $W$ is compared with data
from D0, taken at Run I of the Tevatron \cite{Abbott:2000xv}. The agreement with data is excellent.
It can be recognised that approaching the merging scale from below, the $W$+0jet contribution 
steeply falls and the distribution for larger momenta is mainly covered by the $W$+1jet part, as 
expected. In order to match the measured distribution, the \sherpa\ result has been multiplied by 
a constant $K$-factor of $1.25$.

\begin{figure}[h]
\begin{center}
\begin{tabular}{cc}
\includegraphics[width=8cm]{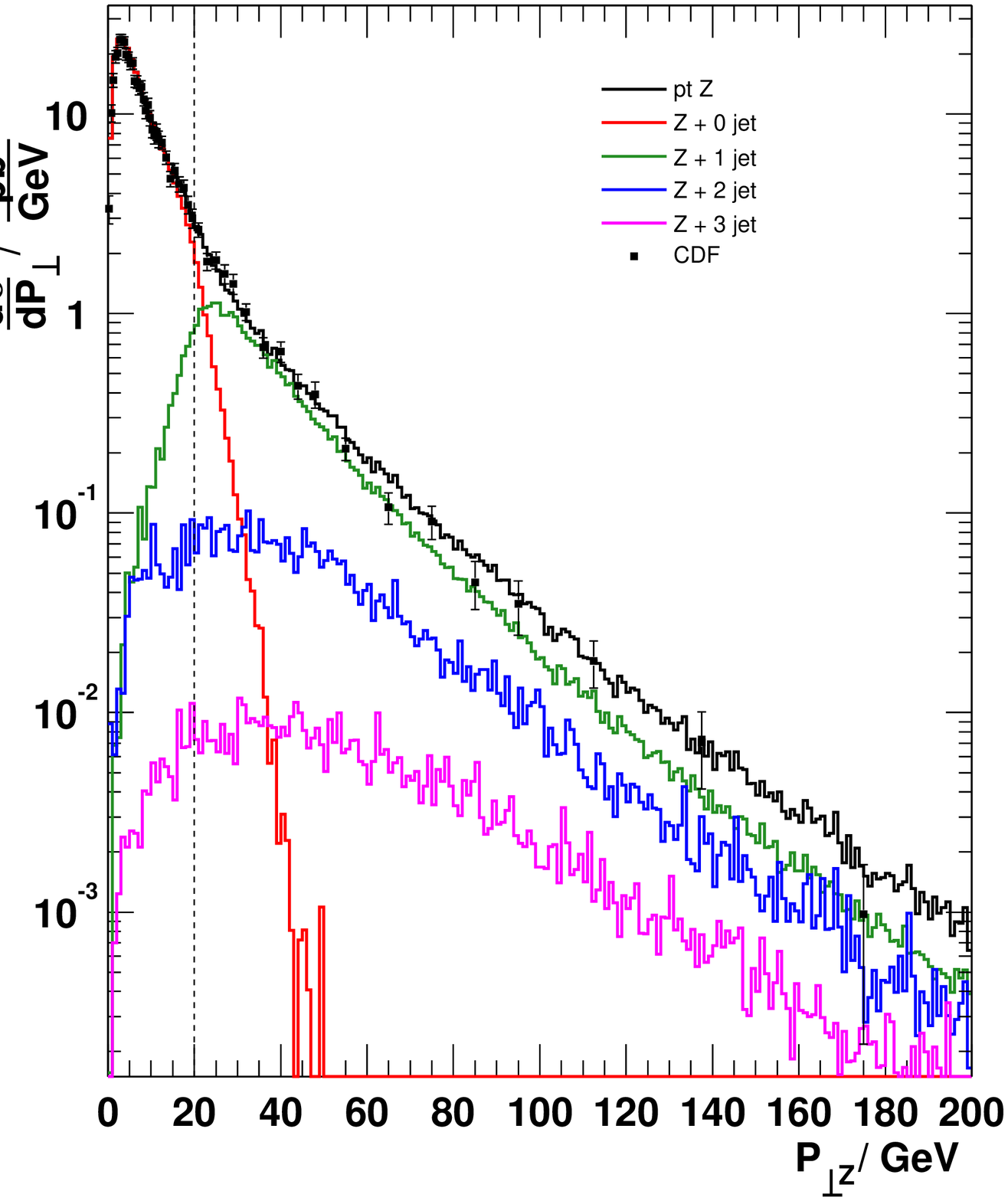}
\includegraphics[width=8cm]{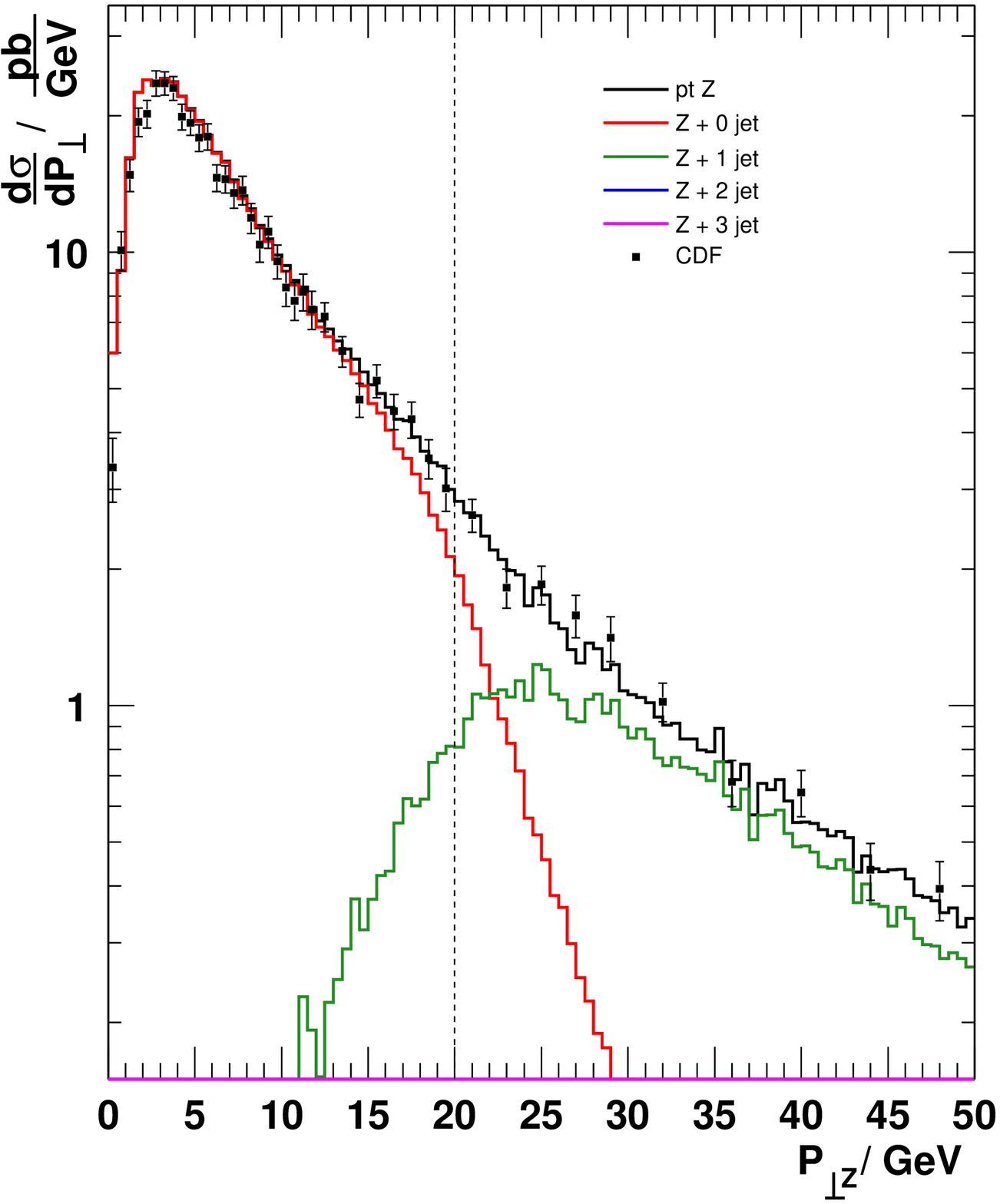}
\end{tabular}
\end{center}
\caption{\label{Z_data} The $p_\perp$ distribution of the $Z$-boson in comparison with data from CDF at
the Tevatron, Run I \cite{Affolder:1999jh}. The total result is indicated by the black line. 
The coloured lines show the contributions of the different multiplicity 
processes. The applied separation cut is $Q_{\rm cut}=20$ GeV. The right plot focuses on the 
low momentum region of the left one.}
\end{figure}

\noindent
Similarly, in Fig.\ \ref{Z_data}, the (inclusive) $p_\perp$ distribution of the $Z$ is compared 
with data, this time taken by CDF at Run I of Tevatron \cite{Affolder:1999jh}. Again the overall 
agreement is excellent. This time the result has been multiplied by a constant $K$-factor of $1.6$
to match the data. The result is perfectly smooth around the merging scale of $Q_{\rm cut}=20$ GeV. 
This is especially highlighted in the left plot of Fig.\ \ref{Z_data}, which  
concentrates on the low momentum region. It is interesting to note that the description of the 
data for momenta smaller than the merging scale is almost only covered by the $Z$+0jet contribution 
and is therefore very sensitive to the details of the parton showers and the treatment of beam 
remnants. A parameter of specific impact on the very low momentum region therefore is the primordial 
(or intrinsic) $k_\perp$ used for the interacting partons. This is modelled through a Gaussian 
distribution with a central value of $0.8$ GeV. Nevertheless, the shower performance of \sherpa\ 
has not been especially tuned; the low momentum behaviour may therefore still be improved once a detailed 
parameter tune is available. 

\section{Conclusion}
\noindent
In this work, predictions for single-boson production processes at the Tevatron have been obtained 
from the new event generator \sherpa\ and compared with the results from other programs and with
data. In all cases an encouraging agreement of results has been found. Especially for the 
description of exclusive multi particle final states, \sherpa\ proved its unique value as the 
simulation tool of choice, limited only by the maximal number of external particles covered
by its intrinsic matrix elements. In practical applications the choice of this input should be 
guided by the question in consideration. For instance, for the simulation of $W+4$jet background 
to top pair production, a highest jet multiplicity of 4 is advisable in order to include most of 
the quantum interferences.

\noindent
Having validated the versatility of \sherpa\ on one of the most important processes at hadron
colliders, its abilities will be further tested in the near future by considering other important 
processes, such as the production of boson pairs, the Higgs boson, heavy quarks, or multijet 
final states.

\section*{Acknowledgements}
\noindent
The authors thank BMBF, DFG, and GSI for financial support. The authors are grateful for
pleasant and fruitful discussions with John Campbell, Stefano Frixione, Michelangelo Mangano and 
Bryan Webber. The organisers of the MC4LHC workshop at CERN, where this work has been started, are 
acknowledged for their kind hospitality.\\[1cm]

\appendix
\section{Input parameters and phase-space cuts}\label{appendix}
\noindent
The PDF set used for all analyses is cteq6l \cite{Pumplin:2002vw}. 
The value of $\alpha_s$ is chosen according to the value taken for 
the PDF, namely $0.118$. For the running of the strong coupling 
the corresponding two-loop equation is used. Jets or initial partons 
are restricted to the light flavour sector, namely $g, u, d, s , c$. 
In fact these flavours are taken to be massless and the Yukawa 
couplings of the quarks are neglected throughout the entire analysis.
\subsubsection{SM input parameters}
\noindent
The SM parameters are given in the $G_{\mu}$ scheme:
\begin{eqnarray}
&&m_W = 80.419\;{\rm GeV}\,,\quad \Gamma_W = 2.06\;{\rm GeV,}\nonumber\\
&&m_Z = 91.188\;{\rm GeV}\,,\quad \Gamma_Z = 2.49\; {\rm GeV,}\nonumber\\
&&G_{\mu} = 1.16639 \times 10^{-5}\; {\rm GeV}^{-2},\nonumber\\
&&\sin^2\theta_W = 1 - m^2_W/m^2_Z = 0.2222,\nonumber\\
&&\alpha_s = 0.118\,.
\end{eqnarray}
\noindent
The electromagnetic coupling is derived from the Fermi constant $G_{\mu}$
according to
\begin{eqnarray}
&&\alpha_{\rm em} = \frac{\sqrt{2}\,G_{\mu}\,M^2_W\,\sin^2\theta_W}{\pi} = 1/132.51\,.
\end{eqnarray}
\noindent
The constant widths of the electroweak gauge bosons are introduced via the 
fixed-width scheme. CKM  mixing of the quark generations is neglected.

\noindent
\subsubsection{Cuts and jet criteria}
\noindent
For all jet analyses the Run II $k_\perp$-clustering algorithm defined in \cite{Blazey:2000qt} is used. 
The parameter of this jet algorithm is a pseudo-cone of size $D$ given below for the Tevatron analysis. 
For the charged leptons the following cuts are applied:
\begin{eqnarray}
p^{\rm lepton}_\perp > 20\;{\rm GeV},\quad |\eta^{\rm lepton}| < 1,\quad m^{ll} > 15\; {\rm GeV}.
\end{eqnarray}
\noindent
For the case of $W$ production an additional cut on missing transverse momentum according to the 
neutrino has been required, namely 
\begin{eqnarray}
p^{\rm miss}_\perp > 20\; {\rm GeV}.
\end{eqnarray}
\noindent
For the jet definition a pseudo-cone size of $D=0.7$ has been used in addition to cuts on 
pseudo-rapidity and transverse momentum:
\begin{eqnarray}
p_\perp^{\rm jet} > 15\; {\rm GeV}, \quad |\eta^{\rm jet}| < 2.
\end{eqnarray}
%

%

\end{document}